\documentclass[aip, jcp,showpacs,amsmath,amssymb]{revtex4-1}
\usepackage{graphicx}
\usepackage{color}
\usepackage{epsfig}
\usepackage{dcolumn}
\usepackage{bm}
\usepackage{mathrsfs}
\usepackage{multirow}
\usepackage{MnSymbol}
\usepackage[all]{xy}
\usepackage{pbox}

\def\(({\left(}
\def\)){\right)}
\def\[[{\left[}
\def\]]{\right]}
\def\dd{\text{d}}
\def\DD{\text{D}}

\def\segment#1#2{
\hspace{0.08cm} 
  \pbox[c]{2cm}{  
    \xymatrix@=0.5pc@C=1.25pc{  *=0{#1} \ar@{-}[r] *{#2} & }
  }
}

\def\dipod#1#2#3{
  \pbox[c]{2cm}{  
    \xymatrix@=0.5pc@C=0.25pc{  & *=0{#1} \ar@{-}[dr] *{#3} \ar@{-}[dl] *{#2} \\ 
      *=0{#2}   & & *{#3} }
  }
}

\def\triangle#1#2#3{
  \pbox[c]{2cm}{       
  \xymatrix@=0.5pc@C=0.25pc{  & *=0{#1} \ar@{-}[dr] *{#3} \ar@{-}[dl] *{#2} \\ 
                *=0{#2} \ar@{-}[rr] *{#3} & & }
  }
}

\def\arch#1#2#3#4{
  \pbox[c]{2cm}{  
  \xymatrix@=1pc{ *=0{#1} \ar@{-}[d] *{#3} \ar@{-}[r] *{#2} & *=0{#2} \ar@{-}[d] *{#4} \\ 
      *{#3} &  *{#4}}         
  }
}

\def\tripod#1#2#3#4{
  \pbox[c]{2cm}{  
    \xymatrix@=0.5pc@C=0.25pc{  & *=0{#1} \ar@{-}[dr] *{#4} \ar@{-}[d] *{#3} \ar@{-}[dl] *{#2} \\ 
      *=0{#2}   & *{#3} & *{#4} }
  }
}

\def\square#1#2#3#4{
  \pbox[c]{2cm}{  
  \xymatrix@=1pc{ *=0{#1} \ar@{-}[d] *{#3} \ar@{-}[r] *{#2} & *=0{#2} \ar@{-}[d] *{#4} \\ 
      *{#3} \ar@{-}[r] *{#4} &  *{#4}}         
  }
}

\def\pentagon#1#2#3#4#5{
 \pbox[c]{2cm}{
  \xymatrix@=0.5pc@C=-0.05pc{ & & *=0{#1} \ar@{-}[dll] *{#2} \ar@{-}[drr] *{#3} & & \\ 
  *=0{#2} \ar@{-}[dr] *{#4} & & & & *=0{#3} \ar@{-}[dl] *{#5} \\ 
     & *{#4} \ar@{-}[rr] *{#5} &  & *{#5} & }
   }
}

\begin{document}

\everymath{\displaystyle}

\title{Dynamical transition of glasses: from exact to approximate. }

\author {Romain Mari, Jorge Kurchan
} \affiliation{ CNRS; ESPCI, 10 rue Vauquelin, UMR 7636, Paris, France 75005,
  PMMH \\
 }

\begin{abstract}
  We introduce a family of glassy models having a 
  parameter,  playing the role of an interaction range, that may be varied   continuously to
  go from  a   system of particles in $d$ dimensions to a mean-field version of it.   The mean-field  limit is exactly described by equations conceptually close,  but different  from, the Mode-Coupling equations. We obtain these by a {\em dynamic virial}
  construction. 
  Quite surprisingly we observe
  that in three dimensions, the mean-field behavior is closely  followed for  ranges as small as one 
   interparticle distance, and still qualitatively for smaller distances. For the original particle model, we expect the present 
   mean-field theory to become, unlike the Mode-Coupling equations, an increasingly good  approximation at higher dimensions.
\end{abstract}

\maketitle

 In the past few years, there has been considerable activity on the application of  Mode-Coupling theory to  liquid systems.
 In its original conception,  Mode Coupling is an approximation for the dynamics in which an (infinite) subset of corrections 
coming from non-linearities is taken into account.  The theory has become popular not so much for the accuracy of its 
predictions -- numerical confirmation often  demands considerably good will to accept -- but because it gives
a unified and qualitative view of the first steps of the slowing down of dynamics, as the system approaches the glass transition. 

Mode-Coupling theory was originally confined to the dynamics in
equilibrium liquid phase.  However, similar approximations may be
applied to the equilibrium statistical mechanics of systems above and
below the glass transition, and to the non-equilibrium ("aging")
dynamics below the glass transition.  Elaborating on an idea of
Kraichnan~\cite{kraichnan1962stochastic1}, Kirkpatick, Thirumalai and
Wolynes~\cite{PhysRevLett.58.2091,PhysRevB.36.5388,kirkpatrick1987connections}
noted that one may view these approximate theories as being the exact
description of the properties of an auxiliary model, different from the
original one. These turn out to be disordered models having nature
that is `mean-field' in the following sense: if a particle (or spin)
$\bf A$ interacts strongly with two other particles $\bf B$ and $\bf
C$, then $\bf B$ and $\bf C$ do not interact strongly with one
another. This may happen either because the network of interactions is
tree-like, or because all individual interactions are weak.

There is a large family of such `mean-field' models, going beyond the
one that leads to the original mode-coupling equations, and they may all be
 treated with the tools developed in the context of spin glass
theory.  When applied to the glass transition, the whole strategy is
referred to as "Random First Order" scenario: a single name for
approximation schemes that may be different is justified because the
expectation is that the nature of the glass transition, of the
equilibrium glass phase, and of the out of equilibrium dynamics, is
qualitatively the same in all these mean-field models. The wider
question whether this scenario holds strictly at finite dimensions is
still very far from established.
 
Once one recognizes that Mode-Coupling is a form of mean-field theory,
the first instinct is to ask under which conditions it becomes exact,
in particular if it does so in high dimensions. The answer for the
latter question is that it does not \cite{PhysRevLett.104.255704}.
This lack of control over the approximation is problematic, because
there is no unambiguous way of relating features of a realistic system
with those of the Mode-Coupling solution -- and we often are not sure
whether some qualitative crossover in the behavior of an experimental
system should be associated with the idealized Mode-Coupling
transition, and in what sense.

In this paper we study an approximation of the same general mean-field
class than, but different from, the one leading to the mode-coupling
equations.  In order to bridge the gap between this limit and reality,
we build explicit models where an interaction range is tuned by some
parameter, thus allowing to go continuously from mean-field to true
finite-dimensions by varying this parameter.  Work in this direction
already exist for spin glasses
~\cite{frohlich1987some,franz2004finite,franz2004kac,sarlat2009these}, where one can consider models with
interactions with tunable range, as originally proposed by Kac.  As we
shall mention below, for particle systems, the usual program \`a la
Kac meets a problem as the interactions are made longer in range and
less strong: at low temperatures and large densities particles tend to
arrange themselves in clusters
\cite{grewe1977kirkwood2,klein1994repulsive,mel1995long}, themselves
arranged in a crystalline or amorphous `mesophase' structure. Thus, it seems
that in order to prevent this, one is forced to add a short-range
hard-core repulsion, thus spoiling the Kac (mean-field) nature of the
model.  In this paper we follow a different path, based on a
suggestion already made by Kraichnan fifty years ago: we study
particles with short-range interactions which are, however, `shifted'
by a random amount having a typical range, which is our parameter.

Within this framework we are able to address  several issues related to the glass transition,
revisiting them via the mean-field model we introduce. As an example, we are able to 
answer questions such as: {\em  "what is the relation between  the point at which the dynamics becomes nonexponential and the dynamic transition"}, because we can continuously take the model from finite dimensional to a mean-field limit, a situation where  both transition points  are well-defined, independently of any fitting procedure.

This paper is divided in two parts, analytic and numeric, which may be
read independently.  Sections II and III are devoted to the analytic
treatment of the statics and the dynamics of the liquid phase,
respectively. The main new result is an equation for the dynamics
that plays the role of the mode-coupling equation, and is exact in the
mean-field limit.  Sections IV and V present the numerical tests for
statics and dynamics, respectively. We are able to compare the results
in the mean-field limit with the ones for finite parameter $\lambda$,
all the way down to the ordinary particle model $\lambda=0$.  In
section VI we discuss an instance where having an approximation with
some limit in which it is well controlled is reassuring: there has
been some doubt whether the so-called "onset temperature" (or
pressure)~\cite{sastry1998signatures}, at which the equilibrium
dynamics becomes nonexponential (and the inherent structures start to
have deep energies), should be identified with the mode-coupling
transition. Here, by taking continuously the parameter $\lambda$ to
infinity, we find that they are in fact two distinct pressures -- at
least in the limit in which they are both well defined.

\section{Model}\label{sec:model}

\subsection{Kac models, clustering and the Kirkwood instability}

One can introduce a mean-field treatment of particle systems in
different ways.  One may, for example, use explicit infinite range
interactions \cite{dotsenko2004infinite,dotsenko2005mean}.
Glassiness is obtained by choosing a potential imposing a strong
frustration, and there is in principle no need for quenched disorder.
Next, one may consider long, but finite ranges, in the spirit of Kac
interactions.  Although quite intuitive, the choice of the potential
is in practice difficult, as crystallization \cite{gils2007absence} or
instabilities in the liquid phase (like the Kirkwood instability
\cite{kirkwood1941statistical,grewe1977kirkwood2,klein1994repulsive,mel1995long,likos2007ultrasoft,fragner2007interparticle})
easily set in as soon as the interaction range is finite.

Consider, for example, the model studied by Dotsenko
\cite{dotsenko2004infinite,dotsenko2005mean}. Particles are in
a confining potential $V_{conf}({\bf x_a})$ (which may be harmonic)
and interact with a long-range, oscillatory potential:
\begin{equation}
H= \frac{1}{\sqrt{N}}\sum_{a \neq b} \cos(|{\bf x_a}-{\bf x_b}|) + \sum_a V_{conf}({\bf x_a})
\label{dots}
\end{equation}
Dotsenko showed that the system indeed has a mean-field glass phase, induced by 
the frustration due to the conflict of attraction and repulsion. 
The next step, in an ordinary Kac program, would be to introduce the model
with a finite range $\gamma$:
\begin{equation}
 H= \frac{1}{\sqrt{\gamma}}\sum_{a \neq b} \cos(|{\bf x_a}-{\bf x_b}|) e^{-|{\bf x_a}-{\bf x_b}|/\gamma}+ \sum_a V_{conf}({\bf x_a})
\label{dots1}
\end{equation}
The result is disappointing: instead of giving a glass, the system
arranges as follows: it forms clusters of many particles, themselves
disposed in a crystalline arrangement, at the optimal distance so that
the interaction between clusters is minimized.  The way to avoid the
crystal-of-clusters mesophase is, of course, to add a hard-core that
hampers the clustering, but then the model is no longer mean field.
Another related difficulty with this strategy is that even the liquid
phase may have a transition to a `liquid' with spatial modulation, the
Kirkwood instability~\cite{kirkwood1941statistical,
  grewe1977kirkwood2, klein1994repulsive, mel1995long}. This phase is
not without interest of its own (see Fig.~\ref{fig:kirkwood} for a
numerical simulation), but it is not what we are wishing to study
here.

Another way to construct a mean-field model is to work with particles
on a Bethe lattice
\cite{PhysRevE.67.057105,rivoire2003glass,PhysRevLett.88.025501,tarzia2004glass,mari2009jamming}.  This introduces quenched disorder and the mean-field nature at the
same time.  By increasing the graph connectivity, one can, at least
formally, recover the original finite dimensional model by setting the
graph connectivity to infinity.

\subsection{Kraichnan's proposal and beyond.}

In this paper, we will follow another route.
We study family of models which are defined through the Hamiltonian:
\begin{equation}
\label{eq:model}
H(\{\mathbf{x}\}, \{\mathbf{A}\})=\sum_{<i,j>} V(\mathbf{x_i}-\mathbf{x_j}-\mathbf{A_{ij}})  
\end{equation}
where $V$ is a short-ranged interaction potential.  The $\mathbf{x}$'s
are the positions of the particles and the $\mathbf{A}$'s are quenched
random variables with a probability distribution $P(|\mathbf{A}|)$,
which has a variance $\lambda^2$. We also impose  that
$\mathbf{A}_{ij}=\mathbf{A}_{ji}$. The model for  $\lambda = \infty$ was
first introduced by Kraichnan some fifty years ago
\cite{kraichnan1962stochastic1}.

\begin{figure}
  \centering
  \includegraphics[width=0.4\columnwidth]{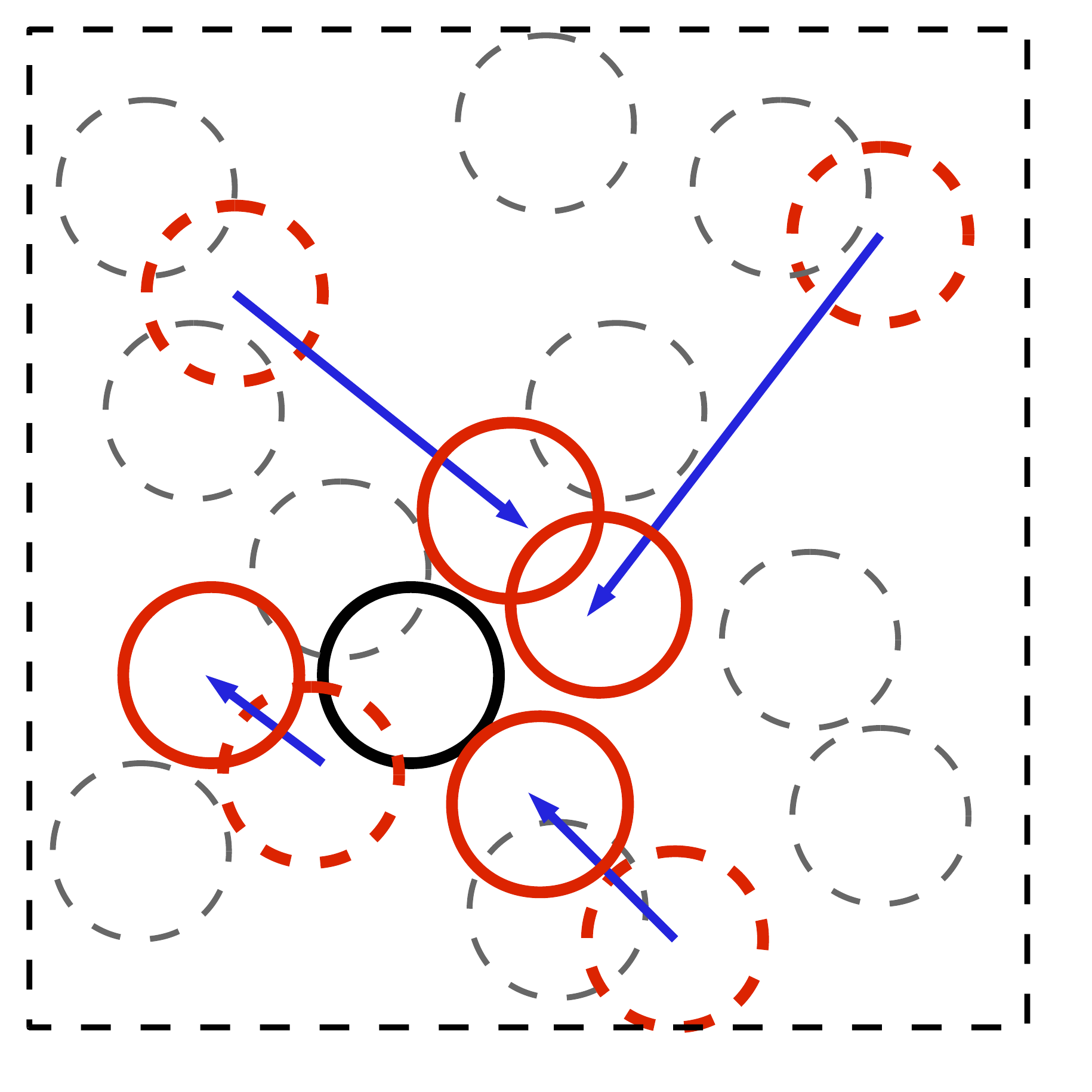}  
  \caption{The black particle effectively interacts with a finite
    number of particles (in red) that have appropriate random shifts
    (blue arrows), but these particles may be anywhere in the sample.}
  \label{fig:shifts}
\end{figure}

When $\lambda=0$,  $P(|\mathbf{A}|)=\delta(|\mathbf{A}|)$, and the
model reduces to an usual $d$-dimensional system. On the other hand,
when $\lambda \to \infty$, one particle $i$ can interact with
particles $j$ which are possibly anywhere in the system, as long as
$|\mathbf{x_i}-\mathbf{x_j}-\mathbf{A_{ij}}|$ is of the order of the
range of the potential $V$. Therefore, the system tends to have a
mean-field nature in this limit even though   one particle
effectively interacts with a \textit{finite} number of other particles, as it
does in a conventional finite-dimensional hard sphere system 
(see  Fig.~\ref{fig:shifts}). Thus by tuning $\lambda$, the
model (\ref{eq:model}) goes from a finite dimensional system ($\lambda =
0$) to a mean-field realization of the same system ($\lambda \to
\infty$).

\begin{figure}
  \centering
  \hspace{0cm}
  \includegraphics[width=0.3\columnwidth]{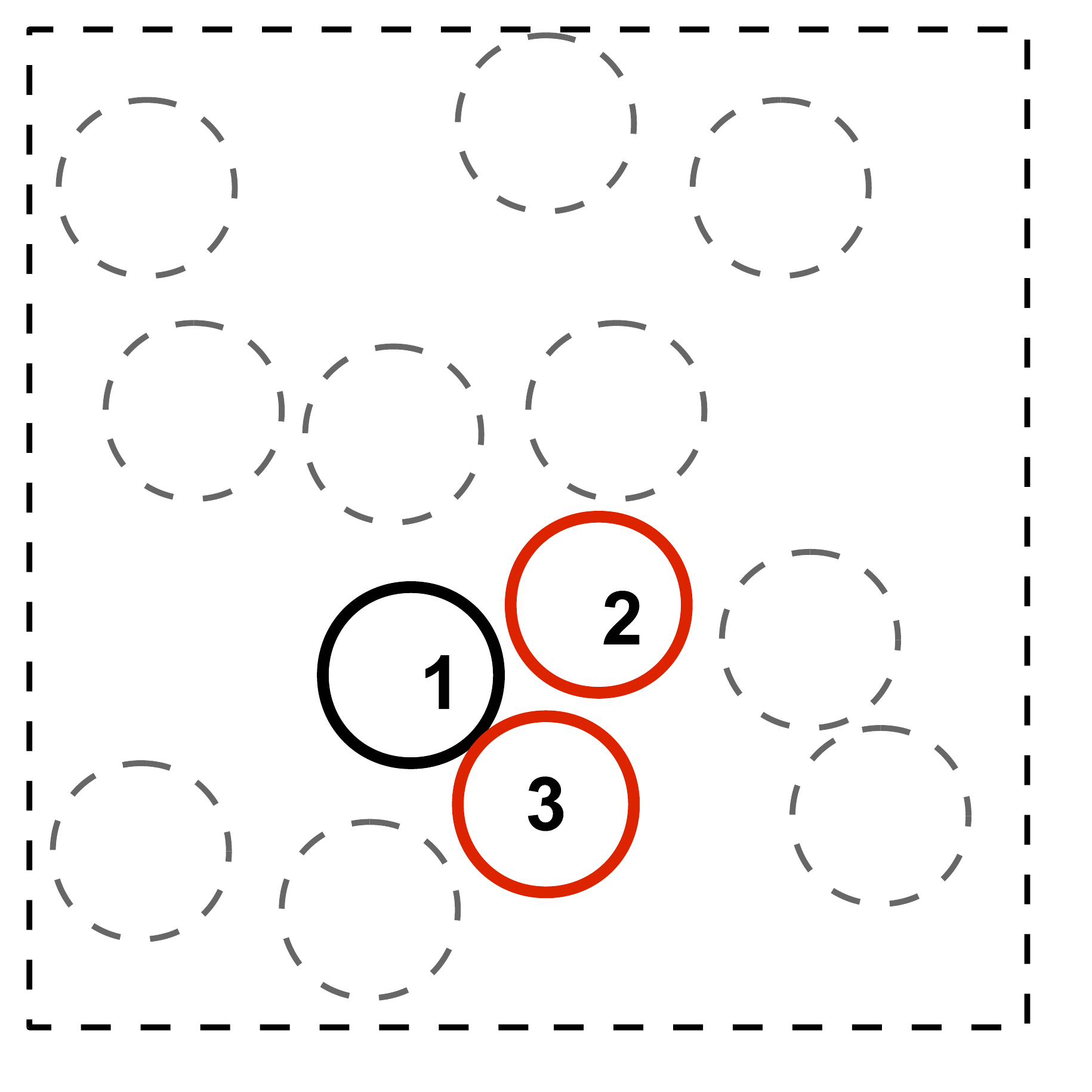}
  \hspace{0.5cm}
  \includegraphics[width=0.3\columnwidth]{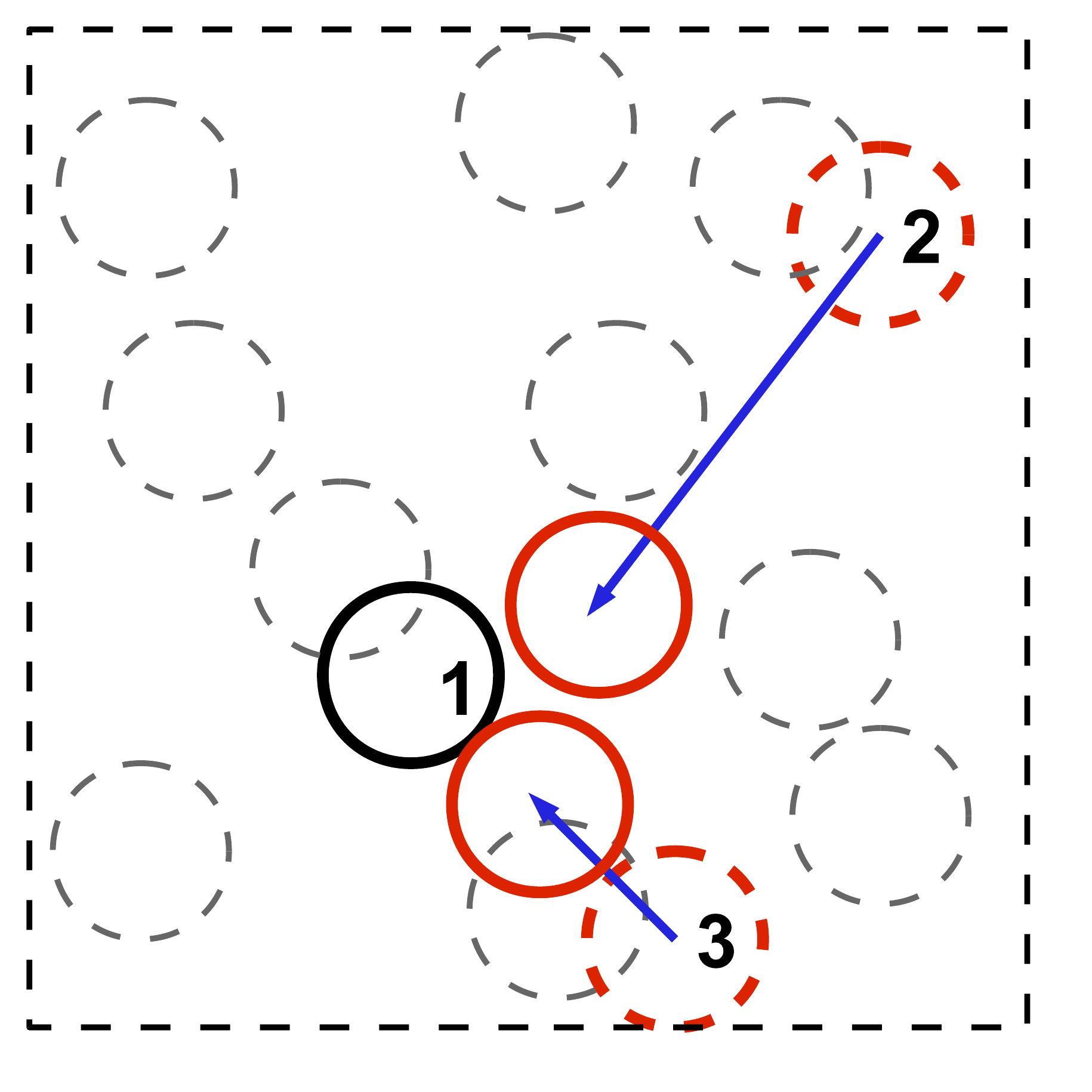}
  \hspace{0.5cm}
  \includegraphics[width=0.3\columnwidth]{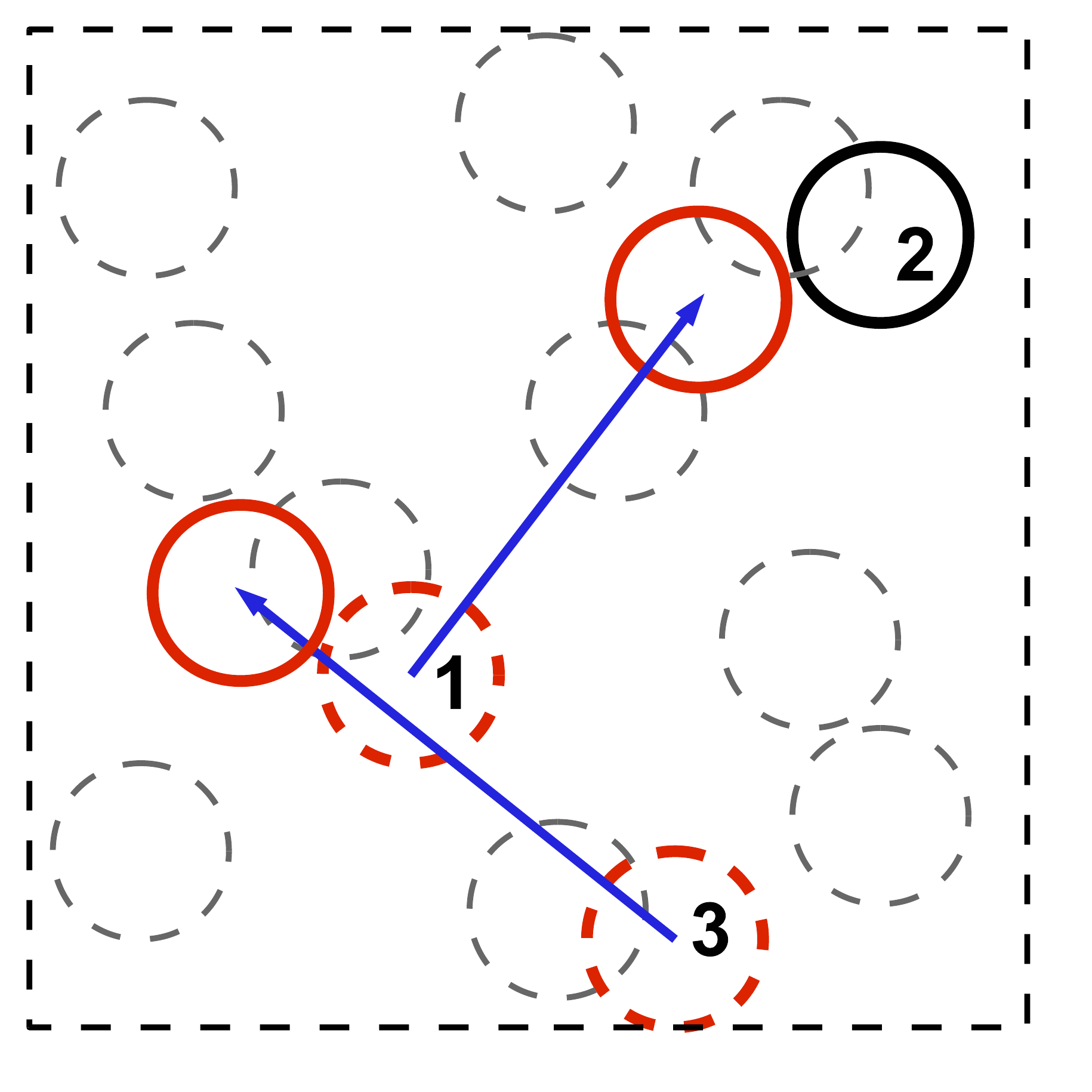}
  \caption{\textbf{Left :} Effective three-body correlation in a
    finite-dimensional hard-sphere system (with no random shifts),
    leading to corrections beyond the first virial
    term. \textbf{Center :} Same situation in the random shift model,
    from the point of view of particle 1. Particle 1 sees particles 2
    and 3 nearby. \textbf{Right :} From the point of view of particle
    2, there is no effective three-body interaction, as the random
    shifts displaces particle 3 very far from particle 2.}
  \label{fig:three_body}
\end{figure}

In the liquid phase, the mean-field limit of the model has an entropy
of an ideal gas plus only the first virial correction, just like a van
der Waals gas. Physically, this comes from the fact that it is very
unlikely that three (or more) spheres effectively interact
simultaneously with one another, as is sketched in
Fig.~\ref{fig:three_body}.  For instance, in the hard-sphere case, one
would need to have, for particles $i$, $j$, and $k$ (having a diameter
$D$):
\begin{equation}
  \begin{array}{c}
  |\mathbf{x_i}-\mathbf{x_j}-\mathbf{A_{ij}}|\sim D  \\
  |\mathbf{x_j}-\mathbf{x_k}-\mathbf{A_{jk}}|\sim D  \\
  |\mathbf{x_k}-\mathbf{x_i}-\mathbf{A_{ki}}|\sim D
  \end{array}
\end{equation}
which requires that the random shifts satisfy
$|\mathbf{A_{ij}}+\mathbf{A_{jk}}+\mathbf{A_{ki}}|\sim D$. This of
course is very unlikely for shifts $|\mathbf{A}|\sim L$, where $L$ is
the linear size of the system.

In this work, we will concentrate mostly on the hard-sphere potential,
but the procedure is a very generic way to obtain a mean-field limit,
and can even be generalized  to objects with rotational
degrees of freedom, where one can introduce a rotational disorder.
We studied both a monodisperse  system, for conceptual simplicity;
and a bidisperse one, to be able to work with arbitrary small $\lambda$ without 
having to deal with crystallisation.

From now on, we set the diameter $D$ of a sphere (in the monodisperse
case) to $D=1$, which will be used as length scale. The temperature
$T$ is set to $1$ almost everywhere, as it is irrelevant for hard
spheres. The only part where we keep it explicit is in the section
dealing with the dynamics of the model.  For simplicity we introduce
the following notations: a sphere with diameter $1$ in dimension $d$
has a volume $2^{-d}v_d$, and a surface $s_d$.

\section{Statics of the liquid phase}\label{sec:eos_liq}

\subsection{Grand-canonical formalism and Mayer expansion}

In this section, we work with a monodisperse system.  We note $V(u)$
the interaction potential:
\begin{equation}
  \begin{array}{r|l}
    \multirow{2}{*}{V(u)=}& \infty \quad \text{if} \quad u>1 \\
    &  0 \quad \text{otherwise}    
  \end{array}
\end{equation}
The canonical partition function of the system reads:
\begin{equation}
Z_{\{\mathbf{A} \}}= \int \prod_{k} d\mathbf{x_k} \exp \(( -\sum_{ij}V(\mathbf{x_i}-\mathbf{x_j}-\mathbf{A_{ij}}) \))    
\end{equation}
One would like to study the entropy of the system averaged over disorder:
\begin{equation}
  S=-\overline{\ln Z_{\{\mathbf{A} \}}}
\end{equation}
In the liquid phase, we can treat this average as an {\em annealed} average over the disorder:
\begin{equation}
  S=S_{an}=-\ln \overline{ Z_{\{\mathbf{A} \}}}
\end{equation}
So the problem reduces to the study of:
\begin{equation}\label{eq:canonical_annealed_partition}
  \overline{Z_{\{\mathbf{A} \}}}= \frac{1}{N!}\int \prod_{lm}P(\mathbf{A}_{lm}) d\mathbf{A}_{lm} \int \prod_{k} d\mathbf{x_k} \exp \(( -\sum_{ij}V(\mathbf{x_i}-\mathbf{x_j}-\mathbf{A_{ij}}) \))    
\end{equation}
We introduced a $1/N!$ prefactor to be able to perform the next step, a
Mayer expansion. We just have to remember to compensate  this factor at
the end of the computation. To do the Mayer expansion, we have to
translate our problem in a grand-canonical formalism. We introduce the
grand-canonical partition function:
\begin{equation}
  \overline{\Theta}=\sum_N z^N \overline{Z_{\{\mathbf{A} \}}}
\end{equation}
where $z$ is the activity, related to the chemical potential $\mu$ by
$z=e^{\mu}$, and we rewrite Eq.~(\ref{eq:canonical_annealed_partition}) as:
\begin{equation}\label{eq:part_func_mayer}
  \overline{Z_{\{\mathbf{A} \}}} = \int \prod_{k} d \mathbf{x_k} \prod_{ij} \[[ \overline{f}(\mathbf{x_i}-\mathbf{x_j})+ 1 \]]  
\end{equation}
where $\overline{f}$ is an annealed Mayer function:
\begin{equation}
  \label{eq:mayer_f}
  \overline{f}(\mathbf{x}-\mathbf{y}) = \int d \mathbf{A} P(\mathbf{A}) \[[ \exp \(( - V(\mathbf{x}-\mathbf{y}-\mathbf{A}) \)) -1 \]] = -\int d \mathbf{A} P(\mathbf{A}) \chi ( |\mathbf{x}-\mathbf{y}-\mathbf{A}|)
\end{equation}
with $\chi(r)$ the step function such that $\chi(r)=1$ if $r<1$ and $\chi(r)=0$ otherwise.
We can then introduce a diagrammatic representation to express the
grand-canonical potential $G=\ln \overline{\Theta}$ as
usual\cite{hansen2006theory}:
\begin{equation}
\begin{array}[b]{lllll}
  \label{eq:G_Mayer}
  G=\ln \overline{\Theta}=\{ \text{ connected diagrams } \}= 
    \xymatrix{\bullet} + &
    \segment{\bullet}{\bullet} + &
    \dipod{\bullet}{\bullet}{\bullet}+ &
    \triangle{\bullet}{\bullet}{\bullet}+ &
    \dots
\end{array}
\end{equation}
where the diagram vertices represent a factor $z$ while the bonds
represent the Mayer function $\overline{f}$.

\subsection{Mean-field equation of state}

As already noted by Kraichnan \cite{kraichnan1962stochastic1}, only the first 
virial correction survives in the mean-field limit. The entropy can then be expressed in term of the density field $\rho( \mathbf{x})$:
\begin{equation}
  \label{eq:F_mean_field}
S_{an}[\rho( \mathbf{x})]  =  -\int \text{d} \mathbf{x} \rho( \mathbf{x}) \left[ \ln \rho( \mathbf{x})-1 \right] + \frac{1}{2} \int  \text{d} \mathbf{x} \text{d} \mathbf{y} \rho( \mathbf{x}) \rho( \mathbf{y}) \overline{f}(\mathbf{x}-\mathbf{y})+N\ln N
\end{equation}
 The $\ln N$ contribution to the entropy density reflects the fact that
particles are not, in this model, truly indistinguishable -- just as
they are not in a system particles of polydisperse sizes.

The fact that the first terms contribute  can be understood by noticing that:
\begin{equation}
  \overline{f}(\mathbf{x}-\mathbf{y})=-v_d/V  
\end{equation}
Thus, in the Mayer expansion Eq~(\ref{eq:G_Mayer}), a diagram with $n$
vertices and $m$ bonds reads $V^{n-m}z^n(-v_d)^m$. If $m \ge  n$, the
diagram vanishes in the thermodynamic limit, because this requires that several random shifts add to zero. Only the diagrams having $m<n$ contribute, and they are the tree ones
(having $m=n-1$):
\begin{equation}
  \label{eq:G_Mayer_tree}
    \ln \overline{\Theta}= 
    \xymatrix{\bullet } + 
    \segment{\bullet}{\bullet} + 
    \dipod{\bullet}{\bullet}{\bullet} + 
    \arch{\bullet}{\bullet}{\bullet}{\bullet} + 
    \tripod{\bullet}{\bullet}{\bullet}{\bullet} + 
    \dots
\end{equation}

The next step is to do the Legendre transform~\cite{hansen2006theory} $z \to \rho$, which leads to:
\begin{equation}\label{eq:saddle_rho_liquid}
  \ln \rho( \mathbf{x})=\ln z + \xymatrix{  *{\circ} \ar@{-}[r] *{\otimes} & }
\end{equation}
where now the $\otimes$ vertex denotes a factor $\rho( \mathbf{x})$,
and ${\circ}$ is $1$.  Eq.~(\ref{eq:saddle_rho_liquid}) is the saddle
point equation which minimizes the grand-canonical potential functional
$G[\rho( \mathbf{x})]$, which can be translated back into the
canonical ensemble, leading to the entropy Eq.~(\ref{eq:F_mean_field}).

In the liquid phase with density $\rho$, the entropy is thus:
\begin{equation}
  \frac{S_{an}}{N}=1 - \ln \rho - \frac{1}{2} \rho v_d + \ln N
\end{equation}
and the system has a van der Waals equation of state:
\begin{equation}
   P = \rho+\frac{1}{2} v_d\rho^2  
\end{equation}
It is remarkable that this equation of state (ideal gas with first
virial correction) is the one of hard spheres when $d \to \infty$
(see \cite{frisch1985classical,frisch1999high,PhysRevE.62.6554}). This is a first indication that  the
limit $\lambda \to \infty$ and the limit $d \rightarrow \infty$ are of the same nature, as we shall see below.

\subsection{Pair correlation function}

As in high-dimensional liquids or in systems with Kac interactions,
the very simple form of the Mayer expansion allows to compute exactly
the two-point correlation functions in the liquid phase.
A `naive' pair correlation function is defined as~\cite{hansen2006theory}:
\begin{equation}\label{eq:pair_correl_def}
  \begin{array}{rl}
  g(\mathbf{x}-\mathbf{y})&=\overline{<\delta(\mathbf{x_1}-\mathbf{x})\delta(\mathbf{x_2}-\mathbf{y})>} \\
  &=\dfrac{N(N-1)}{\rho^2 N!\overline{Z_{\{\mathbf{A} \}}}} \int \prod_{lm}P(\mathbf{A}_{lm}) d\mathbf{A}_{lm} \\
  & \qquad \int \prod_{k} d\mathbf{x_k} \delta(\mathbf{x_1}-\mathbf{x}) \delta(\mathbf{x_2}-\mathbf{y}) \exp \(( -\sum_{ij}V(\mathbf{x_i}-\mathbf{x_j}-\mathbf{A_{ij}}) \)) \\
  &=\dfrac{2}{\rho N} \overline{e}(\mathbf{x_1}-\mathbf{x_2}) \dfrac{\delta \ln \overline{Z_{\{\mathbf{A} \}}}}{\delta \overline{e}(\mathbf{x_1}-\mathbf{x_2})}
  \end{array}
\end{equation}
From Eq.~(\ref{eq:F_mean_field}) we then get:
\begin{equation}\label{eq:pair_correl_non_shifted_result}
  g(\mathbf{x}-\mathbf{y})=1+\overline{f}(\mathbf{x}-\mathbf{y})\simeq 1-v_dP(\mathbf{x}-\mathbf{y})
\end{equation}

\noindent When $\lambda = \infty$ the pair correlation function is simply:
\begin{equation}
  g(\mathbf{x}-\mathbf{y})=1
\end{equation}
This expresses  the absence of structure in the system due to the
quenched disorder, which totally blurs the hard core repulsion when
$\lambda \to \infty$. 
Of course, this does not mean that there are no
real pair correlations  in the system. 
Indeed, a  more interesting
quantity is the pair correlation function `seen from one particle':
\begin{equation}\label{eq:pair_correl_shifted_def}
\begin{array}{rl}
  g_S(\mathbf{x}-\mathbf{y})&=\overline{<\delta(\mathbf{x_1}-\mathbf{x})\delta(\mathbf{x_2}+\mathbf{A}_{12}-\mathbf{y})>} \\
  &=\dfrac{N(N-1)}{\rho^2 N!\overline{Z_{\{\mathbf{A} \}}}} \int \prod_{lm}P(\mathbf{A}_{lm}) d\mathbf{A}_{lm} \\  & \qquad \qquad \int \prod_{k} d\mathbf{x_k} \delta(\mathbf{x_1}-\mathbf{x}) \delta(\mathbf{x_2}+\mathbf{A}_{12}-\mathbf{y}) \exp \(( -\sum_{ij}V(\mathbf{x_i}-\mathbf{x_j}-\mathbf{A_{ij}}) \)) \\
  &=\dfrac{2}{\rho N} \exp \left(-V(\mathbf{x}-\mathbf{y})\right) \dfrac{\delta \ln \overline{Z_{\{\mathbf{A} \}}}}{\delta \overline{e}(\mathbf{x}-\mathbf{y})}
  \end{array}
\end{equation}
Inserting Eq.~(\ref{eq:F_mean_field}), we obtain:
\begin{equation}\label{eq:pair_correl_shifted_result}
  g_S(\mathbf{x}-\mathbf{y})=\exp \left(-V(\mathbf{x}-\mathbf{y})\right)\end{equation}

\noindent This pair correlation function is
identical to the one obtained for a hard sphere system in infinite
dimensions.

\subsection{Relation with a Kac potential in the statics of the liquid phase}

The connection between the two ways of approaching  a mean-field limit
(sending $d \to \infty$ or introducing random shifts with range
$\lambda \to \infty$) is to be compared with a similar connection in
systems with Kac type potentials \cite{klein1986instability}. We can
push forward this analogy by an explicit mapping of our model onto a system with a
Kac potential, valid (only) for static quantities in the low density liquid phase, when $\lambda$ is
large but finite.

In this case, the Fourier transform of the Mayer function defined by
Eq.~(\ref{eq:mayer_f}) is the product of a Bessel function having a
range $\sim 1$ with the Fourier transform $\tilde{P}$ of the shifts
distribution, having a range $\sim 1/\lambda$:
\begin{equation}
  \tilde{f}(k)= -(2\pi)^{d/2} \tilde{P}_{\lambda}(k) \frac{J_{\frac{d}{2}}(k)}{k^{d/2}} 
\end{equation}
Thus, as long as $\lambda\gg1$, we have:
\begin{equation}
  \overline{f}(r) \simeq -v_d P_{\lambda}(r) = - \gamma^{-d} K (\gamma^{-1}r)  
\end{equation}
with $K$ a short ranged bounded positive function.
This function obviously has a range $\gamma \sim \lambda$, as
long as $\lambda\gg1$. Thus, (\ref{eq:F_mean_field}) reads:
\begin{equation}
 S_{an}[\rho( \mathbf{x})]  =  - \int \text{d} \mathbf{x} \rho( \mathbf{x}) \left[ \ln \rho( \mathbf{x}) -1 \right] - \frac{ \gamma^{-d}}{2} \int  \text{d} \mathbf{x} \text{d} \mathbf{y} \rho( \mathbf{x}) \rho( \mathbf{y}) K(\gamma^{-1}|\mathbf{x}-\mathbf{y}|)+N\ln N
\end{equation}

This is exactly the functional obtained in the so-called mean-field
approximation for a potential $K$ (with temperature $\beta=1$), and it
is exact in the Kac limit $\gamma \to \infty$
\cite{grewe1977kirkwood1}. Indeed, the naive pair correlation function
Eq.~(\ref{eq:pair_correl_non_shifted_result}) is equivalent to the
one obtained for a Kac potential~\cite{zachary2008gaussian}.

This mapping concerns the {\em statics} of the {\em liquid}
phase, and   does not mean that the {\em dynamics} of our model has anything to
do with the one of a Kac model, even in the liquid phase. As a matter of fact, in the following
sections we present  results showing that a dynamical glass
transition occurs in our model with a Gaussian distribution for the
shifts, whereas the related Kac model, which is the Gaussian Core
Model~\cite{stillinger1976phase} becomes in the high density limit an
ideal gas~\cite{stillinger1976phase,lang2000fluid}.

\subsection{Avoiding the Kirkwood instability}

\begin{figure}
  \centering
  \includegraphics[width=0.7\columnwidth]{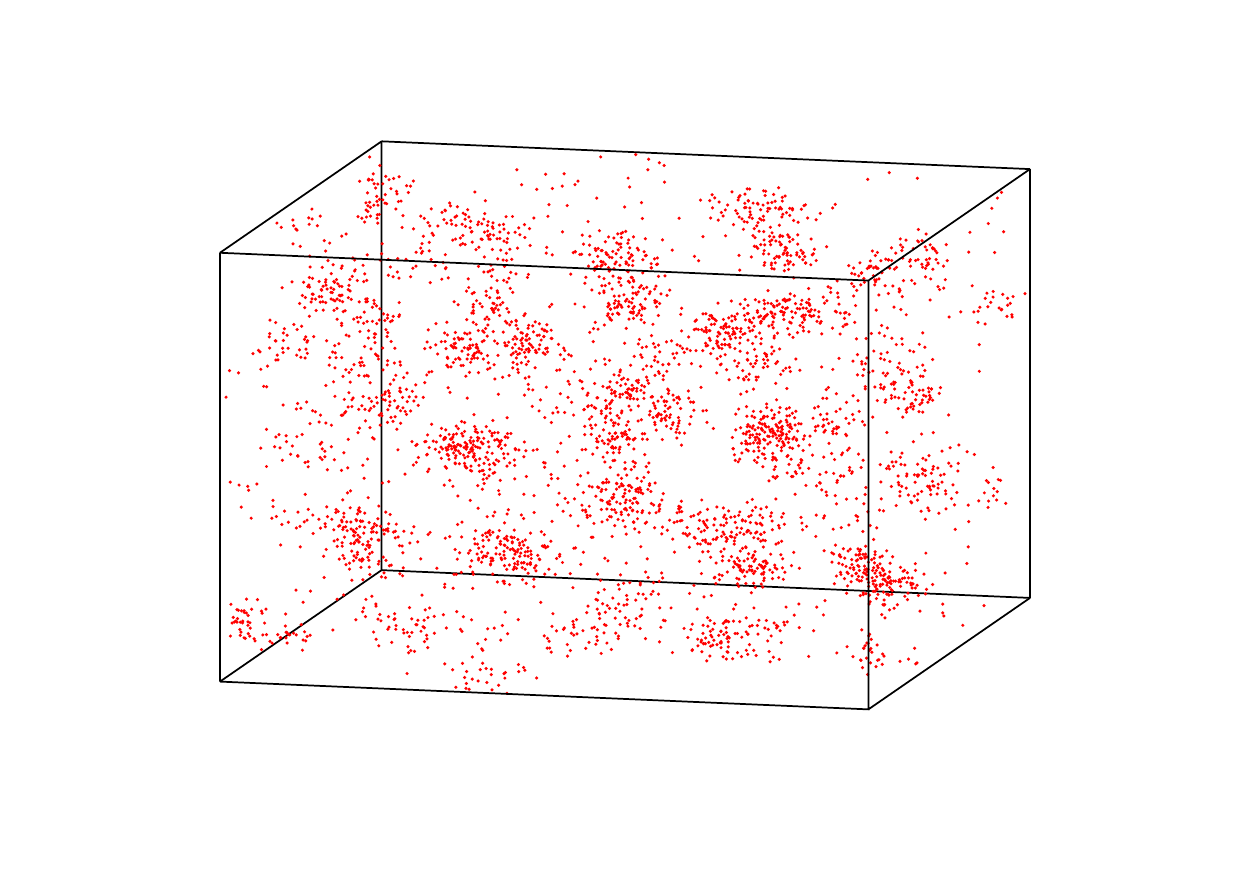}
  \caption{Modulated phase due to a Kirkwood instability. Here the
    random shift distribution is flat within  a sphere with radius
    $\lambda=2$.}
  \label{fig:kirkwood}
\end{figure}

If $\lambda$ is finite but large, we can take
Eq.~(\ref{eq:F_mean_field}) as a good approximation of the liquid
phase, but we have to be careful in our choice of the random shift
distribution $P(\mathbf{A})$. It is well known that the mean-field
entropy functional for Kac models can be unstable above a given
density (the so-called Kirkwood instability~\cite{grewe1977kirkwood2})
towards a phase with spatial density modulations.  This can be seen
via a linear stability analysis of Eq.~(\ref{eq:F_mean_field}). If we
perturb the uniform liquid phase solution $\rho( \mathbf{x})=\rho$
with a small oscillatory term $\rho_{\epsilon}(
\mathbf{x})=(1+\epsilon \exp(i\mathbf{k} \mathbf{x}))\rho$, we get:
\begin{equation}\label{eq:linear_perturb_entopy}
  S_{an}[\rho_{\epsilon}(\mathbf{x})]=S_{an}[\rho] + \frac{\epsilon^2 N \rho }{2} \text{TF} \left[ \dfrac{\delta^2 S_{an}}{\delta \rho(\mathbf{x}) \delta \rho(\mathbf{y})} \right] (\mathbf{k})
\end{equation}
with:
\begin{equation}
  \frac{\delta^2 S_{an}}{\delta \rho(\mathbf{x}) \delta \rho(\mathbf{y})}=-\frac{1}{\rho} \delta(\mathbf{x}-\mathbf{y})+ \overline{f}(\mathbf{x}-\mathbf{y})
\end{equation}
From Eq.~(\ref{eq:linear_perturb_entopy}), it is clear that the uniform liquid phase is stable only if:
\begin{equation}\label{eq:stability_cond1}
  \text{TF} \left[ \dfrac{\delta^2 S_{an}}{\delta \rho(\mathbf{x}) \delta \rho(\mathbf{y})} \right] (\mathbf{k})=-\frac{1}{\rho} + \tilde{f}(\mathbf{k})  < 0 \qquad \forall \mathbf{k}
\end{equation}
If $\tilde{f}(\mathbf{k})$ takes positive values for some wave vector
$k$, there is a value of $\rho$ above which the condition
Eq.~(\ref{eq:stability_cond1}) is not fulfilled: this is the Kirkwood
instability. The only way to avoid this transition is to have a Mayer function
 with a negative Fourier transform:
\begin{equation}
  \label{eq:stability_cond2}
 \tilde{f}(\mathbf{k})
< 0 \qquad \qquad \forall \mathbf{k}
\end{equation}
Since:
\begin{equation}
\tilde{f}(\mathbf{k})= -(2\pi)^{d/2} \tilde{P}_{\lambda}(k) \frac{J_{\frac{d}{2}}(k)}{k^{d/2}}
\end{equation}
we have to tune the distribution of the random shifts to get the
desired property. If $\lambda\gg1$, the range of $\tilde{P}_{\lambda}$
($\sim \lambda^{-1})$ is much smaller than the range of
$J_{\frac{d}{2}}$, therefore $-\tilde{f}(\mathbf{k})= v_d\tilde{P}_{\lambda}(\mathbf{k})$ and taking
$\tilde{P}_{\lambda}(\mathbf{k})>0$ is enough to ensure condition
(\ref{eq:stability_cond2}). 
In this work, we will take a Gaussian
distribution for the shifts:
\begin{equation}
  P_{\lambda}(\mathbf{A})=\frac{1}{(2\pi \lambda)^d}\exp\(( -\frac{|\mathbf{A}|^2}{2\lambda^2}\))  
\end{equation}
As an example of the Kirkwood instability, we show in Fig.~\ref{fig:kirkwood} a dense configuration of the random shift model with a flat density of shifts.

\subsection{Corrections beyond Mean-field: the role of high dimensionality.}

We can consider corrections to the mean-field equation of 
state. As we shall see, for finite $\lambda$ they vanish with dimensionality as $\propto \lambda^{-d}$.
 In the Mayer expansion of the entropy, the dominant correction will 
come from diagrams with $m=n$, which are 
the ring diagrams:
\begin{equation}\label{eq:ring_entropy}
  S_{an,\lambda}= N\ln N -\int \text{d} \mathbf{x} \rho( \mathbf{x}) \left[ \ln \rho( \mathbf{x})-1 \right] + \segment{\otimes}{\otimes} + \triangle{\otimes}{\otimes}{\otimes} + \square{\otimes}{\otimes}{\otimes}{\otimes} + \pentagon{\otimes}{\otimes}{\otimes}{\otimes}{\otimes} + \dots
\end{equation}

\noindent The resummation of these diagrams has been done by 
Montroll and Mayer \cite{mayer1941statistical}:
\begin{equation}\label{eq:S_rings}
\frac{S_{an,\lambda}}{N} = 1 -  \ln \rho - \frac{1}{2} \rho v_d + \frac{1}{2}\frac{1}{(2\pi)^d} \rho^2 \int \Omega_d k^{d-1} \dd k \frac{ \[[\tilde{f}(k)\]]^3 }{ 1-\rho \tilde{f}(k) }+\ln N
\end{equation}
and gives, for $\lambda\gg1$:
\begin{equation}
  \begin{array}{rl}
    \frac{S_{an,\lambda}}{N} & =  1 -  \ln \rho - \frac{1}{2} \rho v_d 
    - \frac{1}{2}\frac{1}{(2\pi)^d} \rho^2 \int \Omega_d k^{d-1} \dd k \frac{ \[[ v_d\tilde{P}_{\lambda}(k)\]]^3 }{ 1+ \rho v_d \tilde{P}_{\lambda}(k) } +\ln N\\ 
    & = 1 -  \ln \rho -  \frac{1}{2}\rho v_d  - \frac{1}{2}\frac{1}{(2\pi)^d \lambda^d} \rho^2 I_{31}(\rho)+\ln N
  \end{array}
\end{equation}
where we have introduced the $\lambda$-independent factor:
\begin{equation}
  I_{ab}(\rho)=\int \Omega_d k^{d-1} \dd k \frac{ \[[ v_d\tilde{P}_1(k)\]]^a }{ \left[1+ \rho v_d \tilde{P}_1(k) \right]^b}
\end{equation}
The equation of state is then:
\begin{equation}
  P=\rho+\frac{1}{2} \rho^2 v_d  + \frac{\rho^3}{(2\pi)^d \lambda^d} \left[ I_{31}(\rho) -\frac{\rho}{2} I_{42}(\rho) \right]
\end{equation}
This expression gives a quantitative estimation of  how the approximation becomes better
at higher dimensions, through the factor  $\lambda^d$. Already in 
$d=3$, the ring corrections are less than 1\% at $\phi=1$ for a range as small f
as $\lambda=1$. When $d>3$, one needs to take a very small range of 
random shifts to feel any finite dimensional effect.

\subsection{Estimation of the  density as $P \rightarrow \infty$}

The model with $\lambda \rightarrow \infty$ has a maximal packing density that diverges in the thermodynamic limit,
a rather awkward property from the thermodynamic point of view.
When $\lambda \to \infty$ we can derive an estimation of the maximum
density of the random shift model.  To do this, we consider the following algorithm:

\noindent {\em i)} Starting a configuration with $N$ spheres in a volume $V$, generate
$N$ random shifts,

\noindent {\em ii)} Locate positions where one can add, without any
overlap, a new sphere
interacting with other spheres via the $N$ random shifts,

\noindent {\em iii)} If such a position exists, add this
$(N+1)$th sphere, and go back to step $1$.

Up to which density such an algorithm can work? When $\lambda \to
\infty$, the $(N+1)$th sphere will see the other spheres as having
completely random positions. Then the probability for a position
$\mathbf{x}$ to satisfy the hard sphere constraints imposed by the $N$
other spheres is simply:
\begin{equation}
  P_N(\mathbf{x})=\left(1-\frac{v_d}{V}\right)^N \simeq e^{-\frac{Nv_d}{V}}=e^{-\phi}
\end{equation}

\begin{figure}
  \centering
  \includegraphics[width=0.7\columnwidth]{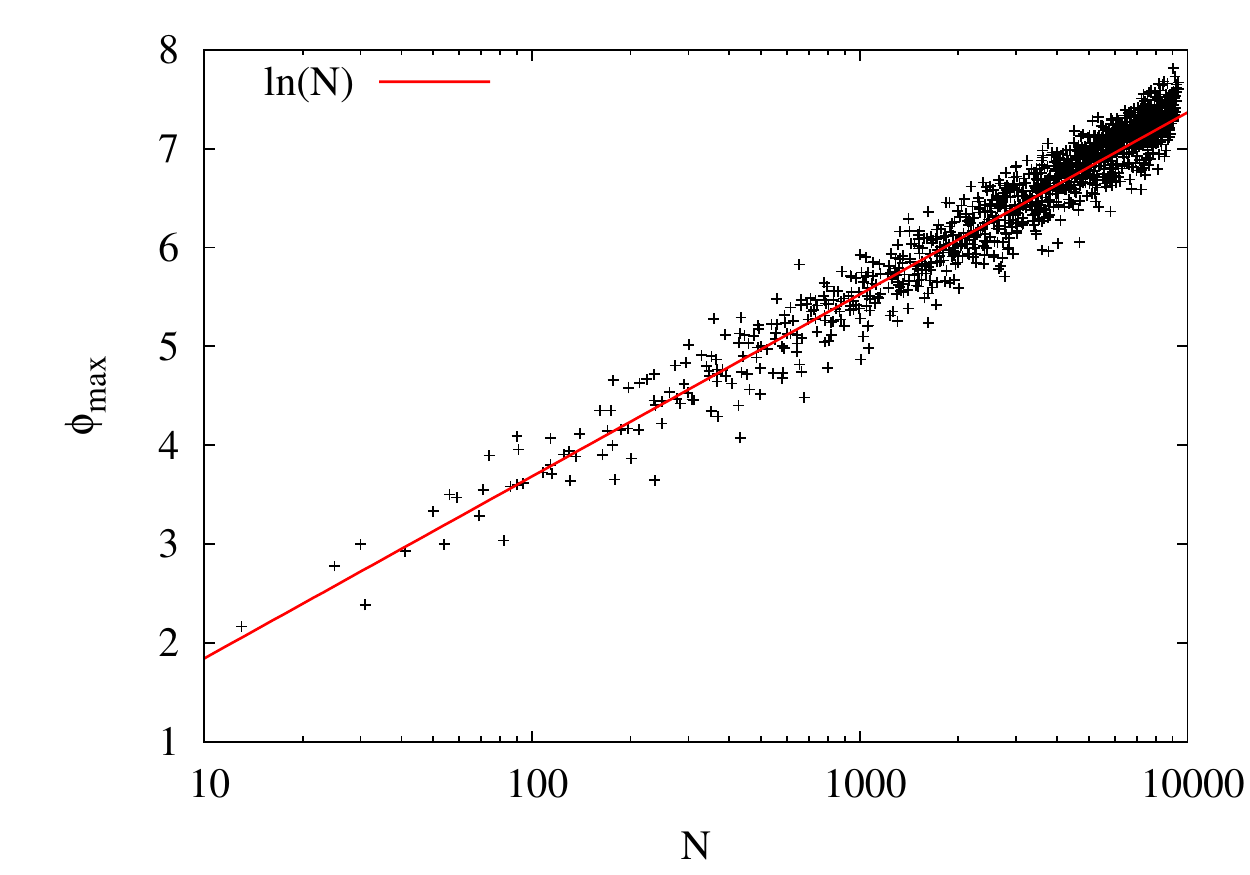}  
  \caption{Maximum density $\phi_{max}$ obtained by the algorithm
    described in the main text, showing the $\ln N$ dependence
   of Eq.~(\ref{eq:phi_max}).}
  \label{fig:sequential_addition}
\end{figure}

Then the available volume to place the $(N+1)$th sphere will be
$Ve^{-\phi}$. The algorithm will not be able to find a solution once
the available volume is of the order of $v_d$.  If we note
$\phi_{max}$ the volume fraction for which this happens, we have:
\begin{equation}
  e^{-\phi_{max}} \sim \frac{v_d}{V}
\end{equation}
which gives:
\begin{equation}\label{eq:phi_max}
  \phi_{max} \sim \ln \frac{V}{v_d} = \ln N -\ln (\ln N) + \ln (\ln (\ln N)) ...
\end{equation}
We implemented this algorithm for the random shift model in
$d=1$. Results are presented in Fig.~\ref{fig:sequential_addition},
showing the $\ln N$ dependence of $\phi_{max}$.


\section{Dynamics in the liquid phase}\label{sec:dynamics}

\subsection{A warming up exercise: computation of the canonical partition function}

The essential steps followed to compute the exact dynamic equation are
similar to the ones performed in making an equilibrium computation in the {\em  canonical} ensemble. It is
thus instructive to understand them in this context first.
Starting again from Eq.~(\ref{eq:part_func_mayer}), and
introducing the density $\rho(\underline{\mathbf{x}})$ by:
\begin{equation}
  \rho(\mathbf{x})=\sum_i \delta(\mathbf{x}-\mathbf{x_i})  
\end{equation}
with ${\bf x_i} $ the coordinate of the $i-th$ particle,
we can rewrite $\overline{Z_{\{\mathbf{A} \}}}$ as:
\begin{equation}
  \overline{Z_{\{\mathbf{A} \}}} =  \int \dd \mathbf{x_i} \int \mathscr{D}[\rho( \mathbf{x})]  \delta \left(\rho( \mathbf{x})-\sum_i \delta(\mathbf{x}-\mathbf{x}_i) \right) 
\exp \left[\frac{1}{2} \int \text{d} \mathbf{x} \text{d} \mathbf{y}  \rho(\mathbf{x}) \rho(\mathbf{y}) \ln\left[ 1+\overline{f}(\mathbf{x_i}-\mathbf{x_j})\right] \right]
\end{equation}
 Now we can exponentiate the $\delta$ constraint using a
second field $\hat{\rho}$, and integrate over the
$\underline{\mathbf{x_i}}$'s:
\begin{equation}
  \begin{array}{rl}
    \overline{Z_{\{\mathbf{A} \}}} =  \int & \mathscr{D}[\rho( \mathbf{x})] \mathscr{D}[\hat{\rho}( \mathbf{x})] \\
& \exp\left\{ i \int \text{d} \mathbf{x} \hat{\rho}( \mathbf{x}) \rho( \mathbf{x}) +  N \ln \int  \text{d} \mathbf{x} e^{-i \hat{\rho}( \mathbf{x})}+ \frac{1}{2} \int \text{d} \mathbf{x} \text{d} \mathbf{y}  \rho(\mathbf{x}) \rho(\mathbf{y}) \ln\left[ 1+\overline{f}(\mathbf{x_i}-\mathbf{x_j})\right] \right\}
  \end{array}
\end{equation}
 Again noticing that:
\begin{equation}
  \overline{f}(\mathbf{x}-\mathbf{y}) = -\frac{v_d}{V} 
\end{equation}
 we can expand the logarithm to get:
\begin{equation}
  \begin{array}{rl}
    \overline{Z_{\{\mathbf{A} \}}} =  \int & \mathscr{D}[\rho( \mathbf{x})] \mathscr{D}[\hat{\rho}( \mathbf{x})] \\
& \exp\left\{ i\int \text{d} \mathbf{x} \hat{\rho}( \mathbf{x}) \rho( \mathbf{x}) + N \ln \int  \text{d} \mathbf{x} e^{-i \hat{\rho}( \mathbf{x})}+ \frac{1}{2} \int \text{d} \mathbf{x} \text{d} \mathbf{y}  \rho(\mathbf{x}) \rho(\mathbf{y}) \overline{f}(\mathbf{x}-\mathbf{y})  \right\}
  \end{array}
\end{equation}
This integral may be evaluated by saddle point with respect to the 
fields $\rho$ and $\hat{\rho}$ (note that each term in the exponential is of order $N$, including the last one, due to the value of $\overline{f}$). This gives:
\begin{equation}\label{eq:saddle_statics}
  \begin{array}{rl}
    \rho( \mathbf{x}) & = N\frac{e^{-i \hat{\rho}( \mathbf{x})}}{\int \text{d} \mathbf{x} e^{-i \hat{\rho}( \mathbf{x})} } \\
    \hat{\rho}( \mathbf{x}) & = i \int \text{d} \mathbf{y} \rho(\mathbf{y}) \overline{f}(\mathbf{x}-\mathbf{y})
  \end{array}
\end{equation}
Thus, the logarithm of the partition function can be written in terms of $\rho$ as:
\begin{equation}\label{eq:replicated_FT_partition0}
  \ln \overline{Z_{\{\mathbf{A} \}}} = - \int \text{d} \mathbf{x} \rho( \mathbf{x}) \ln \rho( \mathbf{x}) + \frac{1}{2} \int  \text{d} \mathbf{x} \text{d} \mathbf{y} \rho( \mathbf{x}) \rho( \mathbf{y}) \overline{f}(\mathbf{x}-\mathbf{y})+N\ln N
\end{equation}
Below, we shall follow the same steps, but with trajectories playing the role of particles.

\subsection{Derivation of the exact dynamics as a partition function of trajectories}\label{sec:RS_dynamics}

We wish to study the  dynamics of the model in the mean-field limit, starting from the Langevin equation:
\begin{equation}\label{eq:langevin}
  \dot{\mathbf{x}}_i(t)=-\sum_{j\neq i} \nabla_i
V(\mathbf{x}_i(t)-\mathbf{x}_j(t)-\mathbf{A}_{ij})+\boldsymbol{\eta}_i(t)+\mathbf{h}_i(t),
\end{equation}
where we have introduced an external field $\mathbf{h}_i(t)$, which
acts as a source term, and will later be set to zero. The vector
$\boldsymbol{\eta}_i(t)$ is a Gaussian white noise with variance $2T$,
where $T$ is the temperature, that we will keep explicit for all the
derivation of the dynamics equations:
\begin{equation}
    <\boldsymbol{\eta}_i(t)\boldsymbol{\eta}_j(t')>=2Td\; \delta_{ij}\; \delta(t-t')
\end{equation}

We denote $P(\{\mathbf{x}^0\},\{\mathbf{x}^\tau\}, \tau)$ the
probability of having particle $i$ in $\mathbf{x}^0_i$ at time $t=0$
and in $\mathbf{x}^\tau_1$ at time $t=\tau$, in the absence of
external field $\mathbf{h}$.  We may express this as a sum over paths
using the Martin-Siggia-Rose/
DeDominicis-Jensen~\cite{zinn2002quantum} formalism. Averaging over
the noise, this gives:
\begin{eqnarray}
  \label{eq:P_MSR}
    P(\{\mathbf{x}^0\},\{\mathbf{x}^\tau\}, \tau)&=&\int \prod_{ij}
    \dd \mathbf{A}_{ij} P(\mathbf{A}_{ij}) \nonumber \\
& & \quad     \int_{\{\mathbf{x}(0)\}=\{\mathbf{x}^0\}}^{\{\mathbf{x}(\tau)\}=\{\mathbf{x}^\tau\}}
    \prod_{i} \DD[\mathbf{x}_i(t),\hat{\mathbf{x}}_i(t)] \exp \left[
      -\mathcal{S}[\{\mathbf{x}_i(t)\},\{\hat{\mathbf{x}}_i(t)\}]
      \right],
\end{eqnarray}
 with
\begin{equation}
  \label{eq:action}
  \begin{array}{rl}
    \mathcal{S}[\{\mathbf{x}_i(t)\},\{\hat{\mathbf{x}}_i(t)\}] &= 
     \int_0^\tau \mbox{d} t \left[\sum_i \left(\dot{\mathbf{x}}_i(t)\hat{\mathbf{x}}_i(t) +
\hat{\mathbf{x}}_i(t)\hat{\mathbf{x}}_i(t)\right)\right] \\
    & \;\;\; +\int_0^\tau \mbox{d} t \left[\frac{1}{2}\sum_{<ij>}\left(
\hat{\mathbf{x}}_i(t) \nabla_i V(\mathbf{x}_i -
\mathbf{x}_j-\mathbf{A}_{ij}))+\hat{\mathbf{x}}_j(t) \nabla_j V(\mathbf{x}_j -
\mathbf{x}_i - \mathbf{A}_{ji}) \right) \right] \\
    &= \sum_i \Phi[\mathbf{x}_i(t),\hat{\mathbf{x}}_i(t)] + \sum_{ij} W_{ij}
[\mathbf{x}_i(t),\hat{\mathbf{x}}_i(t),\mathbf{x}_j(t),\hat{\mathbf{x}}_j(t)]
\end{array}
\end{equation}
where the last two lines define $\Phi$ and $W_{ij}$.
Integrations in Eq.~(\ref{eq:P_MSR}) of the variables
$\hat{\mathbf{x}}_i(t)$ are along the imaginary axis.  The quantity
$P[\{\mathbf{x}_i(t)\},\{\hat{\mathbf{x}}_i(t)\}]=\exp \left[
  -\mathcal{S}[\{\mathbf{x}_i(t)\},\{\hat{\mathbf{x}}_i(t)\}] \right]$
may seem mysterious, because of the variables $\hat{\mathbf{x}}_i(t)$,
which do not have an immediate physical meaning.  In order to
understand them, we consider the probability of paths \textit{in the
  presence } of external fields $\{\mathbf{h}_i(t)\}$:
\begin{equation}
 P[\{\mathbf{x}_i(t)\},\{\hat{\mathbf{x}}_i(t)\}, \{\mathbf{h}_i(t)\}
]=\exp \left\{
  -\mathcal{S}[\{\mathbf{x}_i(t)\},\{\hat{\mathbf{x}}_i(t)\}]-\hat{\mathbf{x}}_i(t)\mathbf{h}_i(t)\right\},
\end{equation}
 Integrating over the `hat' variables, we find that:
\begin{equation}
  \begin{array}{rl}
  P[\{\mathbf{x}_i(t)\}, \{\mathbf{h}_i(t)\}]& =\int \prod_{i}
\DD[\hat{\mathbf{x}}_i(t)] P[\{\mathbf{x}_i(t)\},\{\hat{\mathbf{x}}_i(t)\},
\{\mathbf{h}_i(t)\}]\\
&=
  \int \prod_{i} \DD[\hat{\mathbf{x}}_i(t)]
e^{\hat{\mathbf{x}}_i(t)\mathbf{h}_i(t)}P[\{\mathbf{x}_i(t)\},\{\hat{\mathbf{x}}_i(t)\}]
  \end{array}
\end{equation}
In other words, the `hat' variables $\{\hat{\mathbf{x}}_i(t)\}$ are
the Fourier-transform variables of the fields: the probability of a
trajectory $(\{\mathbf{x}_i(t)\},\{\hat{\mathbf{x}}_i(t)\})$ in the
absence of field, is the Fourier transform of the corresponding
physical probability $\{\mathbf{x}_i(t)\}$ in the presence of an
external field $\{\mathbf{h}_i(t)\}$.

The functional formalism casts the dynamical problem into a form that
resembles a partition function, but with one-dimensional objects (the
trajectories) replacing the point particles.  We may exploit the
analogy to repeat the canonical computation in the preceding
subsection. In particular, we may define the `dynamical Mayer
function' as:
\begin{equation}\label{eq:mayer_f_dyn_def}
1+\overline{f}_d[\mathbf{x}_i(t),\hat{\mathbf{x}}_i(t),\mathbf{x}_j(t),\hat{\mathbf{x}}_j(t)]
=  \int 
    \dd \mathbf{A}_{ij} P(\mathbf{A}_{ij}) \exp \left\{ - W_{ij}
[\mathbf{x}_i(t),\hat{\mathbf{x}}_i(t),\mathbf{x}_j(t),\hat{\mathbf{x}}_j(t)]
\right\} 
\end{equation}

\noindent  and introduce the dynamical `partition function':

\begin{equation}\label{eq:dyn_annealed_partition}
  \begin{array}{rl}
    \overline{Z_{N,\tau}}& = \int \prod_i \dd \mathbf{x}^0_i \dd \mathbf{x}^\tau_i 
P_{\{\mathbf{A}\}}(\{\mathbf{x}^0\},\{\mathbf{x}^\tau\}, \tau)\\
& =  \int \prod_{i} \left(\DD[\mathbf{x}_i(t),\hat{\mathbf{x}}_i(t)]
e^{-\Phi[\mathbf{x}_i(t),\hat{\mathbf{x}}_i(t)]} \right) \prod_{ij}
\left(1+\overline{f}_d[\mathbf{x}_i(t),\hat{\mathbf{x}}_i(t),\mathbf{x}_j(t),\hat{\mathbf{x}}_j(t)]\right)
  \end{array}
\end{equation}
In the above expression the path integral is now performed over paths
with free boundary conditions.  The initial conditions are weighted
with a flat distribution. With these small reinterpretations,
$\overline{Z_{N,\tau}}$ looks like the (static) partition function of $N$
'polymers' $(\mathbf{x}(t),\hat{\mathbf{x}}(t))$, in an external
field $\Phi$, interacting via a potential $\ln \left[1+\overline{f}_d\right]$.

Next we introduce a density field:

\begin{equation}\label{eq:traj_rho}
 \rho[\mathbf{x}(t),\hat{\mathbf{x}}(t)]=\sum_i
\delta[\mathbf{x}(t)-\mathbf{x}_i(t)]
\delta[\hat{\mathbf{x}}(t)-\hat{\mathbf{x}}_i(t)]
\end{equation}
Note that $\delta$ is here a Dirac function in the sense of
trajectories, i.e. a product of ordinary deltas, one for each time.
Inserting this field in the partition function we obtain:
\begin{equation}
\prod_{i} e^{-\Phi[\mathbf{x}_i(t),\hat{\mathbf{x}}_i(t)]}=\exp \left[ -  \int
\DD[\mathbf{x}(t),\hat{\mathbf{x}}(t)] \rho[\mathbf{x}(t),\hat{\mathbf{x}}(t)]
\Phi[\mathbf{x}(t),\hat{\mathbf{x}}(t)] \right],
\end{equation}
and:
\begin{equation}\label{eq:field_introduction}
  \begin{array}{rl}
    \prod_{i\neq
j}\left(1+\overline{f}_d[\mathbf{x}_i(t),\hat{\mathbf{x}}_i(t),\mathbf{x}_j(t),\hat{\mathbf{x}}_j(t)]\right)
  &  =  \exp \bigg[ \frac{1}{2} \int
\DD[\mathbf{x}(t),\hat{\mathbf{x}}(t)]\DD[\mathbf{y}(t),\hat{\mathbf{y}}(t)] 
      \\ 
 &  \hspace{-2.7cm} \rho[\mathbf{x}(t),\hat{\mathbf{x}}(t)]
\rho[\mathbf{y}(t),\hat{\mathbf{y}}(t)] \ln
\left(1+\overline{f}_d[\mathbf{x}(t),\hat{\mathbf{x}}(t),\mathbf{y}(t),\hat{\mathbf{y}}(t)]
\right) \bigg].
    \end{array}
\end{equation}

Note that the right-hand side contains a self-interaction term that
was not present on the left-hand side, but this term is negligible
compared to the interparticle interactions.  

As in the static case, looking more closely to the function
$\overline{f}_d$ defined by Eq.~(\ref{eq:mayer_f_dyn_def}), we see that
in the integral over disorder, the integrand is $1$ if the two
trajectories do not interact. These trajectories do interact if the
random shift is able to bring them close to one  another. If the
trajectories explore a finite volume during between times $0$ and
$\tau$ (\textit{ie} $\tau$ is not too large, at least much smaller
than the ergodic time), this is only possible for a finite volume of
integration on the random shift. As the distribution of the shifts is
$P(\mathbf{A})=1/V$, this means that the function $\overline{f}_d$ is
of order $\Gamma/V$, where $\Gamma$ is the typical volume covered by a
trajectory during an time interval $\tau$. Then, just as in the static calculation, we may use that
$\ln(1+\overline{f}_d)\simeq \overline{f}_d$.  Now, imposing the
condition Eq.~(\ref{eq:traj_rho}) via:
\begin{equation}
\begin{array}{rl}
  & \delta \bigg[ \rho[\mathbf{x}(t),\hat{\mathbf{x}}(t)]-\sum_i
\delta[\mathbf{x}(t)-\mathbf{x}_i(t)]
\delta[\hat{\mathbf{x}}(t)-\hat{\mathbf{x}}_i(t)] \bigg]  = \int 
D\left[ \hat{\rho}[\mathbf{x}(t),\hat{\mathbf{x}}(t)]\right]  \\ & \qquad  \qquad e^{i\int
\DD[\mathbf{x}(t),\hat{\mathbf{x}}(t)]  \rho[\mathbf{x}(t),\hat{\mathbf{x}}(t)]
\hat{\rho}[\mathbf{x}(t),\hat{\mathbf{x}}(t)] -
i\hat{\rho}[\mathbf{x}(t),\hat{\mathbf{x}}(t)] \sum_i
\delta[\mathbf{x}(t)-\mathbf{x}_i(t)]
\delta[\hat{\mathbf{x}}(t)-\hat{\mathbf{x}}_i(t)] }
\end{array}
\end{equation}
and integrating over $\mathbf{x}_i(t)$'s and
$\hat{\mathbf{x}}_i(t)$'s, we get, for a system at equilibrium at time
$0$ (we dropped the time dependence of the paths to simplify the notation):
\begin{equation}\label{eq:field_theory_dyn_part_func}
  \begin{array}{rl}
    Z_{N,\tau}  &= \int D[\rho] D[\hat{\rho}]\exp \bigg\{  i\int
\DD[\mathbf{x},\hat{\mathbf{x}}]\rho[\mathbf{x}, \hat{\mathbf{x}}]
\hat{\rho}[\mathbf{x}, \hat{\mathbf{x}}] + N\ln \int
\DD[\mathbf{x},\hat{\mathbf{x}}] e^{-i\hat{\rho}[\mathbf{x}, \hat{\mathbf{x}}]}
\\ 
& - \int \DD[\mathbf{x},\hat{\mathbf{x}}] \rho[\mathbf{x}, \hat{\mathbf{x}}]
\Phi[\mathbf{x}, \hat{\mathbf{x}}]
 + \frac{1}{2} \int \DD[\mathbf{x},\hat{\mathbf{x}}]
\DD[\mathbf{y},\hat{\mathbf{y}}]  \rho[\mathbf{x}, \hat{\mathbf{x}}] 
\rho[\mathbf{y}, \hat{\mathbf{y}}]
\overline{f}_d[\mathbf{x},\hat{\mathbf{x}},\mathbf{y},\hat{\mathbf{x}}]
\bigg\}
  \end{array}
\end{equation}
For the mean-field dynamics we can take the saddle-point with respect to $\rho$ and $\hat \rho$ of the last
equation. This reads:
\begin{equation}
\begin{array}{rl}  
  \rho[\mathbf{x}, \hat{\mathbf{x}}] & =N\frac{e^{-i\hat{\rho}[\mathbf{x},
      \hat{\mathbf{x}}]}}{\int \DD[\mathbf{x},\hat{\mathbf{x}}]
    e^{-i\hat{\rho}[\mathbf{x}, \hat{\mathbf{x}}]}}, \\
  \hat{\rho}[\mathbf{x}, \hat{\mathbf{x}}] & =-i\Phi[\mathbf{x}, \hat{\mathbf{x}}] + 
  i\int \DD[\mathbf{y},\hat{\mathbf{y}}]  \rho[\mathbf{y}, \hat{\mathbf{y}}]
  \overline{f}_d[\mathbf{x},\hat{\mathbf{x}},\mathbf{y},\hat{\mathbf{y}}].
\end{array}
\end{equation}
which gives a closed equation on the density of paths $\rho[\mathbf{x},
\hat{\mathbf{x}}]$:
\begin{equation}\label{eq:saddle_dynamics}
  \rho[\mathbf{x}, \hat{\mathbf{x}}]  = \frac{1}{\mathcal{N}} e^{-\Phi[\mathbf{x}, \hat{\mathbf{x}}] + \int \DD[\mathbf{y},\hat{\mathbf{y}}]  \rho[\mathbf{y}, \hat{\mathbf{y}}]
    \overline{f}_d[\mathbf{x},\hat{\mathbf{x}},\mathbf{y},\hat{\mathbf{y}}] }
\end{equation}
where $\mathcal{N}$ ensures that the density is normalized to $N$.

\noindent Reinserting this equation in the partition function, we get the functional $\mathcal{S}=- \ln Z$, which reads:
\begin{equation}\label{eq:free_energy_dynamics}
\begin{array}{rl}
\mathcal{S}[\rho] &=   \int \DD[\mathbf{x},\hat{\mathbf{x}}] \rho[\mathbf{x},
\hat{\mathbf{x}}] \ln \left[\rho[\mathbf{x}, \hat{\mathbf{x}}] \right]  + \int
\DD[\mathbf{x},\hat{\mathbf{x}}] \rho[\mathbf{x}, \hat{\mathbf{x}}] \Phi[\mathbf{x},
\hat{\mathbf{x}}]  +\mathcal{S}_{int}[\rho] -N\ln N \\
 & \mathcal{S}_{int}[\rho]=
 - \frac{1}{2} \int \DD[\mathbf{x},\hat{\mathbf{x}}]
\DD[\mathbf{y},\hat{\mathbf{y}}]  \rho[\mathbf{x}, \hat{\mathbf{x}}] 
\rho[\mathbf{y}, \hat{\mathbf{y}}]
\overline{f}_d[\mathbf{x},\hat{\mathbf{x}},\mathbf{y},\hat{\mathbf{y}}]
\end{array}
\end{equation}
Eq.~(\ref{eq:saddle_dynamics})  is the counterpart  of the saddle point
equation~(\ref{eq:saddle_statics}) obtained in the previous subsection for the average density.
As happens often in this kind of problems, even the mean-field equations are hard to solve, in this case
because the complexity of an object like $\rho[\mathbf{x}, \hat{\mathbf{x}}]$
makes the problem  intractable. But, just as in equilibrium calculations, we can
try, as a further approximation, to find extrema of the free energy
Eq.~(\ref{eq:saddle_dynamics}) in a well chosen restricted subspace of
all possible $\rho[\mathbf{x}, \hat{\mathbf{x}}]$. The next subsection
is devoted to the search of such an ansatz.

\subsection{An approximation in terms of two-point functions}

In terms of physical quantities, 
the simplest trial form for the probabilities is to propose a Gaussian form:
\begin{eqnarray}
 & &\ln  P[\{\mathbf{x}(t)\}, \{\mathbf{h}(t)\}] = \nonumber \\
  & & \;\;\;\;\;\; -\frac{1}{2} \int \dd t_a \; \dd t_b \; \{ A_1 (t_a,t_b) {\bf h}(t_a) {\bf h}(t_b) +
 A_2 (t_a,t_b) {\bf h}(t_a){\bf  x}(t_b) +A_3 (t_a,t_b) {\bf x}(t_a) {\bf x}(t_b) \}
\end{eqnarray}
This leads, equivalently, to proposing an ansatz that is Gaussian in the $\{{\bf x,{\hat x}}\}$ variables.
This ansatz has to be invariant with respect to a time-independent translation 
of the trajectory, which may be imposed, in terms of the ${\bf x, {\hat{x}}}$ variables, by the form:
\begin{equation}
  \begin{array}{rl}
 &  P[\{\mathbf{x}(t)\},
  \{      {    \bf  {\hat{x}}(t) }    \}] 
  = \int \dd {\bf \bar x} \; \; \exp \bigg[
    -\frac{1}{2} \int \dd t_a \; \dd t_b \; \{ {\tilde B} (t_a,t_b)  {    \bf  {\hat{x}}}(t_a) {    \bf  {\hat{x}}}(t_b)  \\
  &   \;\;\;\;\;\;\; + {\tilde R} (t_b,t_a)  {    \bf  {\hat{x}}}(t_b) ( {\bf x}(t_a) -{\bf \bar x})+
 {\tilde R} (t_a,t_b)  {    \bf  {\hat{x}}}(t_a) ( {\bf x}(t_b) -{\bf \bar x}) + {\tilde D} (t_a,t_b) ({\bf x}(t_a)-{\bf \bar x})( {\bf x}(t_b)-{\bf \bar x)} \}\bigg]
 \end{array}
\end{equation}
where the integration over ${\bf \bar x}$ is over the whole volume and implements the translational invariance of the
ansatz, i.e. that the quadratic form has a zero-mode in the translations. (A similar strategy has been previously used
in a static replica ansatz, and it is the same idea as the Hill-Wheeler integral that imposes rotational invariance in nuclear theory).
It will turn out that, in equilibrium, the ansatz satisfies the causality and fluctuation-dissipation relations :
\begin{equation}
 \begin{array}{rl}
{\tilde D}(t,t')&=0  \\
{\tilde R(t,t')} &= -\frac{1}{T} \frac{\partial }{\partial t} {\tilde B}(t-t') \Theta(t-t')  
 \end{array}
\end{equation}

The calculation is cumbersome, and we leave it for Appendix B.
The result is expressed in terms of the two-time correlation function:
\begin{equation}
B(t-t') = \frac{1}{N} \sum_a \; \langle ({\bf x}_a(t) - {\bf x}_a(t'))^2 \rangle
\label{bbb}
\end{equation} 
where the average is over the Langevin noise and the initial conditions.
It may be written in the (superficially) Mode-Coupling-like format:
\begin{equation}\label{eq:dynamic_equation}
  \frac{\partial B(t)}{\partial t_a}=\int  \dd t' \Sigma_R[B](t-t') B(t') + 2T
  \end{equation}
  with  $t>0$.
This form is quite general, what defines our equation is the kernel $\Sigma_R$, which in Mode-Coupling equations is a simple function of $B$, and here is computed as follows. Define first the response function $R(t)$, which satisfies the Fluctuation-dissipation theorem:
\begin{equation}
R(t) = -\frac{1}{T} \frac{\partial B(t)} { \partial t} \Theta(-t)
\end{equation}

\begin{figure}
  \centering
  \includegraphics[width=7cm]{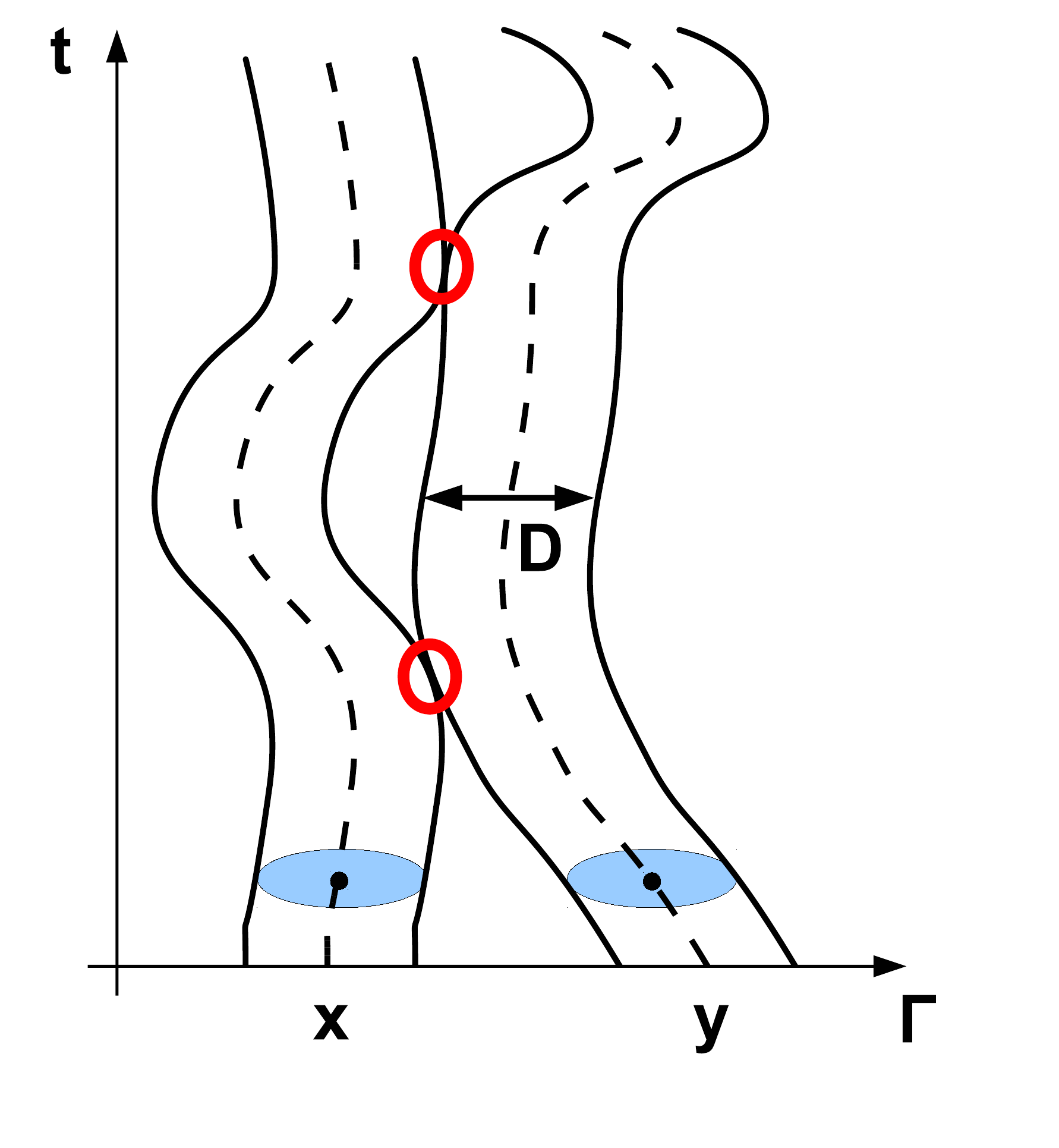}
  \caption{The trajectories contributing to a Mayer diagram are those that come into interaction range at some time.}
\label{fig:dynamics_interaction}

\end{figure}

\noindent Then the kernel is:
\begin{equation}
    \Sigma_R(t_a-t_b)= \frac{2}{T}\int \dd t_{a'} \dd t_{b'} \;  {R}^{-1} (t_a-t_{a'})
\;  \frac{ \partial }{ \partial t_{a'}}\langle    \mathbf{x}(t_{a'}) \mathbf{x}(t_{b'}) \rangle_{int}  \;   {R}^{-1}(t_{b'}-t_b)
\end{equation}
where $R^{-1}$ is the inverse through convolution of $R$:
\begin{equation}
\int \dd t' \; R(t-t') R^{-1}(t') = \delta(t)
\label{inverse}
\end{equation}
and may be obtained from $R$ easily with Laplace transforms.
Self-consistency is imposed by the definition 
\begin{equation}
  \begin{array}{rl}
{\tilde R}&= R^{-1}  \\
{\tilde B} (t)&=  \int \dd t" \dd t' \; R^{-1} (t-t'' )B(t"-t')  R^{-1} (t' )
  \end{array}
\end{equation}
The trajectories that contribute to the averages $\langle \bullet \rangle_{int} $ are those in which two particles
  enter {\em at any intermediate time} within the interaction range, as depicted in figure \ref{fig:dynamics_interaction}, otherwise their contribution vanishes, as in a static Mayer expansion.

\section{Numerical results from $\lambda=\infty$ to $\lambda=0$}\label{sec:num_liq}

In order to investigate the system beyond  mean-field, we studied 
the finite $\lambda$ case by numerical simulations. We used two different
 Monte-Carlo algorithms, an isobaric-isothermal one for the study of the 
equation of state, and an isovolumic-isothermal one for the study of the 
dynamics close to the glass transition.
We worked with systems of $864$ particles. The system we looked at for the 
bidisperse case is a 50:50 mixture of particles 
of diameter $1$ and $1.4$, a common choice for a three dimensional 
bidisperse glass former \cite{o2002random,PhysRevLett.102.085703}.

We equilibrate the system at increasing pressures by annealing simulations. We 
check carefully that equilibrium is reached at every studied pressure by 
performing annealings with several compression rates, and ensuring that they 
give the same values for a given observable, \textit{e. g.} density. Close to 
the glass transition, we also check that no aging is visible in the system.

\subsection{Simulations in the mean-field case}

In the $\lambda \to \infty$ limit, equilibrating the system can be
much easier, thanks to the `planting' technique
\cite{krzakala2009hiding,achlioptas2008algorithmic}: for
mean-field problems with quenched random variables, if the annealed
free energy is exact ($\overline{\ln Z}=\ln \overline{Z}$), one can
create an instance in thermal equilibrium  by
taking a random set of variables (here particle positions), and
looking for a disorder (here random shifts) compatible with this set,
\textit{ie} leading to $H(\{\mathbf{x}\}, \{\mathbf{A}\})=0$. One can show that, within mean-field, this 
creates no bias in the measure.
In our model, this means that we can generate safely an equilibrium
instance as long as the annealed entropy Eq.~(\ref{eq:F_mean_field}) is
valid, \textit{ie} as long as the liquid is the equilibrium phase.
Finding an equilibrium
instance in this ensemble requires at most a few tens of seconds for a
system with $1000$ particles on a desktop computer. This of course, is
very useful as we can get easily equilibrated configurations with
densities close and even \textit{above} the dynamic glass transition
density, as long as we do not reach the Kauzmann transition point, if it
exists.

\subsection{Equation of state}

\begin{figure}
  \centering
  \includegraphics[width=0.7\columnwidth]{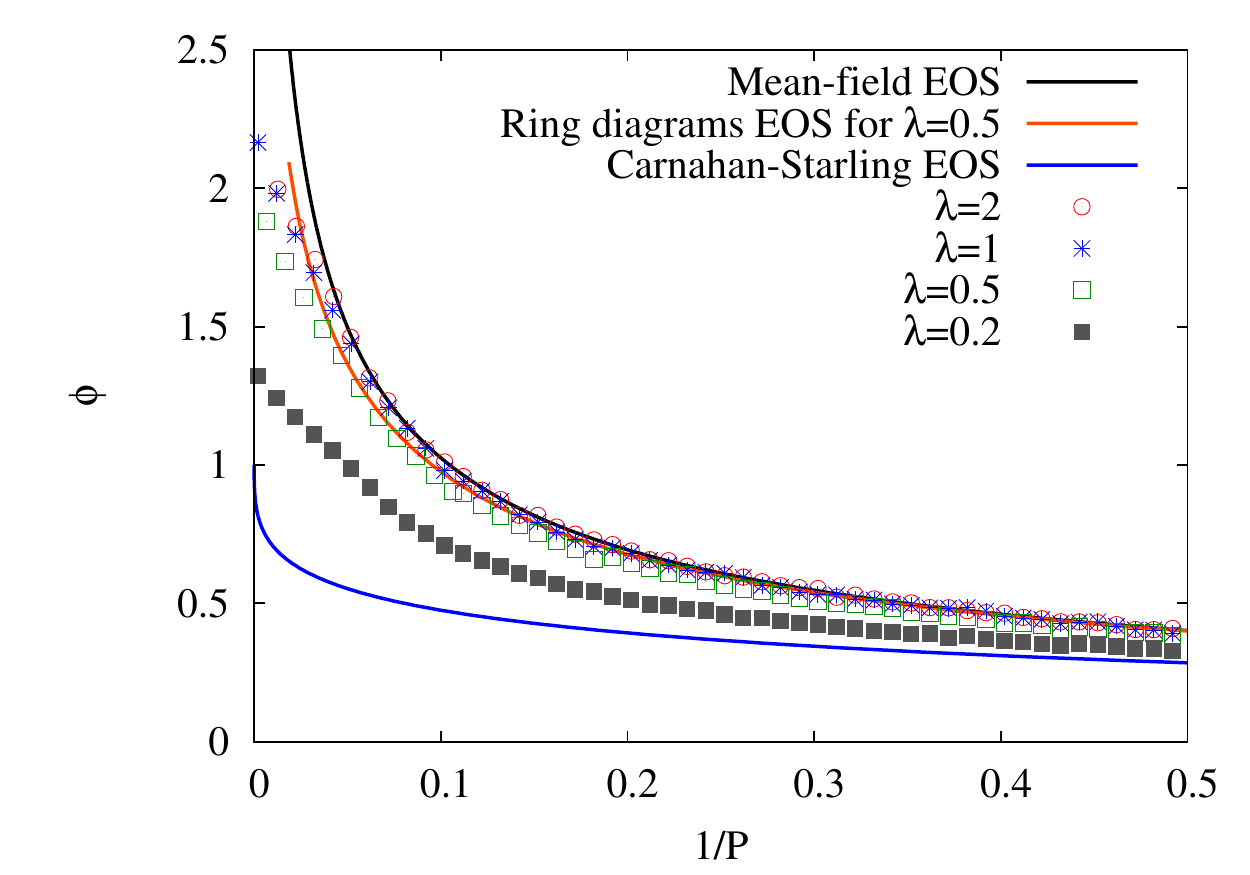}
  \caption{The equation of state of the model for different values of the 
shift range $\lambda$. Solid lines are the analytical equation of state for the mean-field case 
(for which it is exact), and for the monodisperse $3d$ hard-sphere system 
(the  Carnahan-Starling approximation). Points are simulation 
results. The equation of state sticks to the mean-field value for shifts 
larger than $\lambda=1$, except at high pressures where a glass transition prevents 
equilibration.}
\label{fig:EOS}
\end{figure}

Already at the level of the equation of state, we can notice that the
system behaves in a mean-field way even for random shifts as small as
$\lambda=1$. As shown in Fig. \ref{fig:EOS}, all the distance between
the mean-field and the $3d$ hard sphere equation of state is covered by systems
with $0<\lambda<1$.  The  inclusion of ring diagrams does not show any
noticeable difference with the leading order  for $\lambda>1$, but it takes the agreement down to
 $\sim \; \lambda=0.5$, which is far beyond its expected domain of applicability ($\lambda\gg1$). 

The  monodisperse hard-sphere system without disorder undergoes a first
order transition towards a crystal, as does the system with  small
values of $\lambda$. We have checked that nothing occurs in
the equation of state for systems with shifts as small as
$\lambda=0.15$. For smaller shifts, a weak first order transition is
visible, which increases when we decrease $\lambda$.

\subsection{Pair correlation function}

\begin{figure}
  \centering
    \includegraphics[width=0.7\columnwidth]{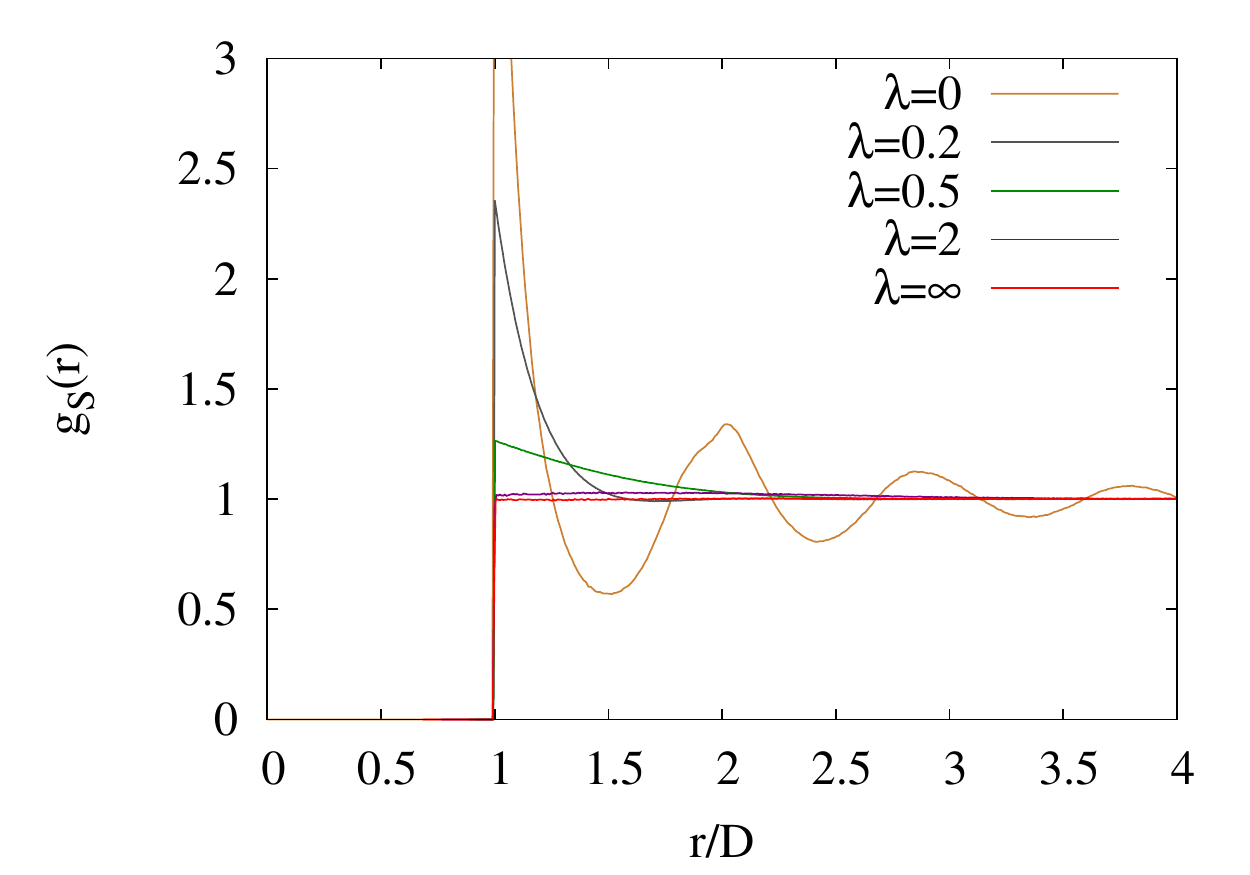}
    \caption{Pair correlation function $g_S(\mathbf{x}-\mathbf{y})$
      (Eq.~(\ref{eq:pair_correl_shifted_def})) of a system  of
      $N=864$ particles with several values of $\lambda$.}
  \label{fig:pair_correlation}
\end{figure}

We compare the pair correlation function obtained by MC simulations
with analytical results of section \ref{sec:eos_liq}. In
Fig.~\ref{fig:pair_correlation} we show that for
$g_S(\mathbf{x}-\mathbf{y})$ (Eq.~(\ref{eq:pair_correl_shifted_def})),
the analytical form derived for large shifts is verified.
For finite $\lambda$, the usual structure of the $3d$ hard-sphere
model appears gradually as we reduce $\lambda$.

\section{Dynamic glass transition}\label{sec:dyn_glass}

In this section, we show that in $d=3$, we observe numerically the
presence of a dynamic glass transition at finite pressure, and we
study some dynamical properties close to the transition. A numerical
simulation cannot exclude that the transition pressure scales with the
system size (for example as $P_c \sim \ln N$), but we have given
arguments in the previous sections that favor the finite-$P_c$
scenario.
Note that, on the contrary, we expect that the Kauzmann pressure $P_K$ might well scale
as $\log N$, but we shall not try to prove this in this paper.

Our model allows to study various interesting aspects of the dynamic
glass transition. First, we can follow the qualitative evolution of
the glass transition from mean-field to finite-dimensions. In
particular we ask if we can find in the mean-field limit generic
features of the mode-coupling theory, and try to see where these
features break down as we move away from mean-field. Secondly, as we
can generate equilibrium mean-field configurations at pressures larger
than the dynamic glass transition, we can numerically access any
desired property in the region between dynamic and equilibrium
transitions $P_c<P<P_K$ (but only at infinite $\lambda$).

\subsection{Dynamic glass transition}

As is well known, mean field approximations give a dynamic transition (e.g. the mode-coupling transition)
which is in fact avoided, thanks to activated processes. Here the situation is conceptually more clear:
for $\lambda=\infty$ we expect activated processes to be absent, and for finite $\lambda$ we expect them
to destroy any pure dynamic transition. For sufficiently large $\lambda$, one may expect that the trace of the 
dynamic transition is quite clear. The (avoided) dynamic transition pressure will still have a dependence on the range,
since even the non-activated dynamics depends on the $\lambda$.

\begin{figure}
  \centering
  \hspace{-1.cm} 
  \includegraphics[width=0.48\columnwidth]{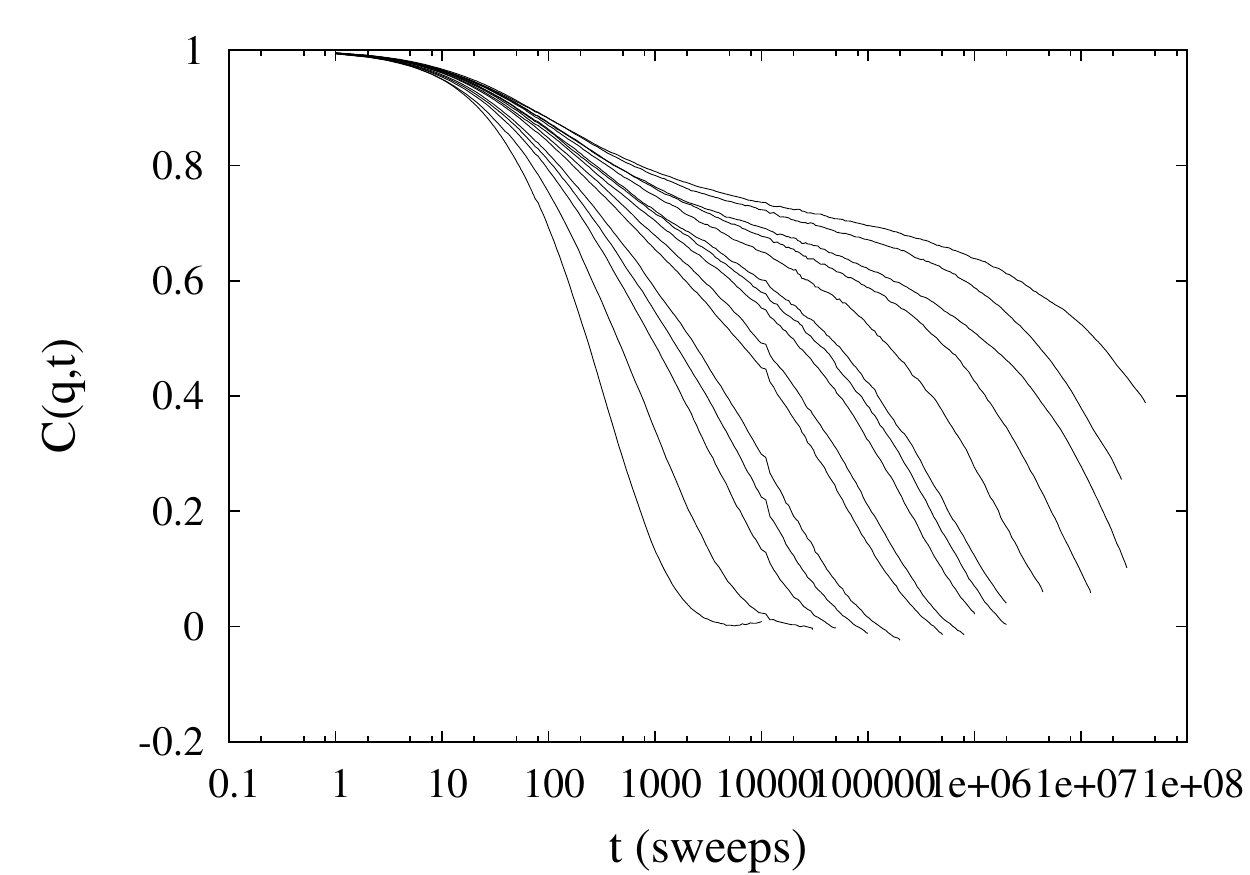}
   \hspace{0.8cm}
  \includegraphics[width=0.49\columnwidth]{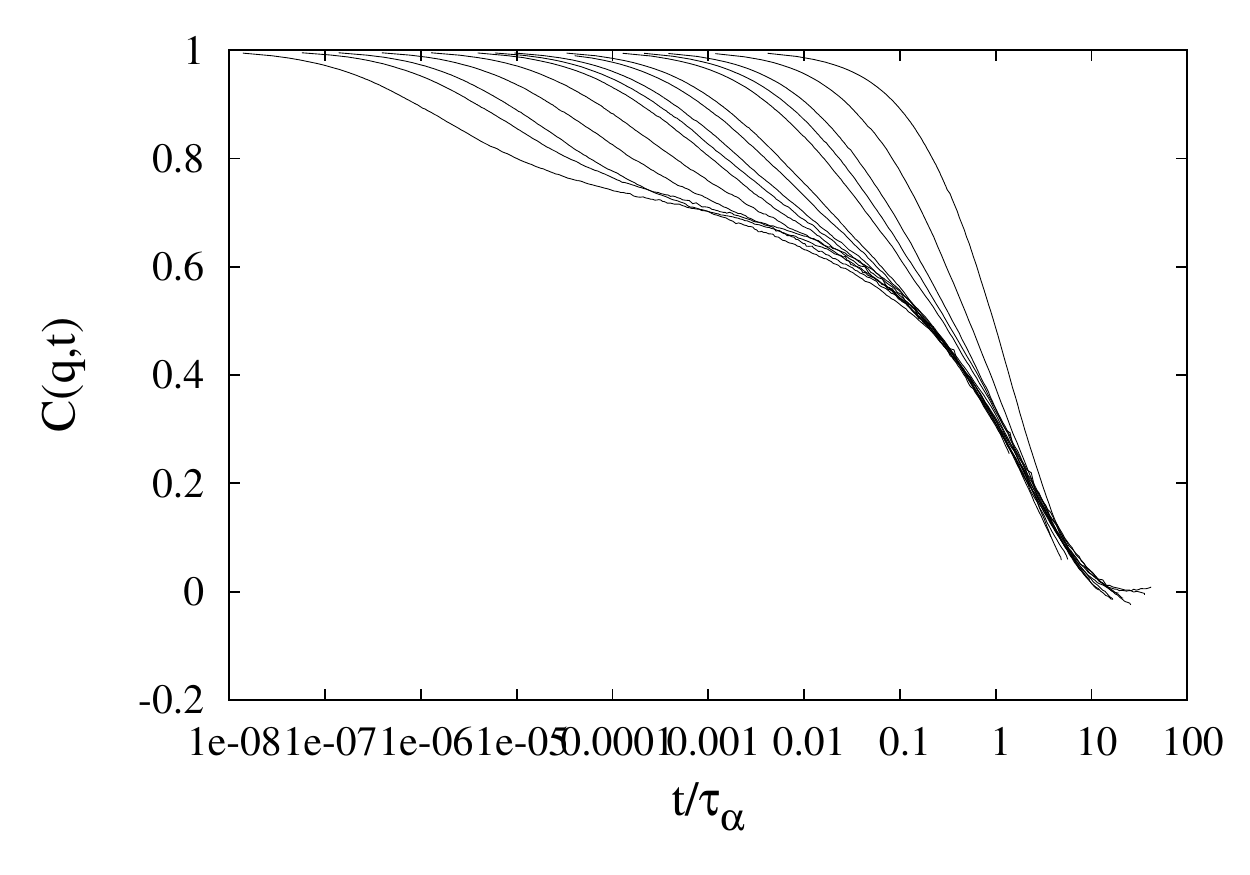}
  \caption{\textbf{Left} Two-time correlation Eq.~(\ref{eq:def_correlation}) for the mean-field model for densities between $\phi=0.531$ and $\phi=1.7478$. There is a plateau appearing upon compression, and the time needed to escape this plateau increases strongly with pressure, signaling the existence of a dynamic glass transition. \textbf{Right}: Same data rescaled by the relaxation time $\tau_{\alpha}$. For high densities, time-density superposition holds.}
\label{correl_RS10}
\end{figure}

To specify this glass transition we look at three standard
quantities: relaxation time, diffusion coefficient and dynamic
susceptibility $\chi_4(t)$.  First we look at the relaxation time of
the system obtained from a two-time correlation function:
\begin{equation}
  \label{eq:def_correlation}
  C(q,t)=\frac{1}{3N} \sum_{i=1}^{N} \sum_{\alpha=1}^{d} \cos \left(q(\mathbf{x}_i^{\alpha}(0)-\mathbf{x}_i^{\alpha}(t)) \right)
\end{equation}
The curves shown is this article were obtained with $q=\pi$.
The relaxation time is defined by the time rescaling that gives the best time-pressure superposition in the $\alpha$ regime as shown in Fig. \ref{correl_RS10}.

\begin{figure}
  \centering
  \hspace{-1.cm} 
  \includegraphics[width=0.48\columnwidth]{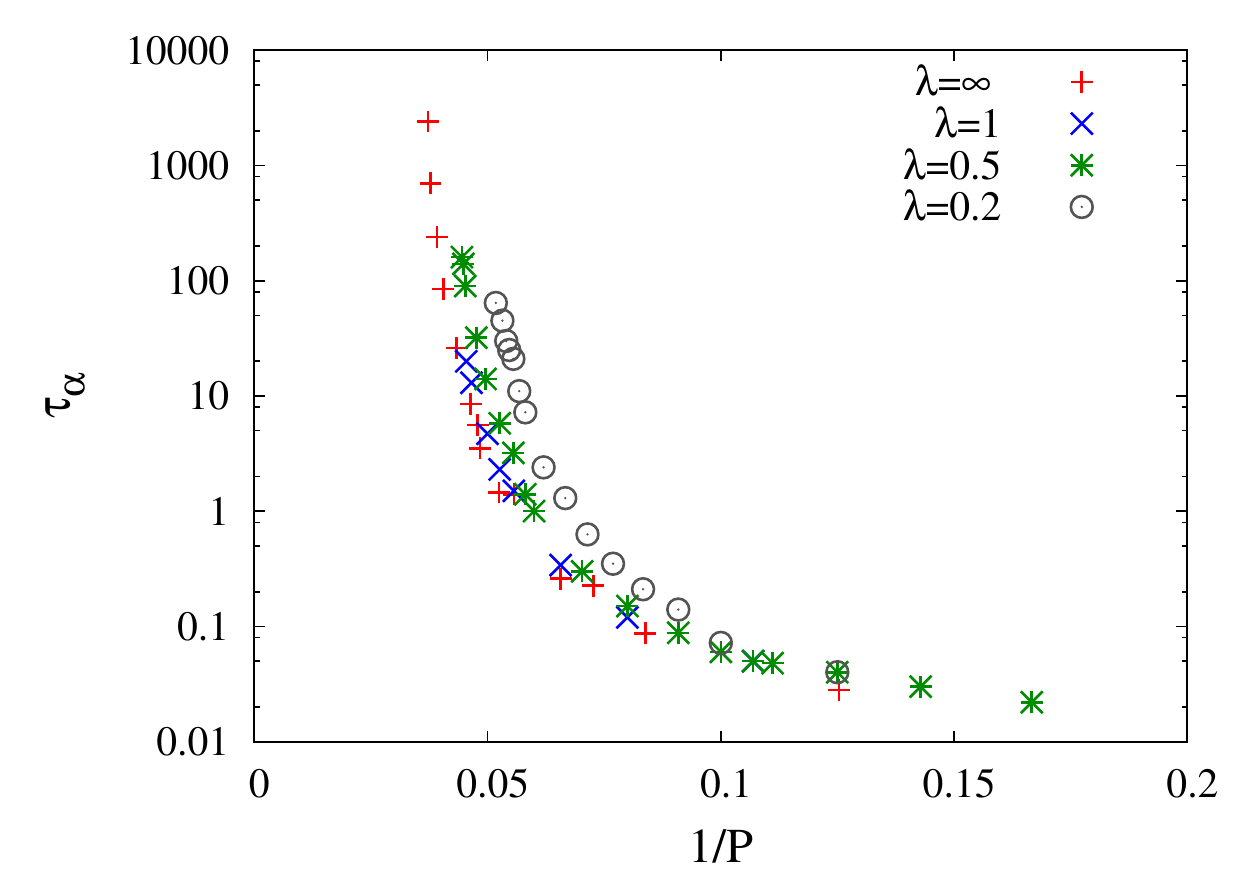}
   \hspace{0.8cm}
  \includegraphics[width=0.48\columnwidth]{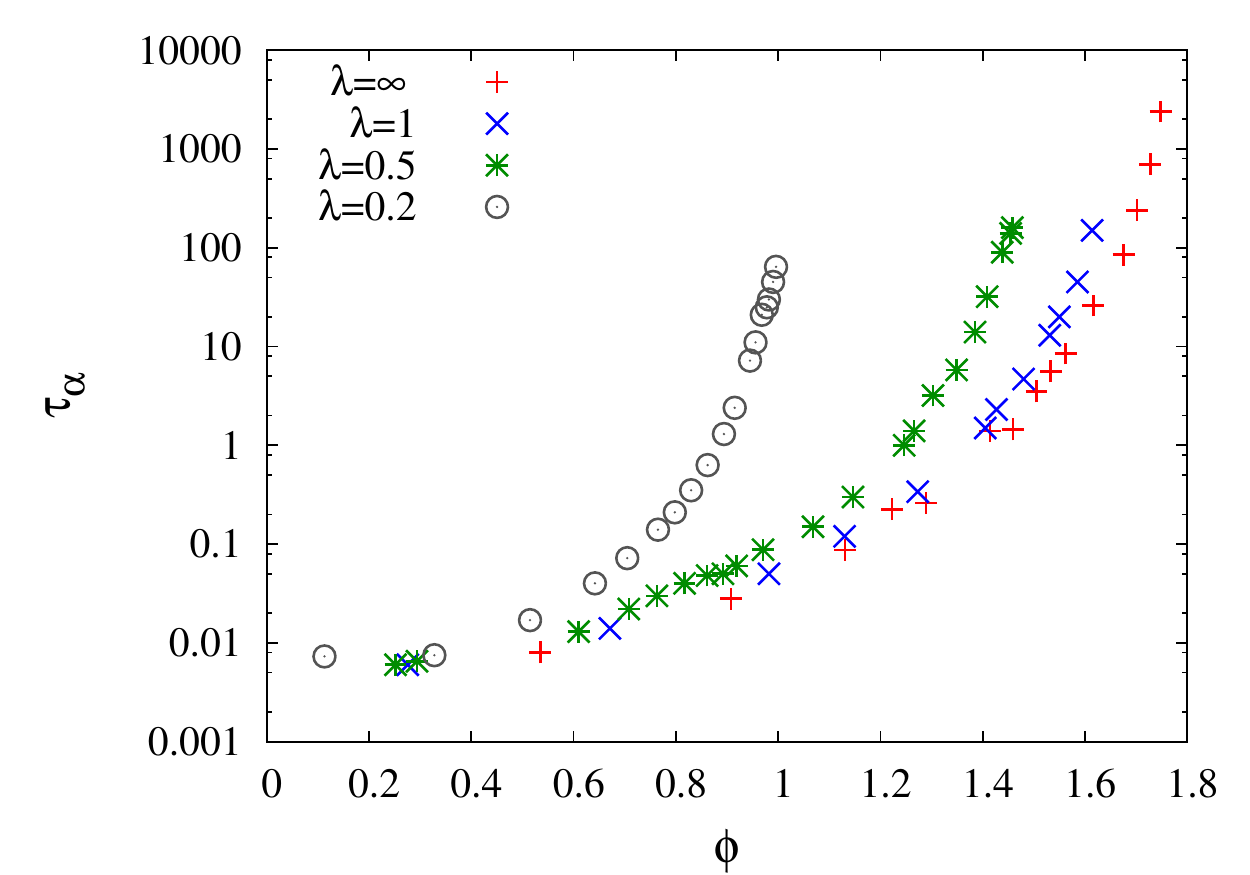}
  \caption{\textbf{Left:} Relaxation time $\tau_{\alpha}$ as a function of the pressure for several values of $\lambda$. The relaxation is clearly super-Arrhenius, and there is a  divergence at a finite pressure for $\lambda=\infty$. For finite $\lambda$,
  the (apparent) dynamic transition pressure  (the analogue of the Mode-coupling transition)  slowly decreases  when $\lambda$ decreases. \textbf{Right:} Relaxation time $\tau_{\alpha}$ as a function of the density. The glass transition density shows a much broader change with the range of the disorder.}
\label{fig:tau_vs_P_RS10}
\end{figure}

The relaxation time shows a super-Arrhenius dependence with pressure
(Fig. \ref{fig:tau_vs_P_RS10}), and seems to diverge for a finite
pressure (or density) value, which is characteristic of a fragile
glass former.  This is independent of the range of the
disorder. Varying $\lambda$ from $\infty$ to $0.2$ increases the glass
transition pressure $P_d$ by  at most 50\%, while the transition
density $\phi_d$ drops by almost a factor 2. This is to be related to
the large shift in the equation of state as we go from a mean-field to a $3d$ system, as shown in the previous
section .

\begin{figure}
  \centering
  \hspace{-0.5cm}
  \includegraphics[width=0.48\columnwidth]{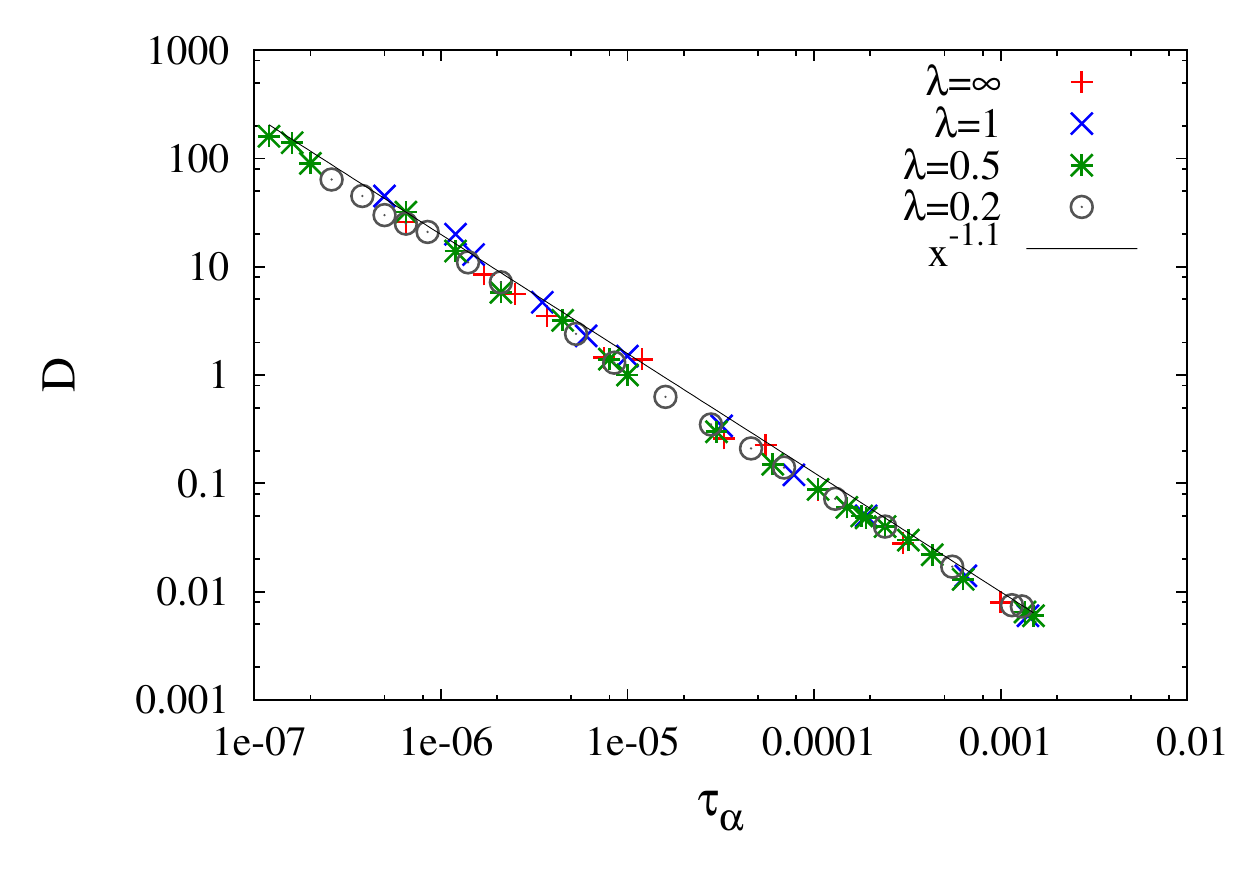}
  \hspace{0.5cm}
  \includegraphics[width=0.48\columnwidth]{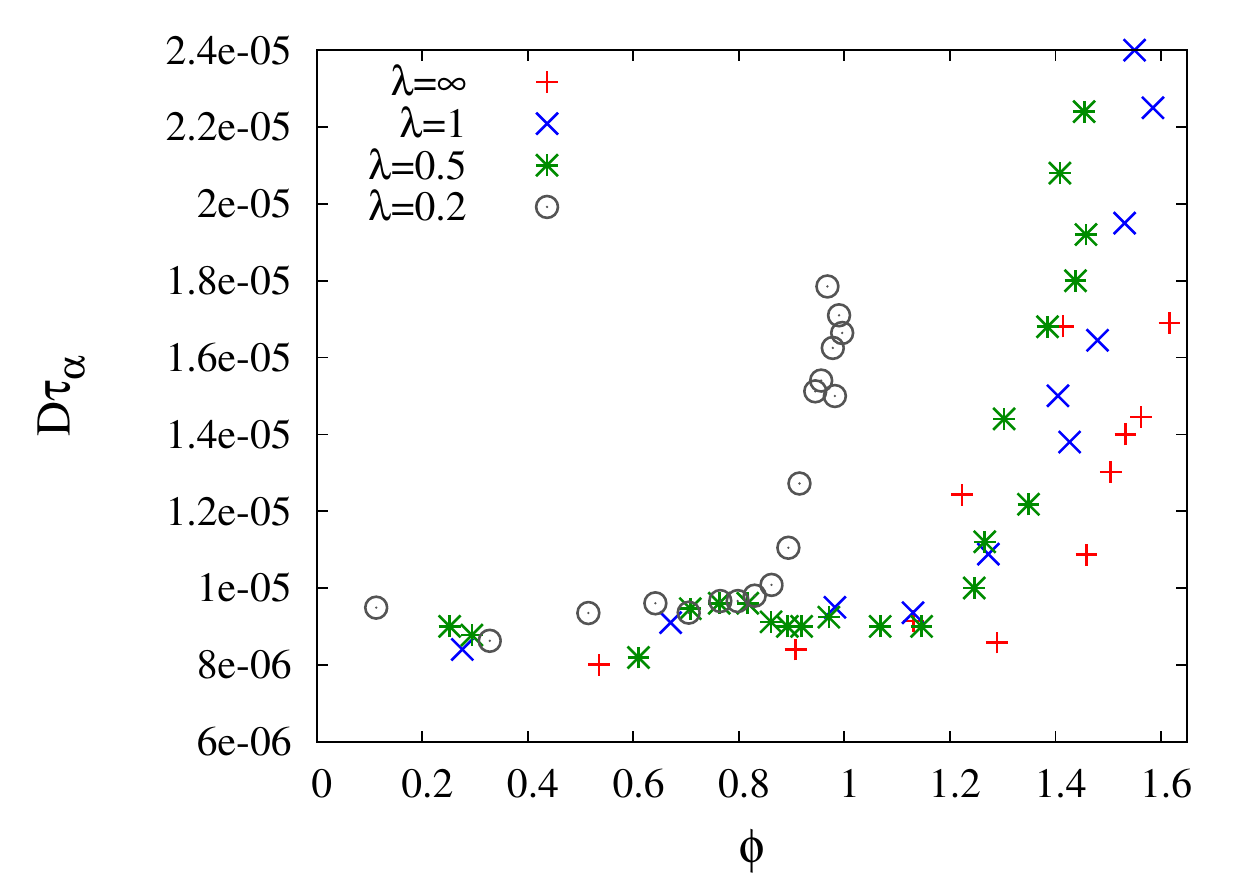}
  \caption{\textbf{Left} Diffusion coefficient as a function of the
    relaxation time for several values of $\lambda$. The diffusion
    coefficient indicates a glass transition at the same density than
    the relaxation time. \textbf{Right} Evolution of the product $D
    \tau_{\alpha}$ with density. The Stokes-Einstein relation predicts a
    constant $D\tau_{\alpha}$.}
\label{fig:D_vs_tau}
\end{figure}

The same divergence is found for the diffusion coefficient $D$. In
Fig.~\ref{fig:D_vs_tau}, we plot the relation between $D$ and
$\tau_{\alpha}$, showing a weak violation of the Stokes-Einstein
relation, which is commonly observed for supercooled liquids. Note
that this curve shows no dependence on $\lambda$, and extrapolates smoothly 
to the mean-field limit, which is quite
remarkable, as the violation of Stokes-Einstein relation is often
explained by the existence of dynamic heterogeneities, which depend
on $\lambda$.

A natural quantity to study in a fragile glass former is the $\chi_4(t)$ susceptibility, a measure
of dynamic heterogeneities. It is defined as the variance of the correlation $C$:
\begin{equation}\label{eq:def_X4}
  \chi_4(t)=N(<C(q,t)^2>-<C(q,t)>^2)
\end{equation}
For fragile glass former, this quantity exhibits a peak
on a time scale $\sim \tau_{\alpha}$ with a  height that sharply
increases when approaching the glass transition from the liquid (low pressure)
side. For pressures above the dynamic glass transition (if there is one),
we expect $\chi_4(t)$ to  saturate to a plateau for a time $\sim
\tau_{\beta}$, the characteristic time of the $\beta$-relaxation, with a 
height that increases closer to the dynamic transition. When $\lambda$
is finite, we cannot reach the glass  region,  as it is impossible to
equilibrate a system with an infinite relaxation time
$\tau_{\alpha}$. However, when $\lambda \to \infty$, we can use the
'planted-configuration' ensemble mentioned above to generate an
equilibrium configuration even at a pressure larger than the dynamic
glass transition pressure.

\begin{figure}
  \hspace{-2cm}
  \includegraphics[width=0.5\columnwidth]{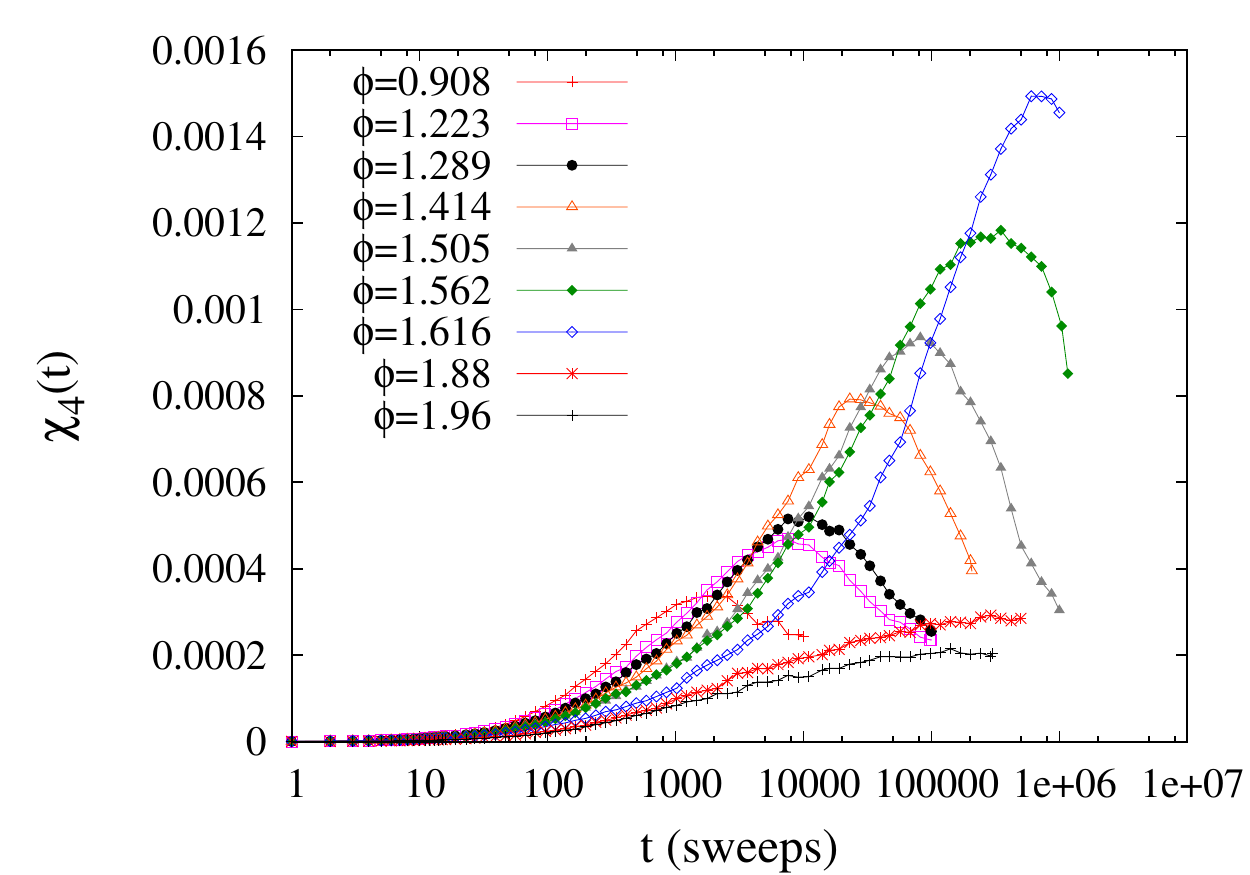}
  \includegraphics[width=0.5\columnwidth]{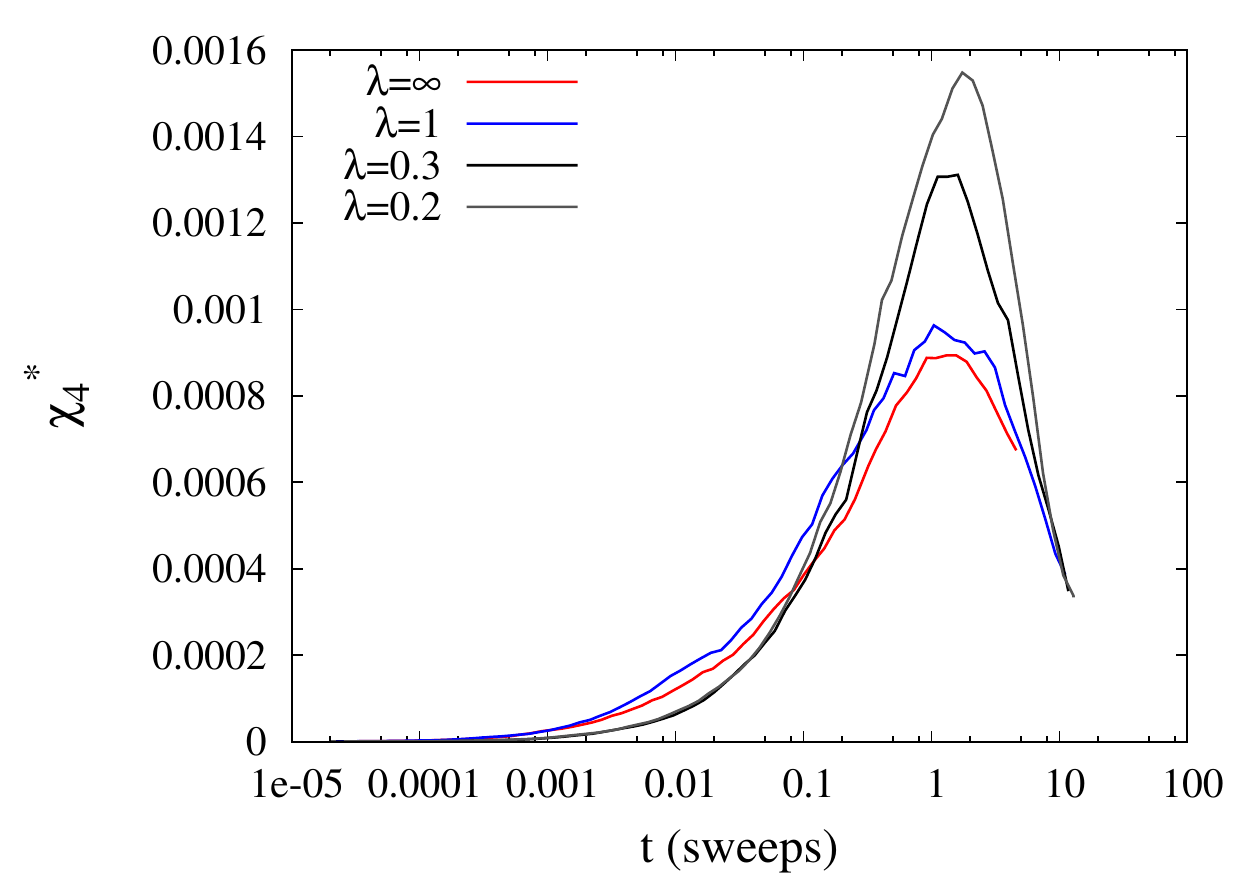}
  \caption{\textbf{Left:} Dynamic susceptibility $\chi_4(t)$ (\ref{eq:def_X4}) for the mean-field model at several densities. The peak position and value grow  
rapidly on approaching  to the glass transition.
\textbf{Right:} $\chi_4(t)$ with different values of the range of random 
shifts $\lambda$, for equal relaxation times. Going from $\lambda=\infty$ to 
$\lambda=1$ does not affect the dynamic heterogeneities, while for smaller values of $\lambda$ (closer to 
finite a dimensional system), heterogeneities  increase strongly.}
\label{fig:X4}
\end{figure}

A typical result in the mean-field limit of our model for $\chi_4(t)$
is shown in Fig.~\ref{fig:X4}. The two expected behaviors -- growth from above and from below -- are observed,
indicating that we crossed a dynamic glass transition. The location of
this qualitative change coincides with the point where the relaxation
time seems to diverge, and also to the analytic estimate of the transition pressure we give below.

The dependence of $\chi_4(t)$ on $\lambda$ (at a given relaxation
time) again shows little difference between the $\lambda=1$ and the
mean-field cases (see Fig.~\ref{fig:X4}). Further from mean-field,
the height of $\chi_4(t)$ depends on $\lambda$, indicating a clear
enhancement of the heterogeneities when we get close to the $3d$
system. This is consistent with the idea that a long ranged disorder 
correlates large regions and thus inhibits heterogeneities occurring on 
smaller scales.

\begin{figure}
  \centering
  \includegraphics[width=0.7\columnwidth]{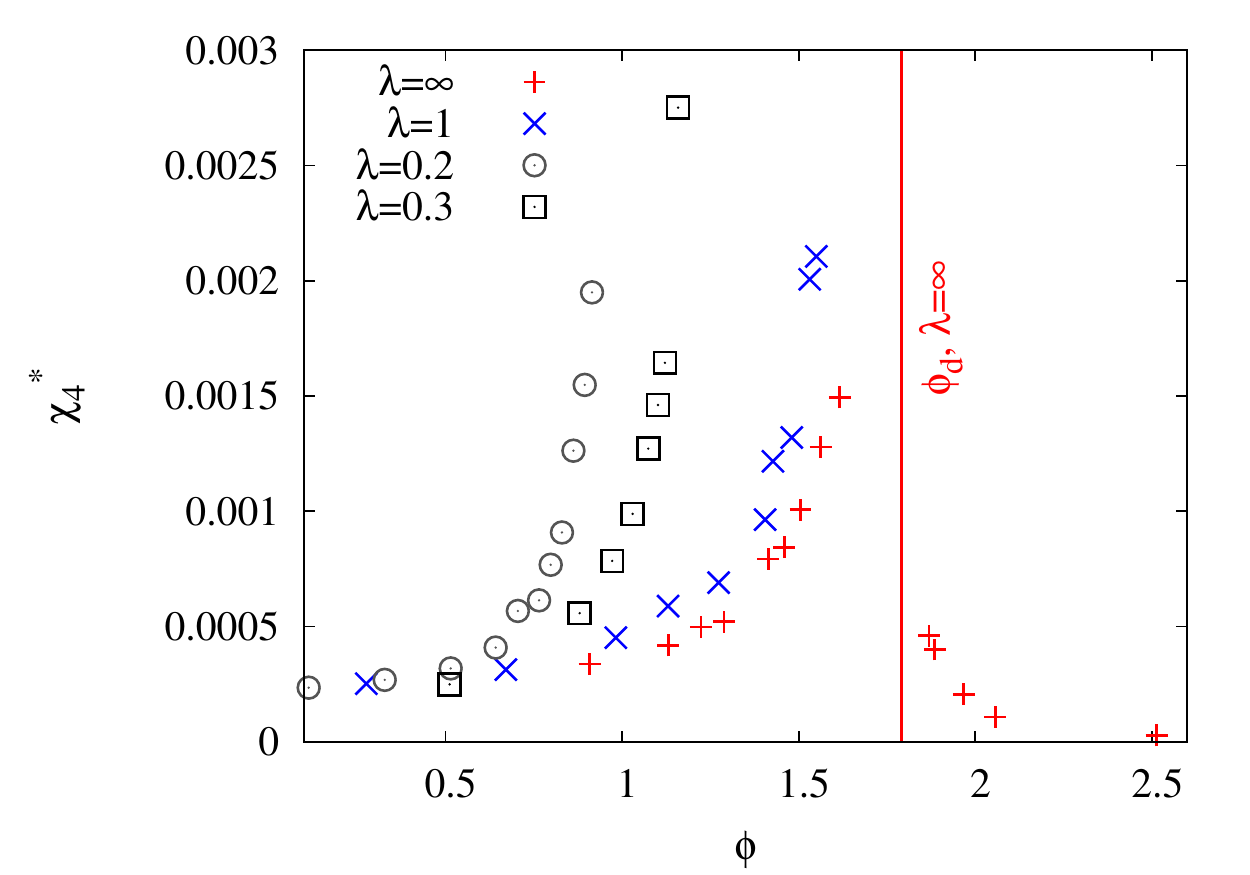}
  \caption{Peak of the dynamic susceptibility $\chi_4^*$ as a function of the density for several values of $\lambda$. This quantity shows a slight difference between $\lambda=1$ and the mean-field model.}
\label{fig:X4_vs_phi}
\end{figure}

We can infer the location of the transition by considering the  divergence of the
peak value $\chi_4^*$ of the dynamic susceptibility. This leads to
the results compatible with those obtained from  the relaxation time. We plotted the density
dependence of the peak value $\chi_4^*$ in
Fig.~\ref{fig:X4_vs_phi}. Note that in the mean-field case, the
divergence can be observed on both sides of the glass transition.

It is striking that, for all these quantities, the mean-field
behavior seems to be quite close to the $3d$ hard-sphere model. There
is no dramatic change between $\lambda \to \infty$ and
$\lambda=1$. This is also probably true for other choices of the
potential $V$, suggesting  that one can create very simply a mean-field
caricature  of any finite-dimensional glass former by adding random shifts
with a range of the order of the range of the potential.

\subsection{The approach to the dynamic transition}

The mode-coupling approximation predicts
that the timescale 
$\tau_{\alpha}$ diverges algebraically with the 
distance to the glass transition density $\phi_{c}$ with an exponent 
$-\gamma$. 
\begin{equation}
  \tau_{\alpha} \sim \left(\frac{\phi - \phi_{c}}{ \phi_{c}} \right)^{-\gamma}
  \label{mctt}
\end{equation}

In real life, the mode-coupling transition becomes at best a crossover, and the question
is to what extent should one believe, and in what temperature-pressure range, extrapolations
within  mode-coupling functional forms.
In our case, we also expect a divergence   in the limit $\lambda=\infty$. What is interesting about this model, is that we
may make $\lambda$ gradually smaller and follow the transition as it becomes a crossover, and keep track on
the interpolations as they become less and less obvious, right down to the original particle model.

In order to test the behavior (\ref{mctt}), we have to fit our simulation data
with two free parameters $\phi_{c}$ and $\gamma$. In practice, this
is a delicate task as one needs to have a relaxation time running
over many decades to be able to chose unambiguously the couple $\{
\phi_{c},\gamma \}$ (see for example
\cite{PhysRevLett.105.199605}). A way to help this procedure is to
look at the four-point dynamic susceptibility $\chi_4(t)$. Within MCT,
the maximum $\chi_4^*$ should diverge~\cite{berthier2007spontaneous2} like $\chi_4^*(\phi)\sim
\left(\frac{\phi-\phi_{c}}{\phi_{c}}\right)^{-1}$
. This gives us  an independent
measure of the $\gamma$ exponent, using the relation $\chi_4^*\sim
\tau_{\alpha}^{1/\gamma}$ if there is a region where mode-coupling
scaling holds. We can then look for the pair of parameters  $\{\phi_{c},\gamma
\}$ giving the best fit for both measures.
\begin{figure}
  \centering
  \hspace{-1.cm}
  \includegraphics[width=0.48\columnwidth]{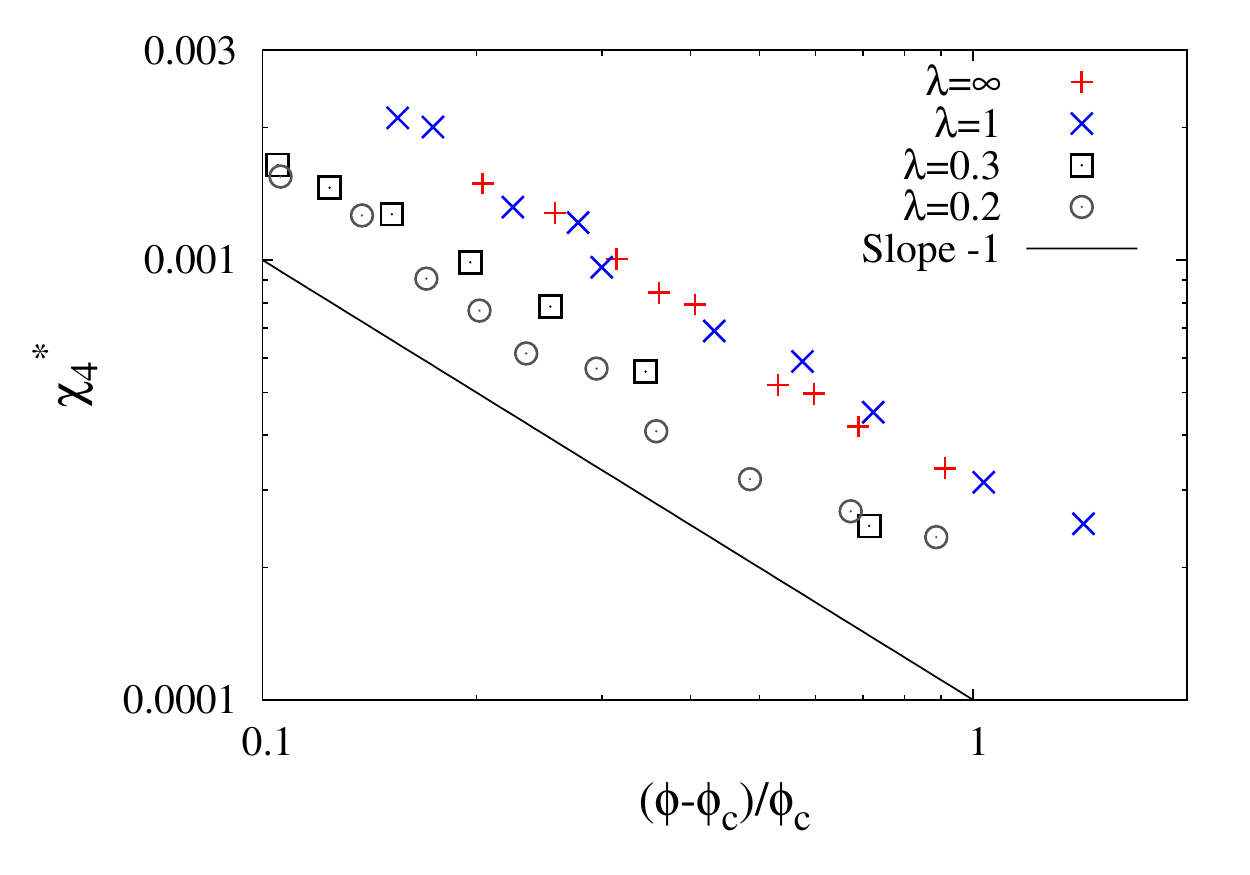}
  \hspace{1.cm}
  \includegraphics[width=0.48\columnwidth]{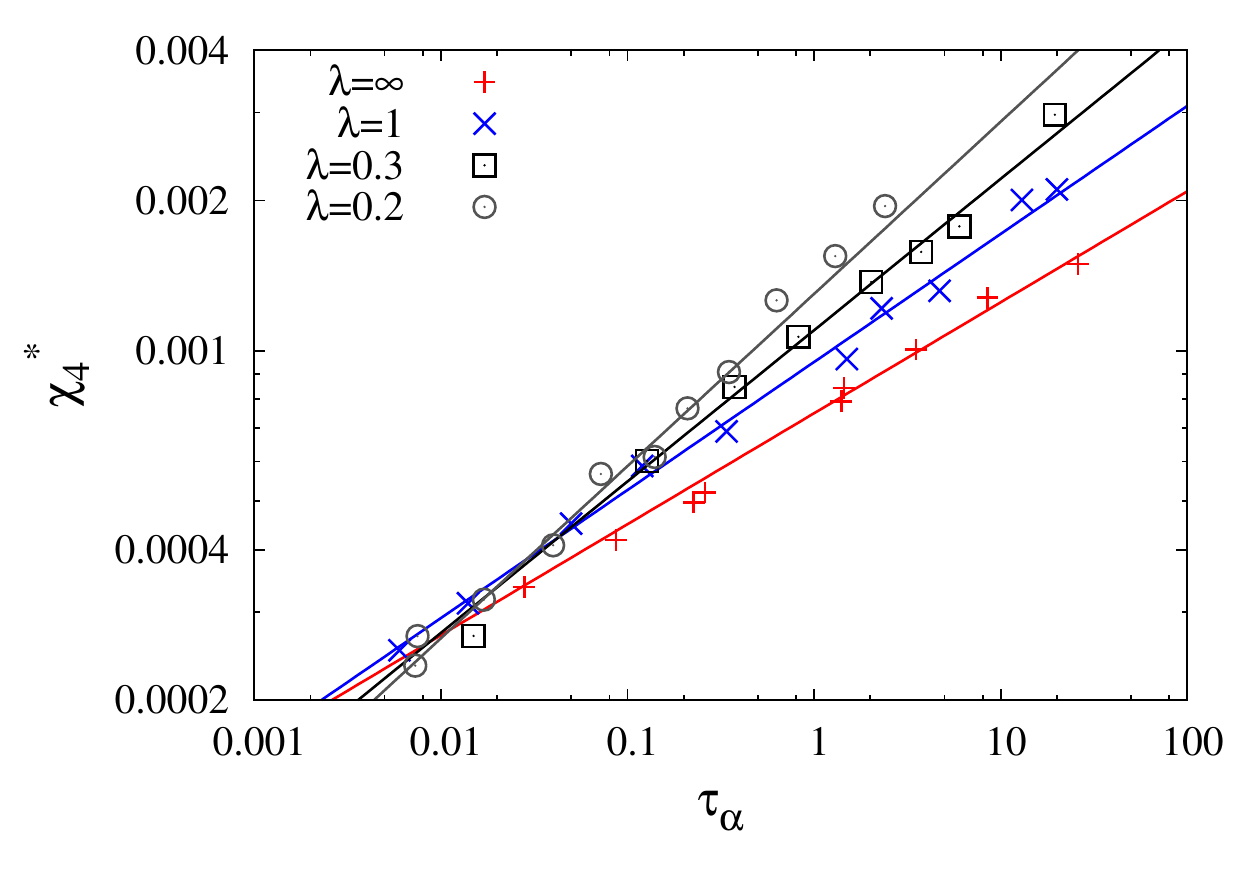}
  \caption{\textbf{Left:} Peak of the dynamic susceptibility $\chi_4^*$ as a function of the distance to the dynamic glass transition for different values of $\lambda$. The MCT-like  scaling $\left(\frac{\phi-\phi_{c}}{\phi_{c}}\right)^{-1}$ is clearly visible. \textbf{Right:} $\chi_4^*$ as a function of $\tau_{\alpha}$ in the same regime. The power law relation obtained in MCT  also seems to hold here for every values of $\lambda$. }
\label{fig:X4_vs_tau}
\end{figure}
\begin{figure}
  \includegraphics[width=0.7\columnwidth]{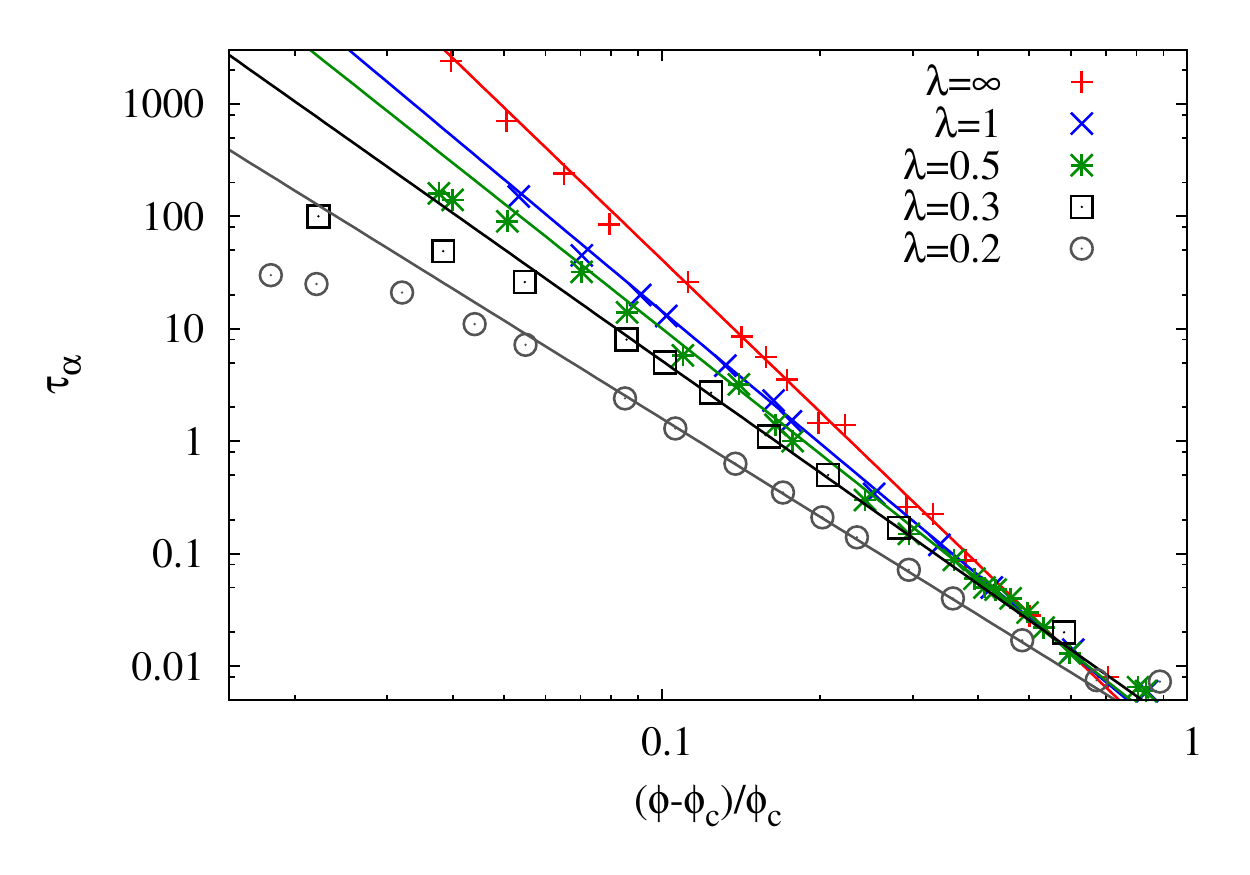}
  \caption{$\tau_{\alpha}$ as a function of $\frac{\phi-\phi_{c}}{\phi_{c}}$ for $\phi_{c}$ giving the broadest range of densities verifying MCT-like  behaviour. The agreement is perfect for values of $\lambda$ down to $0.5$, 
whereas for $\lambda=0.2$, the dynamic transition seems to be missed.}
\label{fig:tau_MCT}
\end{figure}

The results are shown in Fig~\ref{fig:X4_vs_tau} and Fig.~\ref{fig:tau_MCT}.
We get an excellent agreement with power law divergence on 6 decades of relaxation time
for the mean-field model, as expected. For small values of $\lambda$, however, we
observe a deviation to the power law scaling when we get close to the
transition, presumably due to activated processes, a fact that is well attested  for a $3d$ binary hard sphere
system~\cite{PhysRevLett.102.085703}. This seems to confirm the idea
that mean field theories give  correct qualitative features for the relaxation time,
but  only over a limited range of density not too close to  the glass transition. Quite interestingly,
intermediate values of $\lambda$ show a good  power-law scaling up to
$\lambda=1$, and for $\lambda>1$  we were not able to observe any evidence of activation 
 within the range
of relaxation times that are reachable with our simulations. 

\begin{figure}
  \centering
  \hspace{-1.cm}
  \includegraphics[width=0.48\columnwidth]{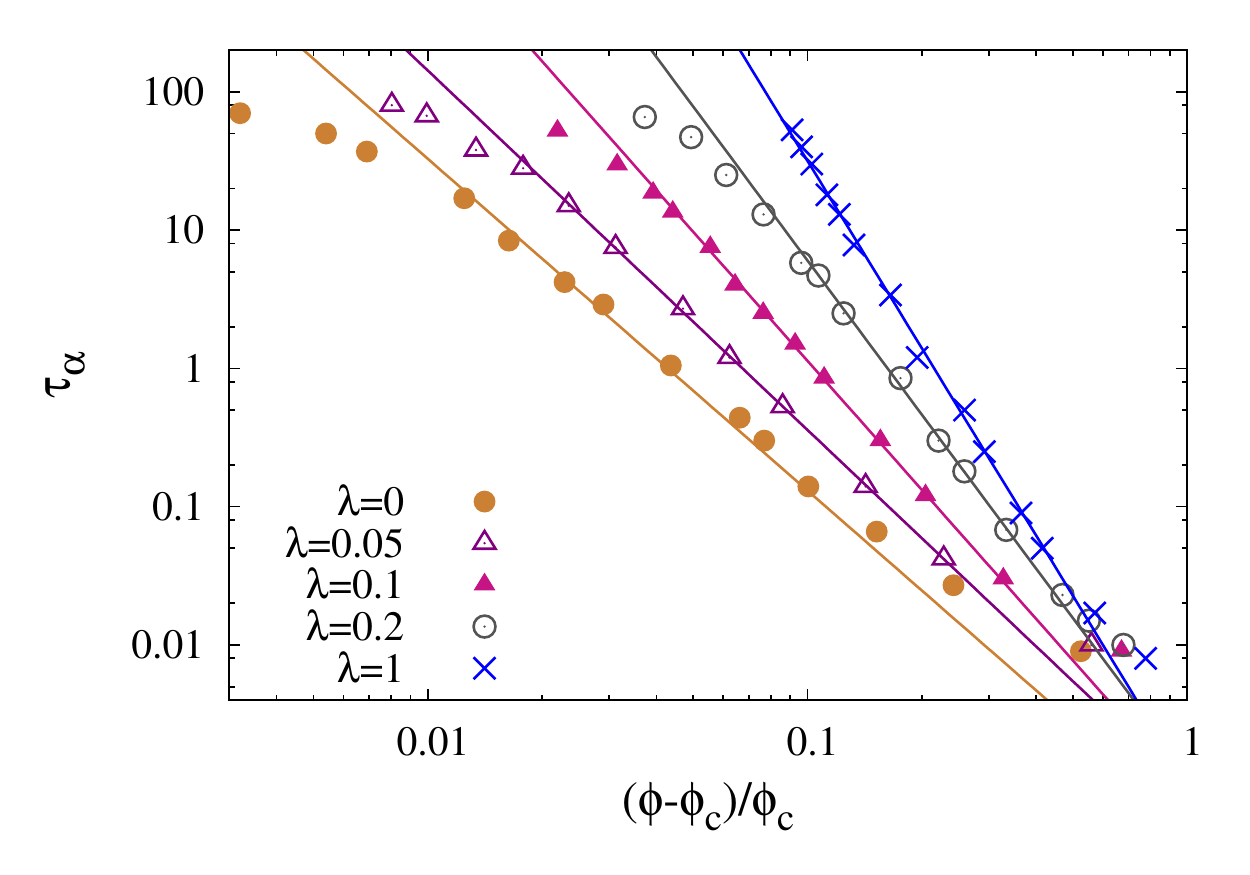}
  \hspace{1.cm}
  \includegraphics[width=0.48\columnwidth]{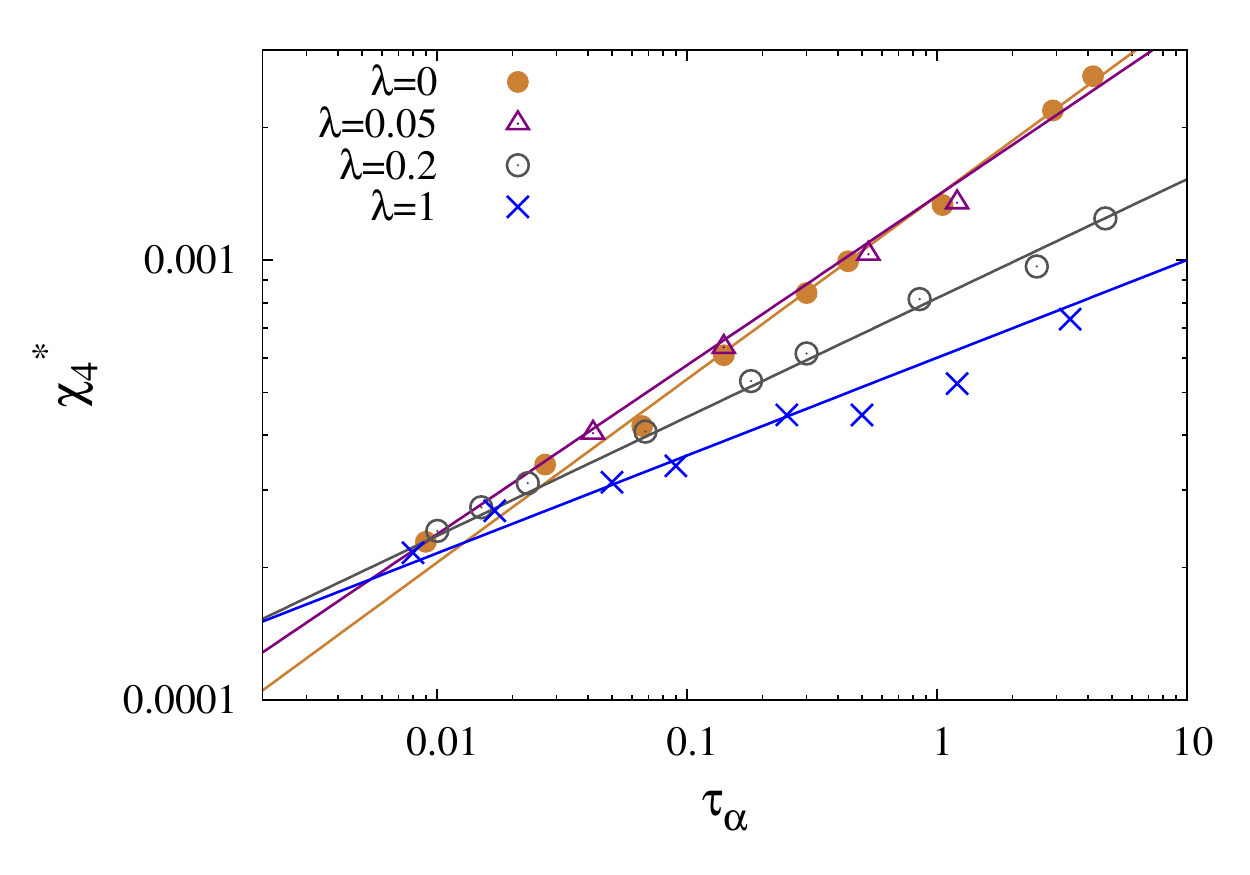}
  \caption{\textbf{Left:} $\tau_{\alpha}$ as a function of
    $\frac{\phi-\phi_{c}}{\phi_{c}}$ for $\phi_{c}$ giving the
    broadest range of densities verifying MCT-like  behaviour. The agreement
    is perfect for values of $\lambda$ down to $0.5$, whereas for
    $\lambda=0.2$, the dynamic transition seems to be missed.}
\label{fig:tau_MCT_bi}
\end{figure}

\begin{figure}
  \centering
  \hspace{-0.9cm}
  \includegraphics[width=0.48\columnwidth]{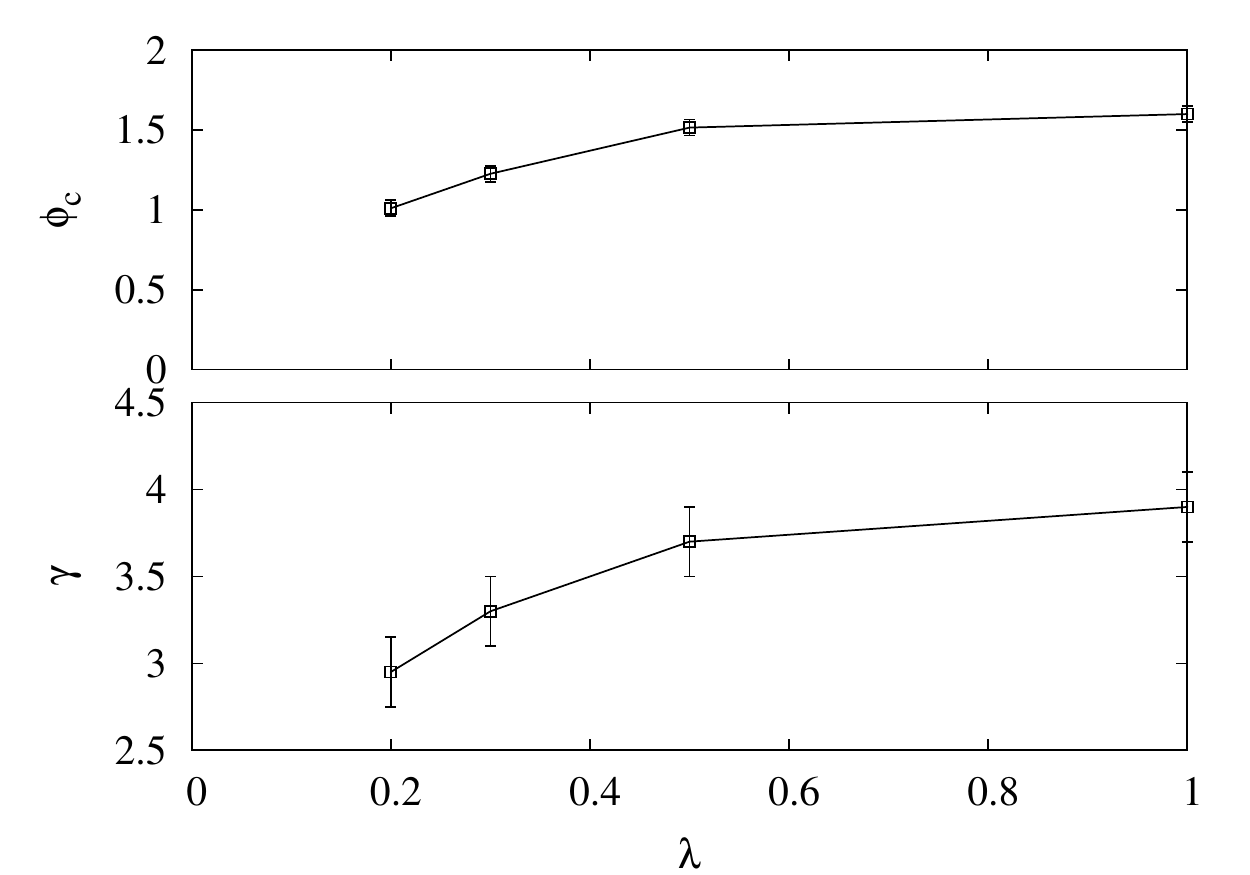}
  \hspace{1.cm}
  \includegraphics[width=0.48\columnwidth]{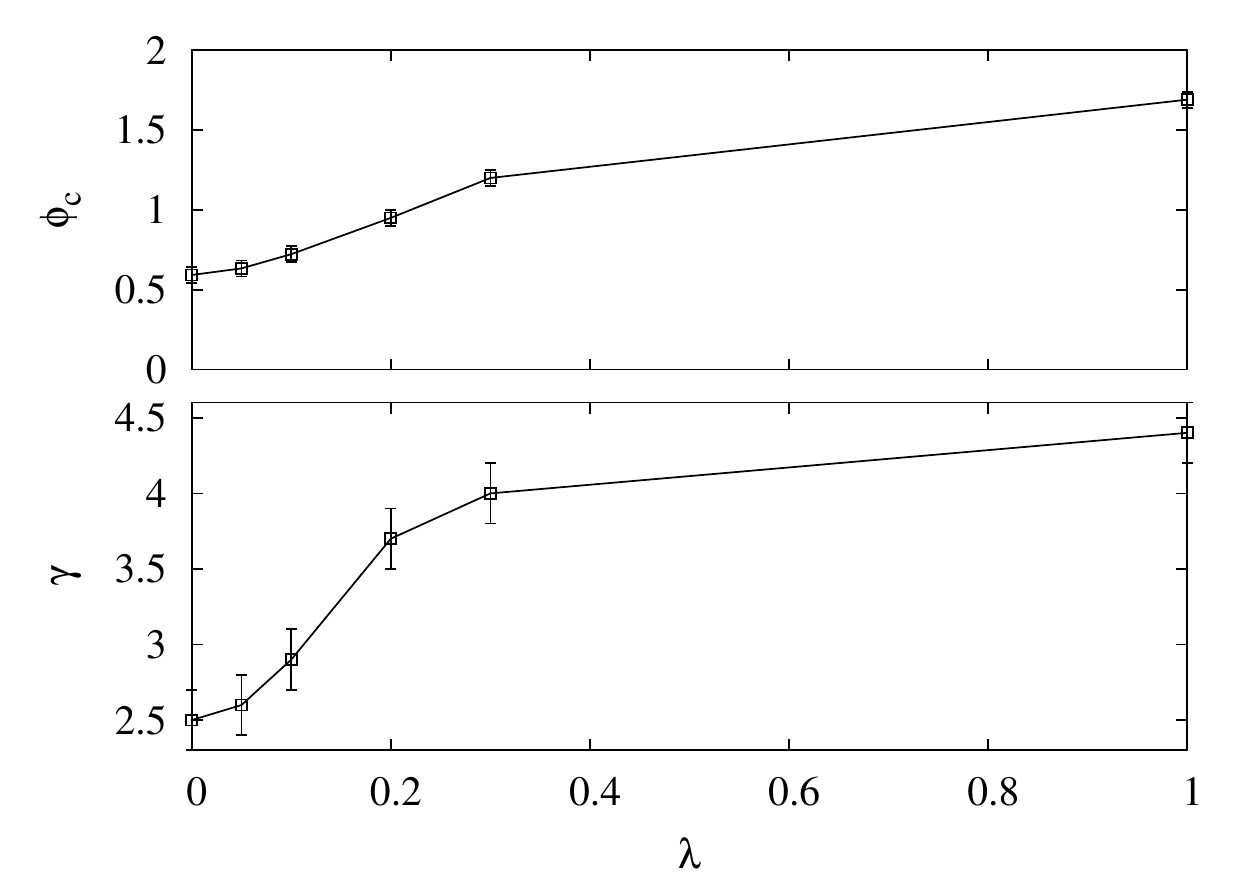}
  \caption{The values of the parameters of the MCT-like divergence,
    for different values of $\lambda$, in the monodisperse (left) and
    bidisperse (right) cases. These quantities show a continuous
    behavior from the $3d,\; \lambda=0$ system to the mean field
    $\lambda=\infty$ one.}
\label{fig:gamma_vs_lambda}
\end{figure}

One would like to take these simulations  all the way  down to
$\lambda=0$. Unfortunately, this is not possible with a monodisperse
system, but one can use a bidisperse system and follow the same
steps. This is what is done in Fig.~\ref{fig:tau_MCT_bi}. 
Here we see that the effect is more clear:
in term of relaxation time, the algebraic divergence is followed on
two decades when $\lambda=0$, but extends to more than three decades
when $\lambda=1$.
In figure
\ref{fig:gamma_vs_lambda} we show the behavior of the extrapolated dynamical  transition density and
the $\gamma$ exponent as a function of  $\lambda$. We observe that as 
$\lambda$ is lowered,  these values continuously approach the known values for $3d$
systems, both in the monodiperse~\cite{van1994glass} and in the
bidisperse case~\cite{PhysRevLett.102.085703}.

\begin{figure}
  \centering
  \hspace{-0.5cm}
  \includegraphics[width=0.48\columnwidth]{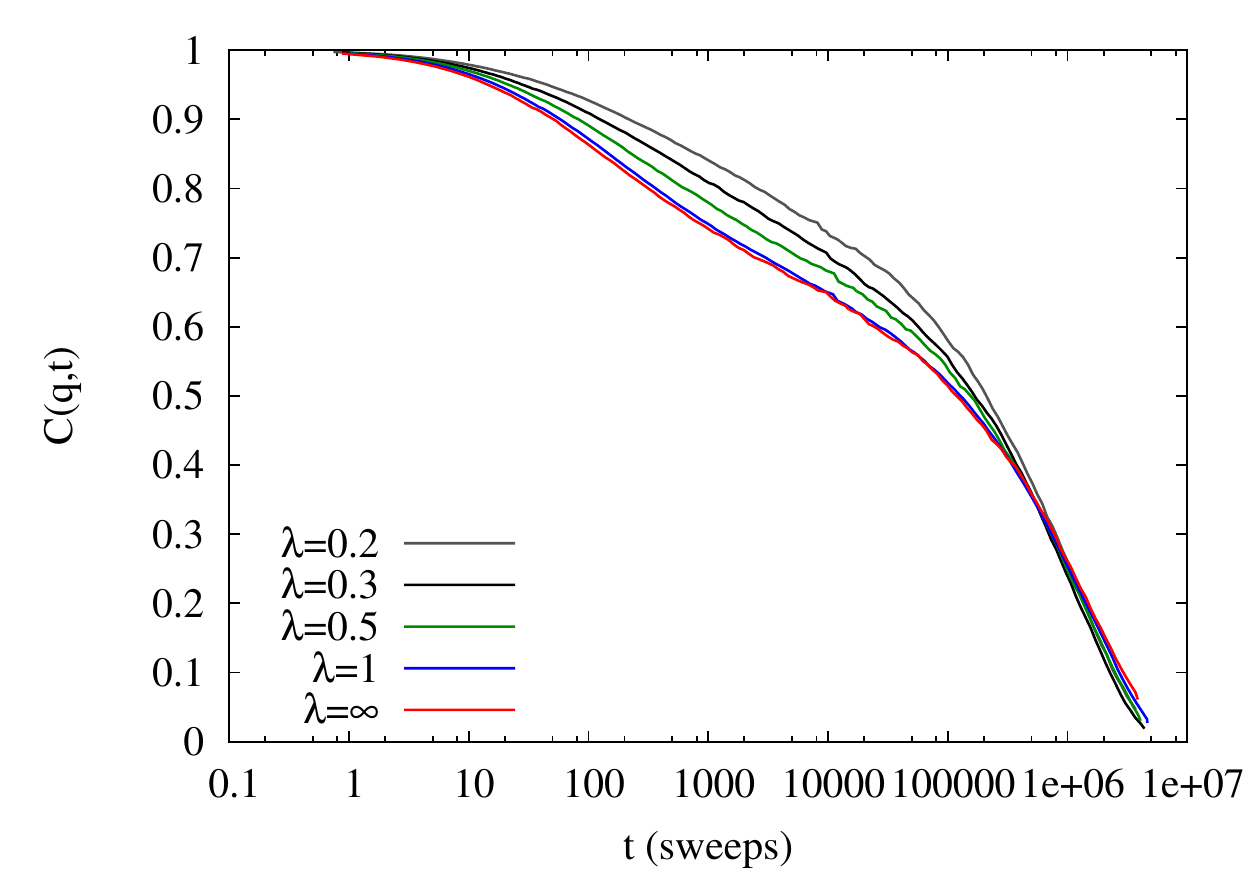}
  \hspace{0.5cm}
  \includegraphics[width=0.48\columnwidth]{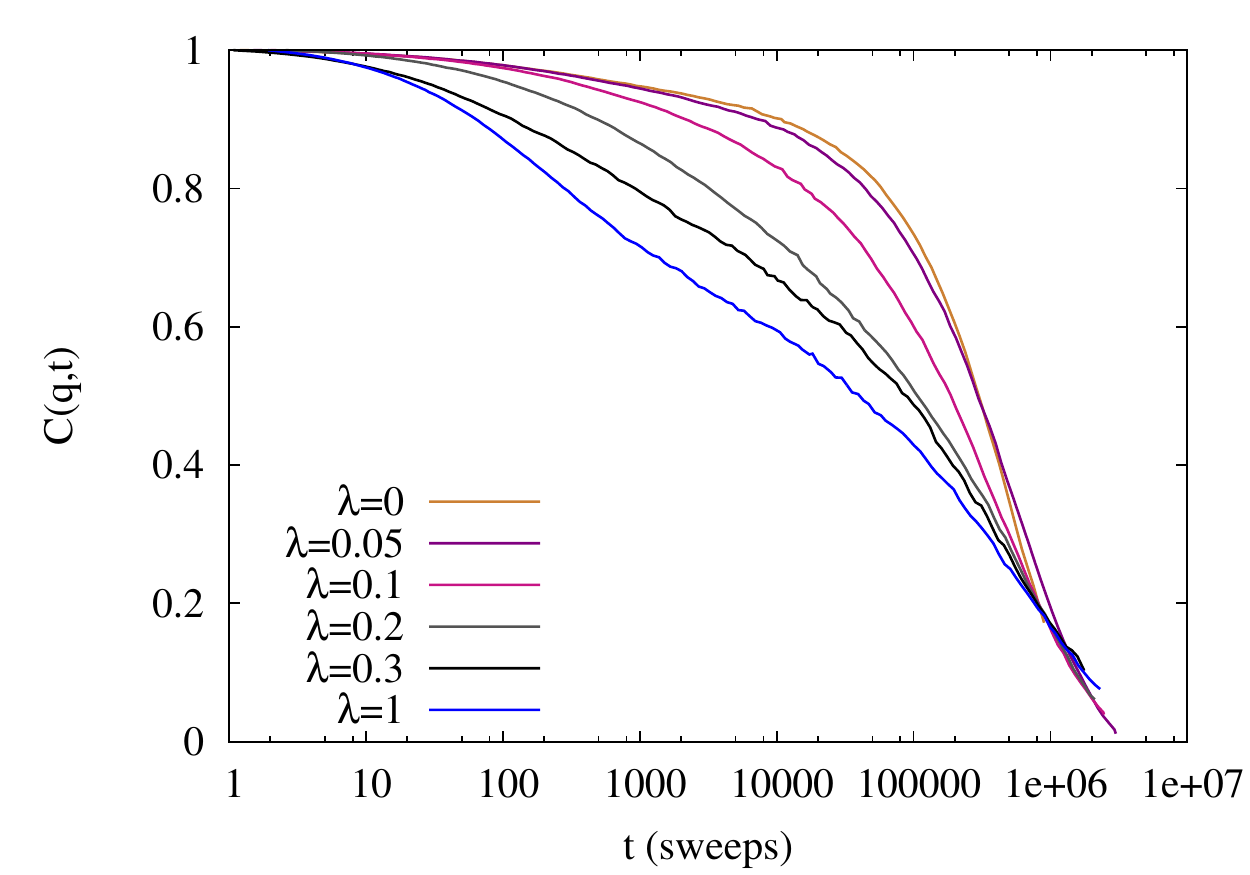}
  \caption{Relaxation curves $C(q,t)$ with different values of the
    range of random shifts $\lambda$, for equal relaxation
    times. \textbf{Left:} monodisperse case.  The main evolution while
    tuning $\lambda$ is a lengthening the
    $\beta$-relaxation. \textbf{Right:} Bidisperse case. The evolution
    is emphasized here. The shape of the $\alpha$-relaxation depends
    varies strongly between the mean-field case and small values of
    $\lambda$.}
\label{fig:correl_compare}
\end{figure}

The relaxation curves for the correlations contain  information beyond that of the timescale $\tau_\alpha$.
The shape of the relaxation curves does depend on the range of the
disorder, as is presented on Fig. \ref{fig:correl_compare}, with the
$\beta$-relaxation part becoming slower  when $\lambda$ decreases, 
both in the case of monodisperse and of bidisperse systems.  There is no visible
difference between $\lambda=1$ and $\lambda=\infty$, yet  another
indication that for this range of disorder, the system behaves in a
mean-field manner. 
For small $\lambda$, we do not observe a complete
separation between $\alpha$ and $\beta$-relaxations, and the plateau is
not well defined. These features however should strongly depend on
the microscopic dynamics, and it is known that Monte-Carlo dynamics
leads to a longer $\beta$-relaxation than molecular dynamics (see for example
\cite{kob2002supercooled}).

 Remarkably, the $\alpha$-relaxation 
shape is almost insensitive to the value of $\lambda$ in the
monodisperse case. We can just barely  notice an increase of the slope when
$\lambda$ gets smaller, a feature that appears much more clearly in the
binary mixture case, as we can access lower $\lambda$ values.

\subsection{Small or large exponent?}

We wish to stress again  that our goal is this work is to
perform a data analysis of the same type as performed with MCT, and in
particular we wanted to have an exponent $\gamma$ that fits the
divergence of $\tau_{\alpha}$ on the broadest possible range
\textit{and} satisfies the relation between $\chi_4^*$ and
$\tau_{\alpha}$. It is clear that by relaxing the constraint given by $\chi_4^*$ , we can fit the divergence of $\tau_{\alpha}$ on a
broader range, and possibly on the whole available range, just
by taking an  exponent of the order of four, and a larger density for the
divergence point (see for instance~\cite{PhysRevLett.105.199605}).

 Indeed, if one looks at the relation $\chi_4^*$ versus $\tau_{\alpha}$
very close to the transition (something we have not done here but
can be found  in Fig.~3 of Brambilla \textit{et
  al.}~\cite{PhysRevLett.102.085703}), one can fit the relation with a
larger $\gamma$ on a restricted range.

Then, one can legitimately think that the exponent $\gamma \simeq 4.5$
we find in the mean-field limit  is also the one of the $3d$ hard sphere system, as long
as one accepts to relax the constraint on the (mode-coupling inspired) relation of  $\chi_4^*$ versus
$\tau_{\alpha}$. 
Therefore, what our work proves is that,
provided we restrict ourselves to orthodox MCT fitting, the
glass transition is more and more mode-coupling-like when we approach
mean-field and the part of mode-coupling divergence which is observed
in finite dimension is somehow a shadow of the mean-field one. What is
appealing in this point of view is that the analysis gives for the
$3d$ bidisperse case an exponent $\gamma \simeq 2.5$, which is
compatible with what has already been found in experiments and
simulations of comparable
systems~\cite{PhysRevE.58.6073,PhysRevLett.102.085703,el2009dynamic},
and is reasonably close to the real MCT (not only its phenomenology)
for a monodisperse system~\cite{van1993glass}.

\subsection{Analytic estimation  of the dynamic transition}

One quick way to determine dynamic (mode coupling-like) transition
points, is to make a static calculation of the equilibrium state,
considered as an ensemble of metastable ergodic components; or,
equivalently, to look for the lowest pressure at which the {\em
  effective potential} (free-energy at fixed distance between
configurations) still has two minima.
  
In Appendix A we do this, leading to the determination in the figure
below:
\begin{figure}
  \centering
  \includegraphics[width=0.48\columnwidth]{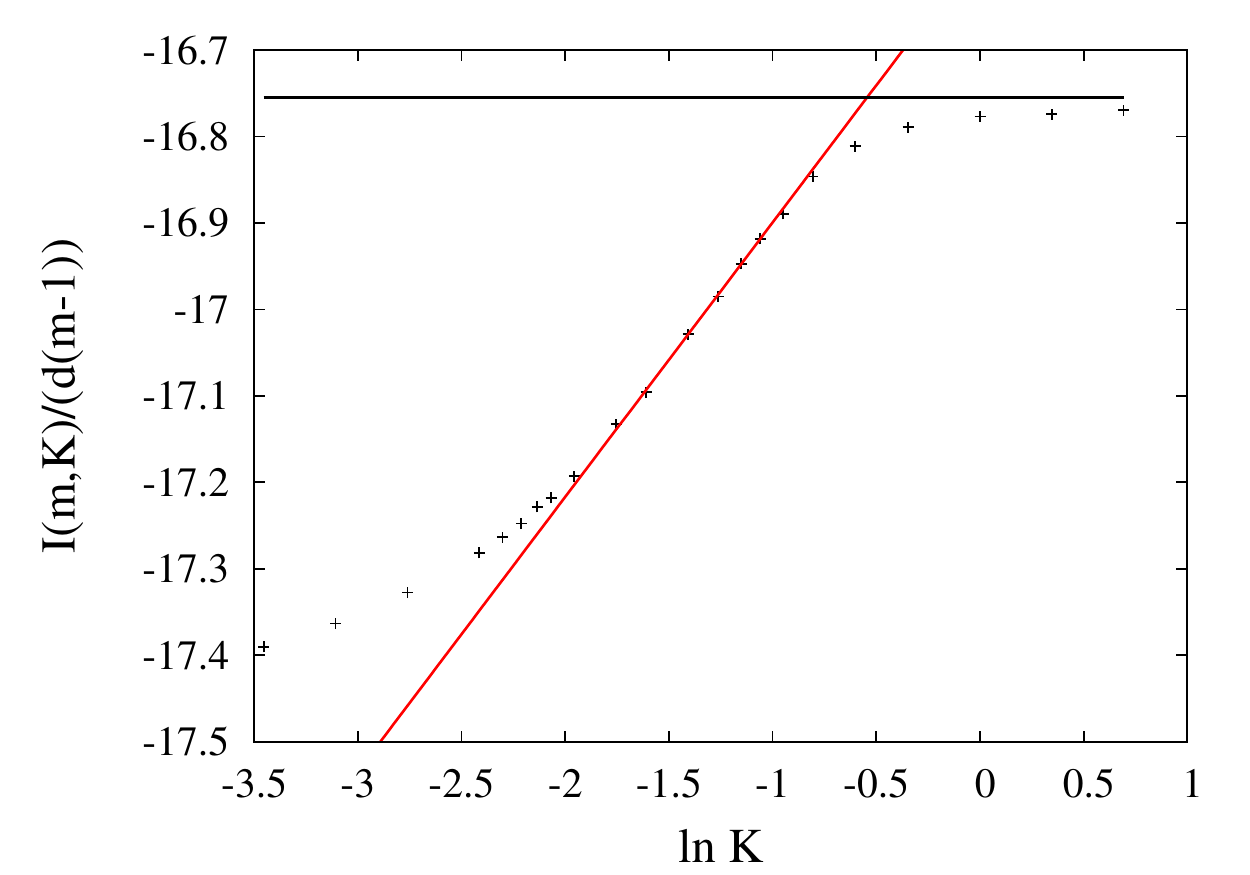}  
    \includegraphics[width=0.48\columnwidth]{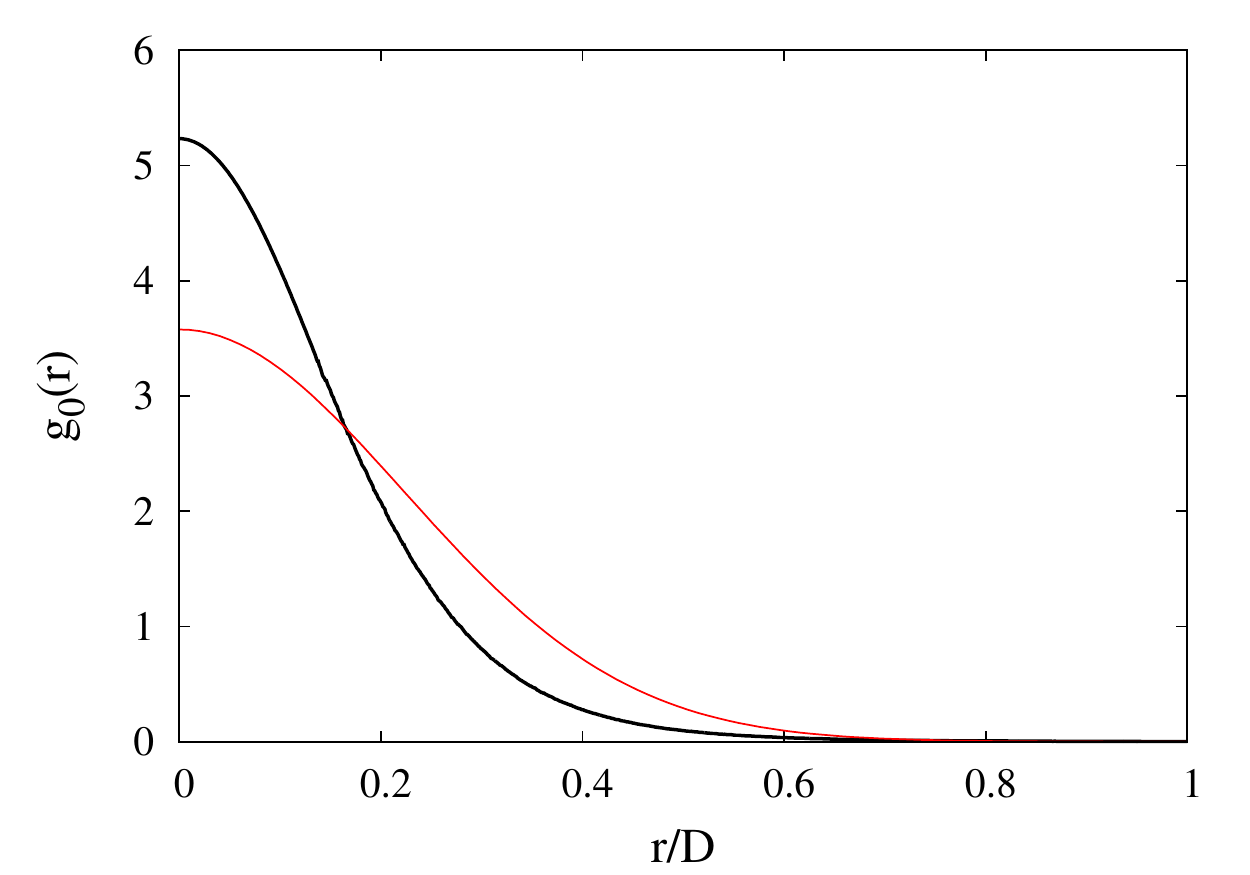}
    \caption{\textbf{Left:} The function defined in Eq.(\ref{bla}), in
      terms of the "cage size" parameter $K$, for $m=0.96$.  The
      solution for the transition pressure is obtained by searching
      for the point with the largest gradient, which is $1/\rho_d$.
      One also obtains the estimated cage size at this
      point. \textbf{Right:} Density profiles of a single particle in
      its cage, defined by Eq.~(\ref{eq:cage_density}), obtained from
      simulations (in black), and from analytical approach with a
      Gaussian ansatz (in red).}
  \label{fig:dynamic_transition_gaussian}
\end{figure}
The estimated transition density is $\phi_d=2^{-d}\rho_d= 1.65$ (see
left-hand side of figure~\ref{fig:dynamic_transition_gaussian}) to be
compared with the value estimated numerically $\phi_d=1.82$. Note that
the analytic computation is not exact because it relies on a Gaussian
approximation.

The estimated cage size is also consistent with the simulated
values. In the right-hand side figure~\ref{fig:dynamic_transition_gaussian}, we plot the
quantity:
\begin{equation}\label{eq:cage_density}
  g_0(\mathbf{r})= \frac{1}{N} <\sum_i \delta\left[(\mathbf{x}_i-<\mathbf{x}_i>)-\mathbf{r}\right]>
\end{equation}
\noindent which is as the density profile of a single particle in its
cage, and is the long time limit of the so-called van Hove
self-correlation function.

\section{Onset Pressure}\label{sec:onset}

A decade ago, Sastry \textit{et al.} introduced a  temperature
scale $T_{on}>T_g$ in supercooled liquids, the so-called onset
temperature~\cite{sastry1998signatures}. They noticed that in a
Lennard-Jones binary mixture, the energy of the inherent structures
(the configurations reached after a quench at $T=0$) associated with
equilibrium configurations at a temperature $T$ shows a crossover from
a roughly constant value above $T_{on}$ to a regime where it decreases
when $T$ decreases. They argued that this temperature is the one at
which the dynamics becomes landscape-influenced, with a
super-Arrhenius dependence of the relaxation time.  It was a few years
later argued~\cite{brumer2004mean} that there is a connection between
$T_{on}$ and the computed mode-coupling density temperature $T_c$,
opening the possibility that in mean-field systems $T_{on}$ and
$T_g(=T_c)$ might coincide. The question is legitimate, since one expects that 
for the spherical $p$-spin glass, the two temperatures do indeed coincide.

\begin{figure}
  \centering
  \includegraphics[width=0.7\columnwidth]{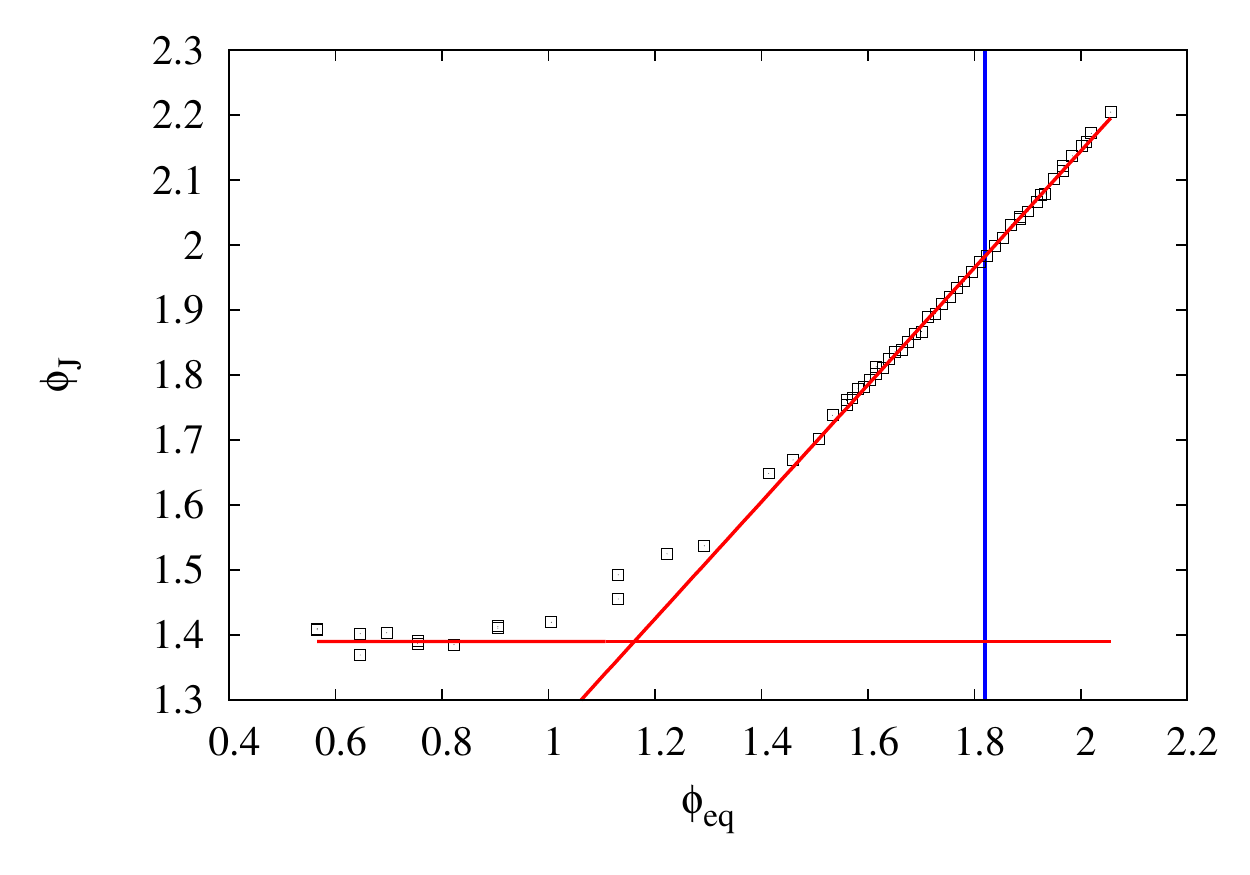}
  \caption{Inherent structure density $\phi_J$ as a function of the
    equilibrium liquid density $\phi_{eq}$. The crossover density is
    clearly different from the dynamic glass transition point,
    indicated by the vertical blue line.}
\label{fig:onset}
\end{figure}

In this section, we show that this connection does not exist for our
model.  Because we are dealing with hard spheres, an inherent
structure is the (infinite pressure) configuration reached after a
rapid compression  process, like the one introduced
in~\cite{stillinger1964systematic}. In figure ~\ref{fig:onset} we
plot the inherent structure density reached after such a process, in
terms of the initial density, {\em for the model with
  $\lambda=\infty$} . The crossover, corresponding to the `onset
density' (which is the relevant quantity for hard spheres, instead of
temperature, see previous section) is clearly visible.  The dynamic
transition in this mean-field limit ($d=3$) occurs at density
$\phi_g=1.82$, while we find here
$\phi_{on}\simeq 1.15$.  Because $\phi_g$ is a well defined quantity
here, we see beyond doubt that both densities do not coincide.
As expected, the crossover density coincides with  the point where
the relaxation (\ref{eq:def_correlation}) starts to show a
shoulder. Thus, also in the mean-field limit, $\phi_{on}$ marks the
onset of a qualitative change in the dynamics (see
Fig.\ref{correl_RS10}), which becomes landscape influenced.

\section{Conclusions}

We have introduced an approximation scheme for particle systems that
is close in spirit to the Mode-Coupling approximation, but has the
advantage that one may construct a continuous range of models, with at
one end the original one, and at the other end one for which the
approximation is exact.  The approximation becomes better at higher
dimensionality of space, everything else remaining equal.

As it stands, the present scheme is derived from the microscopic
model. This was originally the case also with the Mode-Coupling
equations, although the standard practice has become to modify freely
the interactions in such a way as to obtain the observed static
structure factor and transition temperature or pressure.  In our case,
the analogous procedure would be to substitute the true potential by
one based on the pair correlation function
(\ref{eq:pair_correl_shifted_def})
\begin{equation}
V_{eff} = - T \ln[g(r)]
\end{equation}
We have not tried this strategy.

In this paper we have not discussed the possible static Kauzmann transition, and we suspect that  it might happen at divergent pressures, of the order of $\ln \lambda$, \textit{i.e.} divergent in the mean-field limit.

By following the model from the original problem to its mean-field
limit, we have used the present construction to give new arguments on
the existence of a vestige of the genuine mean-field dynamic
transition, following a mode-coupling-like behavior. We also argued
conclusively that the `onset' temperature (or pressure) should not be
identified with the dynamic transition.  More generally, one may
follow this strategy to decide whether features found in true system
that one `explains' within random first order theory, really
extrapolate to the corresponding feature in the limit in which the
theory is exact.

\section*{Acknowledgements}

We would like to thank Ludovic Berthier, Florent Krzakala, Marco
Tarzia and Francesco Zamponi for discussing with us various points of
this work.

\section*{Appendix A: An estimation of the dynamical transition pressure from replicas}\label{sec:ideal_glass}

\subsection*{Physical discussion}

The dynamic glass transition is related to the existence of an
exponential number of amorphous metastable states. Above a given
dynamic glass transition density $\phi_d$, the liquid phase can be
seen as the sum of all these states. 

To derive the properties of the model at high density in order to test
the above scenario, we will study the partition function of $m$ copies
of the original system:
\begin{equation}\label{eq:m_replicated_partition}
  Z_m=\int  \prod_{\alpha=1}^{m} \prod_{k} \dd \mathbf{x_k}^{\alpha} \exp \(( -\sum_{\alpha=1}^{m}\sum_{ij}V(\mathbf{x_i}^{\alpha}-\mathbf{x_j}^{\alpha}-\mathbf{A_{ij}}) \))    
\end{equation}

The idea is the following~\cite{monasson1995structural,
  mezard1996tentative, mezard1999thermodynamics, mezard1999first}: if
we force the $m$ copies to be close one another  by adding a
small coupling term between them, we can expect that if we study the
system at a density higher than $\phi_d$, and switch off the coupling
after taking the thermodynamic limit, the $m$ copies will be confined in
the same metastable state.

This is done in practice by looking at the entropy of the replicated system:
\begin{equation}\label{eq:replicated_entropy_def}
  S=\overline{\ln Z_m}=\lim_{n\to 0} \frac{1}{n}\left(\overline{Z_m^n}-1 \right)
\end{equation}
where we use the replica trick to average over disorder the logarithm of $Z_m$.
Looking more closely to $\overline{Z_m^n}$, we see that it can be interpreted as
the partition function of $N$ `molecules'
$\underline{\mathbf{x_i}}=\{\mathbf{x}_i^1,...,\mathbf{x}_i^{nm}\}$ made
of $nm$ spheres:
\begin{equation}\label{eq:mn_replicated_partition}
\begin{array}{rl}
  \overline{Z_m^n} & =\int \prod_{ij} \dd \mathbf{A_{ij}} P(\mathbf{A_{ij}}) \int  \prod_{\alpha=1}^{mn} \prod_{k} \dd \mathbf{x_k}^{\alpha} \exp \(( -\sum_{\alpha=1}^{mn}\sum_{ij}V(\mathbf{x_i}^{\alpha}-\mathbf{x_j}^{\alpha}-\mathbf{A_{ij}}) \))\\
& =\int \prod_{k} \dd \underline{\mathbf{x_k}} \prod_{ij}\int \dd \mathbf{A_{ij}} P(\mathbf{A_{ij}}) \exp \(( -\sum_{\alpha=1}^{mn}\sum_{ij}V(\mathbf{x_i}^{\alpha}-\mathbf{x_j}^{\alpha}-\mathbf{A_{ij}}) \))
  \end{array}
\end{equation}
Then,  using standard liquid theory techniques, we are able to write the
entropy~Eq.~(\ref{eq:replicated_entropy_def}) as a functional of the
density of molecules $\rho(\underline{\mathbf{x}})$. The metastable
states of the system are then the maxima of this entropy with respect
to $\rho(\underline{\mathbf{x}})$ when $m=1$, which we expect to
couple replicas.

\subsection*{Replicated canonical formalism}

We have to study the following partition function
Eq.~(\ref{eq:mn_replicated_partition}).
Introducing $\underline{\mathbf{x_i}}=\{\mathbf{x_i}^1,...,
\mathbf{x_i}^{nm}\}$, we can see
Eq.~(\ref{eq:mn_replicated_partition}) as the partition function of a
$N$ molecules interacting via a potential $\tilde{V}$ such that:
\begin{equation}\label{eq:replicated_mayer}
 \left[ 1+ \overline{f}_r(\underline{\mathbf{x}}-\underline{\mathbf{y}})\right]=\exp\left(-\tilde{V}(\underline{\mathbf{x}}-\underline{\mathbf{y}})\right)=\int P(\mathbf{A}) \dd \mathbf{A} \exp \left( -\sum_{\alpha=1}^{mn}\sum_{ij}V(\mathbf{x}^{\alpha}-\mathbf{y}^{\alpha}-\mathbf{A}) \right)
\end{equation}
Each `molecule' $\underline{\mathbf{x_i}}$ is made of $n$ independent
sets of $m$ coupled (in the same state) original particles.
Then, Eq.~(\ref{eq:mn_replicated_partition}) reads:
\begin{equation}\label{eq:mn_replicated_partition2}
  \overline{Z_m^n} = \int \prod_{i} \dd \underline{\mathbf{x_i}} \prod_{ij}\left[ 1+ \overline{f}_r(\underline{\mathbf{x_i}}-\underline{\mathbf{x_j}})\right]
\end{equation}

We wish  to do a Mayer expansion like in the non-replicated case,
but we cannot add a convenient combinatoric prefactor in
(\ref{eq:mn_replicated_partition2}), as we do not know what it should be
\textit{a priori}. However the specific properties of the
random-shift model allows us  to do a canonical treatment of the
partition function as we show in the following.
Introducing the density of molecules $\rho(\underline{\mathbf{x}})$ by:
\begin{equation}
  \rho(\underline{\mathbf{x}})=\sum_i \delta(\underline{\mathbf{x}}-\underline{\mathbf{x_i}})  
\end{equation}
 we can rewrite $\overline{Z_m^n}$ as:
\begin{equation}
  \overline{Z_m^n} =  \int \dd \underline{\mathbf{x_i}} \int \DD[\rho( \mathbf{x})]  \delta \left(\rho( \mathbf{\underline{x}})-\sum_i \delta(\mathbf{\underline{x}}-\mathbf{\underline{x}}_i) \right) 
\exp \left[\frac{1}{2} \int \text{d} \mathbf{\underline{x}} \text{d} \mathbf{\underline{y}}  \rho(\mathbf{\underline{x}}) \rho(\mathbf{\underline{y}}) \ln\left[ 1+ \overline{f}_r(\underline{\mathbf{x}}-\underline{\mathbf{y}})\right] \right]
\end{equation}
We may  now exponentiate the $\delta$ constraint at the cost of adding a second 
field $\hat{\rho}$, and integrating over the $\underline{\mathbf{x_i}}$'s:
\begin{equation}
  \begin{array}{rl}
    \overline{Z_m^n} =  \int & \DD[\rho( \underline{\mathbf{x}})] \DD[\hat{\rho}( \underline{\mathbf{x}})] \\
& \exp\left\{ i \int \text{d} \underline{\mathbf{x}} \hat{\rho}( \underline{\mathbf{x}}) \rho( \underline{\mathbf{x}}) + N \ln \int  \text{d} \underline{\mathbf{x}} e^{-i \hat{\rho}( \underline{\mathbf{x}})}+ \frac{1}{2} \int \text{d} \underline{\mathbf{x}} \text{d} \underline{\mathbf{y}}  \rho(\underline{\mathbf{x}}) \rho(\underline{\mathbf{y}}) \ln\left[ 1+ \overline{f}_r(\underline{\mathbf{x}}-\underline{\mathbf{y}})\right] \right\}
  \end{array}
\end{equation}

 The next important step is to look more closely at the
function
$\overline{f}_r(\mathbf{\underline{x}}-\mathbf{\underline{y}})$, as we did 
in the non-replicated case. In the integral of
Eq.~(\ref{eq:replicated_mayer}), the exponential of  the potential
$V$ vanishes $1$ whenever the two molecules in
$\mathbf{\underline{x}}$ and $\mathbf{\underline{y}}+\mathbf{A}$
overlap, and gives $1$ otherwise. Then we have:
\begin{equation}
  -nm\frac{v_d}{V} \leq  \overline{f}_r(\underline{\mathbf{x}}-\underline{\mathbf{y}})\leq -\frac{v_d}{V} 
\end{equation}
Thus, by expanding the logarithm, we get:
\begin{equation}
  \begin{array}{rl}
    \overline{Z_m^n} =  \int & \DD[\rho( \underline{\mathbf{x}})] \DD[\hat{\rho}( \underline{\mathbf{x}})] \\
    & \exp\left\{  i\int \text{d} \underline{\mathbf{x}} \hat{\rho}( \underline{\mathbf{x}}) \rho( \underline{\mathbf{x}}) + N\ln \int  \text{d} \underline{\mathbf{x}} e^{-i \hat{\rho}( \underline{\mathbf{x}})}+ \frac{1}{2} \int \text{d} \underline{\mathbf{x}} \text{d} \underline{\mathbf{y}}  \rho(\underline{\mathbf{x}}) \rho(\underline{\mathbf{y}}) \overline{f}_r(\underline{\mathbf{x}}-\underline{\mathbf{y}}) \right\}
\end{array}
\end{equation}
We wish then to evaluate this integral by   saddle point  with
respect to the fields $\rho$ and $\hat{\rho}$ (because each term in
the exponential is of order $N$, including the last one, due to the
value of $f$). This gives:
\begin{equation}\label{eq:saddle}
  \begin{array}{rl}
    \rho( \mathbf{\underline{x}}) & = N\frac{e^{-i \hat{\rho}( \mathbf{\underline{x}})}}{\int \text{d} \mathbf{\underline{x}} e^{-i \hat{\rho}( \mathbf{\underline{x}})} } \\
    \hat{\rho}( \mathbf{\underline{x}}) & = i  \int \text{d} \mathbf{\underline{y}} \rho(\mathbf{\underline{y}})\overline{f}_r(\mathbf{\underline{x}}-\mathbf{\underline{y}})
  \end{array}
\end{equation}
 Thus, the logarithm of the partition function is:
\begin{equation}\label{eq:replicated_FT_partition}
  \ln \overline{Z_m^n} = - \int \text{d} \mathbf{\underline{x}} \rho( \mathbf{\underline{x}}) \ln \rho( \mathbf{\underline{x}}) + \frac{1}{2} \int  \text{d} \mathbf{\underline{x}} \text{d} \mathbf{\underline{y}} \rho( \mathbf{\underline{x}}) \rho( \mathbf{\underline{y}}) f(\mathbf{\underline{x}}-\mathbf{\underline{y}})+N\ln N
\end{equation}
 We wish to stress the similarity of
Eq.~(\ref{eq:replicated_FT_partition}) with
Eq.~(\ref{eq:F_mean_field}). In fact the average over disorder does
exactly the same job for the replicated liquid than for the bare
non-replicated liquid: it disallows `three-molecule effective
interactions', just like in a high-dimensional system.

\subsection*{Location of the dynamic transition, within the Gaussian  ansatz}

We may obtain an approximation for $\rho$ by assuming it has a Gaussian
form.  Of course, this ansatz will not  be a solution of the full saddle
point equation Eq.~(\ref{eq:saddle}), but we can still extremize the
entropy Eq.~(\ref{eq:replicated_FT_partition}) within this ansatz. A
natural choice is the  ansatz\cite{parisi2010mean}:
\begin{equation}
  \label{eq:ansatz}
  \rho( \mathbf{\underline{x}})=\frac{N}{V^{n}}\prod_{\gamma=1}^n
\int \dd \mathbf{X_{\gamma}} \frac{1}{(2\pi K)^{md/2}}\exp \left[ \sum_{\alpha=\gamma (m-1)+1}^{\gamma m} \frac{\left(\mathbf{x^{\alpha}} - \mathbf{X_{\gamma}}\right)^2}{2K} \right]
\end{equation}
Note the similarity with the dynamic ansatz.

This is a 1-step replica symmetry breaking (1-RSB) ansatz.  Injecting
Eq.~(\ref{eq:ansatz}) in the partition function
Eq.~(\ref{eq:replicated_FT_partition}) leads to integrals exactly
similar to the ones one has compute in the hard sphere system (without
random shifts).  This has been done in \cite{parisi2010mean}.
Following the computations along the lines of \cite{parisi2010mean},
one finds:
\begin{equation}\label{eq:action_gaussian}
  \begin{array}{rl}
   \frac{S[\rho( \mathbf{\underline{x}})]}{N}= \ln N+ 1-\ln \rho - \frac{d}{2}(1-m) \ln (2\pi K) + \frac{d}{2}\ln m - \frac{d}{2}(1-m) - \frac{\rho}{2} I(m,K)
  \end{array}
\end{equation}
where $I(m,K)$ is the integral:
\begin{equation}
  I(m,K)=\int \dd \mathbf{X} \left[\int \dd \mathbf{x} \dd \mathbf{y} \frac{1}{(2\pi K)^{d}}\exp \left(\frac{\left(\mathbf{x} - \mathbf{X}\right)^2}{2K} \right)\exp \left(\frac{\mathbf{y}^2}{2K} \right) \chi(\mathbf{x}-\mathbf{y}) \right]^m
\end{equation}
Then, the saddle point equation on $K$ reads:
\begin{equation}
  \frac{d(m-1)}{\rho}= \frac{\partial I(m,K)}{\partial \ln K}
\end{equation}

 Thus, there exist metastable states in the liquid phase
(which is recovered in the limit $m\to 1$) if there is a cage size $K$
which verifies:
\begin{equation}
  \frac{1}{\rho}=\lim_{m \to 1}\frac{1}{d(m-1)} \frac{\partial I(m,K)}{\partial \ln K}
  \label{bla}
\end{equation}
The dynamic glass transition density is $\rho_d$, beynd  which no
solution to this equation can be found.  It is possible to compute
$I(m,K)$ numerically, and we find $\phi_d=2^{-d} \rho_d \simeq 1.65$
for $d=3$ (see Fig.~\ref{fig:dynamic_transition_gaussian}), not far
from the value ($1.82$) found in numerical simulations. The value of
$K$ at the transition is also comparable with what is found in the
simulations (see Fig.~\ref{fig:dynamic_transition_gaussian}).

\section*{Appendix B : Gaussian ansatz for the mean-field dynamics}

\subsection{Notation}

The calculation we are going to follow is quite heavy. It may be made
somewhat more compact, and one may follow the analogy with the static
treatment better, by using the supersymmetric
notation~\cite{zinn2002quantum} (see
\cite{kurchan1992supersymmetry,semerjian2004stochastic}).  One
introduces two extra Grassmann variables $\theta$ et
$\overline{\theta}$. Denoting $ a=(t,\theta,\overline{\theta})$, the
trajectories $\mathbf{x}(t)$ and $\hat{\mathbf{x}}(t)$ may be encoded
in a superfield $\psi(a)$:
\begin{equation}
  \psi(a)=\mathbf{x}(t)+ \theta \overline{\theta} \hat{\mathbf{x}}(t)
\end{equation}
Defining the operator  $D_a$  as
\begin{equation}
  D_a=T\frac{\partial^2}{\partial \theta  \partial \overline{\theta}}+\theta\frac{\partial^2}{\partial \theta  \partial t}-\frac{\partial}{ \partial t}
\end{equation}
we have:
\begin{equation}
\Phi[\mathbf{x}, \hat{\mathbf{x}}] =  \int \dd a \dd b \; \delta(b-a) D_a  \left(\psi(a)-\psi(b)\right)^2 
\end{equation}
Similarly:
\begin{equation}
  \begin{array}{rl}
  1+\overline{f}_d[\mathbf{x},\hat{\mathbf{x}},\mathbf{x}',\hat{\mathbf{x}}']&= 
  \int \dd \mathbf{A} P(\mathbf{A}) \exp \left\{ - \frac{1}{2} \int_0^{\tau} \dd t\; \hat{\mathbf{x}}(t) \nabla_{\mathbf{x}} V(\mathbf{x} - \mathbf{y}-\mathbf{A}))+\hat{\mathbf{y}}(t) \nabla_{\mathbf{y}} V(\mathbf{y} - \mathbf{x} - \mathbf{A})  \right\} \\
&=\int 
    \dd \mathbf{A} P(\mathbf{A}) \exp \left\{-\frac{1}{2} \int \dd a\;  V\left(\psi(a)-\psi'(a)-\mathbf{A} \right) \right\}\\
&=1+\overline{f}_d[\psi,\psi']
  \end{array}
\end{equation}
 The action can be written in the following compact way:
 \begin{equation}\label{eq:susy_action}
  \begin{array}{rl}
   \mathcal{S}\left[\rho [\psi ] \right] =  &\int \DD \psi \rho[\psi] \ln \left[\rho[\psi] \right] +\int \DD \psi \rho[\psi] \int \dd a\; \psi(a) D_a \psi(a) \\
& \qquad - \frac{1}{2} \int \DD[\psi,\psi'] \rho[\psi] \rho[\psi']\overline{f}_d[\psi-\psi']-N\ln N
\end{array}
\end{equation}
By analogy with the statics, one may make a `Gaussian' ansatz
$\rho$~\cite{semerjian2004stochastic}:
\begin{equation}
  \rho[\psi] = \frac{\sqrt{2}}{V\; \det^{\frac 1 2}(\mathcal{B})} \int d{\bf{\bar x}} \; \exp \left[ - \int d a d b  \; {\mathcal B}^{-1}(a,b) \;  (\psi(a)-\mathbf{\bar{x}})(\psi(b)-\mathbf{\bar{x}})  \right]
\end{equation}
with
\begin{equation}
\mathcal{B} (a,a)=0  \;\;\; \forall \;\;\; a
\end{equation}
In components, the super-correlators read:
\begin{eqnarray}
 \mathcal{B}(a,b)&=&{ B}(t_a,t_b) - \overline{\theta}_a \theta_a { R} (t_b,t_a)
  -  {\bar \theta}_b  \theta_b { R}(t_a,t_b) + { D} (t_a,t_b) {\bar \theta}_a \theta_a {\bar \theta}_b \theta_b\nonumber \\
  \mathcal{B}^{-1}(a,b)&=&{\tilde B}(t_a,t_b) - \overline{\theta}_a \theta {\tilde R} (t_b,t_a)
  -  {\bar \theta}_b  \theta_b {\tilde R}(t_a,t_b) + {\tilde D} (t_a,t_b) {\bar \theta}_a \theta_a {\bar \theta}_b \theta_b
  \label{inverse1}
\end{eqnarray}
and they are related through:
\begin{equation}
\int \dd{\bar \theta}_b \dd \theta_b \dd t_b \;  \mathcal{B}(a,b) \mathcal{B}^{-1}(b,c)= \delta({\bar \theta_a}-{\bar \theta_c})
\delta( \theta_a-{\theta_c}) \delta(t_a-t_c)
\end{equation}
This inversion formula may be developed, to obtain the "tilde" variables in terms of the ones without tilde.

\noindent In equilibrium, the fluctuation-dissipation theorem implies that ${\cal B}$ and ${\cal B}^{-1}$ take the form:
\begin{eqnarray}
 \mathcal{B}(a,b)&=&{ B}(t_a-t_b) - \overline{\theta}_a \theta_a { R} (t_b-t_a)
  -  {\bar \theta}_b  \theta_b { R}(t_a-t_b) \nonumber \\
  \mathcal{B}^{-1}(a,b)&=&{\tilde B}(t_a-t_b) - \overline{\theta}_a \theta_a {\tilde R} (t_b-t_a)
  -  {\bar \theta}_b  \theta_b {\tilde R}(t_a-t_b) 
\end{eqnarray}
where 
\begin{equation}
  R(t) = -\frac{ 1}{T} \frac{\partial B}{\partial t} \;\;\;  ; \;\;\; {\tilde R}(t)= -\frac{ 1}{T} \frac{\partial{\tilde  B}}{\partial t}
\end{equation}
where $R$ and ${\tilde R}$ are zero for negative time-differences, due to causality.

\noindent It is easy to show that:
\begin{equation}
 \Phi[\mathbf{x}, \hat{\mathbf{x}}] =  \frac{1}{V} \int \dd {\bf{\bar x}} \;  \int \dd a \dd b \; \delta(b-a) D_a  (\psi(a)-\mathbf{\bar{x}})(\psi(b)-\mathbf{\bar{x}}) 
\end{equation}
The super-correlator  $\mathcal{B}$ is given by:
\begin{equation}
  \mathcal{B}(a,b)=\int \text{D}\psi  \left(\psi(a)-\psi(b)\right)^2\rho[\psi]
\end{equation}

\noindent Inserting this ansatz in Eq.~(\ref{eq:susy_action}), we get:
\begin{equation}
 \mathcal{S}\left( \mathcal{B} \right)=-\frac{1}{2} \text{Tr}\ln \mathcal{B}+ \int \dd a \dd b\; \delta(a-b)D_a\mathcal{B}(a,b)-\mathcal{S}_{int}(\mathcal{B})
\end{equation}
with
\begin{equation}
  \begin{array}{rl}
    \mathcal{S}_{int}(\mathcal{B})&=  \frac{1}{V^2 \det \mathcal{B}} \int \DD [\psi, \psi'] \int \dd \bar{\mathbf{x}} \dd \bar{\mathbf{y}} \\
    & \quad \exp \left[  - \int \dd a \dd b \;  \mathcal{B}^{-1}(a,b) \left[(\psi(a)-\bar{\mathbf{x}})(\psi(b)-\bar{\mathbf{x}})+(\psi'(a)-\bar{\mathbf{y}})(\psi'(b)-\bar{\mathbf{y}})\right] \right]\;  \overline{f}_d[\psi,\psi'] 
  \end{array}
\end{equation}
The integrand is invariant with respect to independent translations of ${\bf x} $ and $ {\bf y}$, because they may be absorbed into the shift $\mathbf{A}$, so we may write:
\begin{equation}
  \mathcal{S}_{int}(\mathcal{B})=   \frac{1}{\det \mathcal{B}} \int \DD [\psi, \psi'] \; \exp \left[  - \int \dd a \dd b \; \mathcal{B}^{-1}(a,b) \left[\psi(a)\psi(b)+\psi'(a)\psi'(b)\right] \right]\; \overline{f}_d[\psi,\psi']
\end{equation}
The saddle point equation gives:
\begin{equation}\label{eq:dynamics_saddle}
  0=\frac{\delta \mathcal{S}}{\delta \mathcal{B}(a,b)} = -\frac{1}{2}\mathcal{B}^{-1}(a,b)+ \delta(a-b)D_{a}- \frac{\delta \mathcal{S}_{int}}{\delta \mathcal{B}(a,b)}
\end{equation}
with:
\begin{equation}
  \frac{\delta \mathcal{S}_{int}}{\delta \mathcal{B}(a,b)}= \frac{1}{2}\mathcal{B}^{-1}(a,b) - \left[ \mathcal{B}^{-1}\otimes\langle \psi(a')\psi(b') \rangle_{int}\otimes  \mathcal{B}^{-1} \right](a,b)
\end{equation}
 and:
\begin{equation}
   \langle \bullet \rangle_{int}  = \int \DD \psi \int \DD \psi' \bullet \rho[\psi]
\rho[\psi']\; \overline{f}_d[\psi,\psi']
\end{equation}
Making the convolution product with $\mathcal{B}(b,c)$, we obtain:
\begin{equation}\label{eq:super_dynamics}
  0=    D_a\mathcal{B}(a,b) + \int \dd c \; \Sigma(a,c)\mathcal{B}(c,b),
\end{equation}
where $\Sigma$ is given by 
\begin{equation}
  \Sigma(a,b)= - 2\left[ \mathcal{B}^{-1}\otimes \langle \psi(a') \psi(b')\rangle_{int} \otimes  \mathcal{B}^{-1} \right](a,b),
\end{equation}
 also of the form (\ref{inverse1}):
\begin{equation}
 {\Sigma}(a,b)={ \Sigma_B}(t_a-t_b) - \overline{\theta}_a \theta_a { \Sigma_R} (t_b-t_a)
  -  {\bar \theta}_b  \theta_b { \Sigma_R}(t_a-t_b) 
\end{equation}
$\Sigma_B$ and $\Sigma_R$  satisfy also a fluctuation-dissipation relation:
\begin{equation}
\Sigma_R(t) = -\frac{ 1}{T} \frac{\partial \Sigma_B}{\partial t} 
\end{equation}
Because correlation and response satisfy a fluctuation-dissipation relation, we may write everything
exclusively in terms of correlations:
\begin{equation}
  \frac{\partial B(t_a-t_b)}{\partial t_a}=-TR(t_b-t_a)+\int \dd t_c \; \Sigma_R(t_a-t_c) B(t_c-t_b)-\left.\Sigma_C(t_b-t_c) B(t_b-t_c)\right|_{-\infty}^{t_b},
\end{equation}
with:
\begin{equation}
   \Sigma_R(t_a-t_b)= \frac{2}{T}\int \dd t_{a'} \dd t_{b'} \;  {R}^{-1} (t_a-t_{a'})
\;  \frac{ \partial }{ \partial t_{a'}}\langle    \mathbf{x}(t_{a'}) \mathbf{x}(t_{b'}) \rangle_{int}  \;   {R}^{-1}(t_{b'}-t_b)
\end{equation}
Given that, as $t_a \rightarrow t_b$, $B(t_a-t_b) \sim 2T  |t_a-t_b|$, we have that:
\begin{equation}
  \begin{array}{rl}
  \frac{\partial B(t_a,t_b)}{\partial t_a}&=-TR(t_b,t_a)+\int  \dd t_c
\; \Sigma_R(t_a - t_c) B(t_c - t_b) +2T \\
2T&=-\left.\Sigma_B(t_b- t_c) B(t_b- t_c)\right|_{-\infty}^{t_b},
  \end{array}
\end{equation}
which is the result (\ref{eq:dynamic_equation}).


%

\end{document}